\documentclass[preprint,12pt,authoryear]{elsarticle}

\usepackage{amssymb}
\usepackage{amsmath}
\usepackage{amsthm}
\usepackage{booktabs}
\usepackage{algorithm}
\usepackage{algpseudocode}
\usepackage{multirow}
\usepackage{xcolor} 

\usepackage{amsmath}
\usepackage{amsfonts}
\usepackage{graphicx}
\usepackage{natbib}
\usepackage{graphicx}
\usepackage{subcaption}
\usepackage{graphicx}      
\usepackage{subcaption}    
\usepackage{float}         
\definecolor{ForestGreen}{RGB}{34,139,34}

\begin{document}

\begin{frontmatter}


\title{Multi-Hypothesis Prediction for Portfolio Optimization: A Structured Ensemble Learning Approach to Risk Diversification} 




\author[addr1,addr2]{Alejandro Rodriguez Dominguez}
\ead{arodriguez@miraltabank.com,a.j.rodriguezdominguez@pgr.reading.ac.uk}

\author[addr1]{Muhammad Shahzad\corref{mycorrespondingauthor}}
\ead{m.shahzad2@reading.ac.uk}

\author[addr1]{Xia Hong}
\ead{xia.hong@reading.ac.uk}

\address[addr1]{University of Reading, Department of Computer Science, Whiteknights House, Reading, RG6 6UR, United Kingdom}
\address[addr2]{Miralta Finance Bank S.A.,28020, Madrid, Spain }

\begin{abstract}

This work proposes a unified framework for portfolio allocation, covering both asset selection and optimization, based on a multiple-hypothesis predict-then-optimize approach. The portfolio is modeled as a structured ensemble, where each predictor corresponds to a specific asset or hypothesis. Structured ensembles formally link predictors' diversity, captured via ensemble loss decomposition, to out-of-sample risk diversification. A structured data set of predictor output is constructed with a parametric diversity control, which influences both the training process and the diversification outcomes. This data set is used as input for a supervised ensemble model, the target portfolio of which must align with the ensemble combiner rule implied by the loss. For squared loss, the arithmetic mean applies, yielding the equal-weighted portfolio as the optimal target. For asset selection, a novel method is introduced which prioritizes assets from more diverse predictor sets, even at the expense of lower average predicted returns, through a diversity–quality trade-off. This form of diversity is applied before the portfolio optimization stage and is compatible with a wide range of allocation techniques. Experiments conducted on the full S\&P 500 universe and a data set of 1,300 global bonds of various types over more than two decades validate the theoretical framework. Results show that both sources of diversity effectively extend the boundaries of achievable portfolio diversification, delivering strong performance across both one-step and multi-step allocation tasks.

\end{abstract}






\end{frontmatter}


\section{Introduction}

The problem of portfolio allocation with a quantitative approach was introduced by Harry Markowitz in 1952 in the so-called Modern Portfolio Theory (MPT) \citep{https://doi.org/10.1111/j.1540-6261.1952.tb01525.x}. It consists of two phases: asset selection and portfolio weight optimization. The asset selection phase decides which assets from a larger set, such as a market, will be included in the portfolio, which has a fixed number of assets but not predetermined choices or weights. The portfolio weight optimization phase determines the percentage of capital allocated to each selected asset. It also has the merit of introducing the concept of diversification as a fundamental component in risk management for investments. In this specific case, diversification is induced by the covariance matrix \citep{https://doi.org/10.1111/j.1540-6261.1952.tb01525.x}. Since then, a large number of methods known as "plug-in" (or data-driven) approaches have been developed which use optimization inputs with improved estimates of future returns  \citep{b8473306-c76a-342a-a678-7bb18db4be6a, 10.1007/BF02282040, annurev:/content/journals/10.1146/annurev-financial-101521-104735}. 

The standard data-driven approach to portfolio choice optimization replaces population parameters with their sample estimates, which leads to poor out-of-sample performance due to parameter uncertainty \citep{article2018}. This issue is typically addressed by deriving expected loss functions that quantify the impact of using sample means and covariance matrices to estimate the optimal portfolio \citep{Kan_Zhou_2007}. Other researchers have focused on portfolio optimization using robust estimators, which can be computed by solving a single non-linear programming problem \citep{DeMiguel2007PortfolioSW, articleLassance}. Shrinkage techniques adjust extreme coefficients toward more central values in the sample covariance matrix to reduce estimation error, often incorporating a parameter to control the level of shrinkage \citep{Ledoit2003,JMLR:v25:22-1337}. 

Some researchers have also employed the use of predictive models to optimize decision-making processes, often with the objective of minimizing decision-related costs.
Such a conventional "predict-then-optimize" paradigm decouples prediction and decision-making into two distinct stages: firstly, a predictive model is developed to maximize predictive accuracy; secondly, decisions are derived based on the model's outputs and associated cost functions. A significant limitation of this approach is its failure to incorporate cost considerations during the predictive modeling phase \citep{articleDonti}. To address this shortcoming, recent advances have introduced integrated methodologies such as "predict-and-optimize" \citep{articleWilder, Vanderschueren2022PredictthenoptimizeOP} and Smart "Predict, then Optimize" (SPO) \citep{Elmachtoub2017SmartT}. These approaches bridge the gap between structured prediction and cost-sensitive optimization, offering a robust solution for enhancing decision-making processes \citep{kotary2023predictthenoptimizeproxylearningjoint}. These approaches, however, are computationally expensive and mostly focus on solving the classification problems \citep{Goh2016StructuredPB}.

While the aforementioned solutions reduce parameter uncertainty and can improve out-of-sample performance, they fail to establish a connection between in-sample and out-of-sample diversification, which is a key feature addressed in this work.
In this context, a framework is proposed for portfolio optimization using structured ensembles for prediction with multiple hypotheses. Each portfolio component corresponds to a hypothesis, with its return series modeled by an individual predictor. This approach differs from return forecasting \citep{MA2021113973} or scoring models \citep{NGUYEN2012407,10.3389/fenrg.2022.852520}, where models are used to score assets and predict their returns before applying optimization techniques for portfolio allocation. 
Building on prior research, the degree of diversity in ensemble learning is parametrically controlled during predictor training and encoded in a structured data set of predictions \citep{10.1007/978-3-031-77915-2_7}. This data set serves as input for an ensemble model designed to predict portfolio returns. The ensemble model is optimized as a supervised prediction problem, using as a target a portfolio whose weights are chosen to match the ensemble combiner rule derived from the Bias-Variance-Diversity decomposition \citep{wood2023unified}. In the case of squared loss, the corresponding combiner is the arithmetic mean, implying that the equal-weighted portfolio should be selected. The optimal portfolio weights are then obtained by normalizing the optimal ensemble parameters. To the best of the authors' knowledge, this is the first method which connects ensemble learning diversity with portfolio risk diversification, enabling its parametric control before the optimization stage.


Additionally, this work also investigates the hypothesis that out-of-sample portfolio diversification relates to the asset or hypothesis selection. Diversity in asset selection is controlled parametrically, independently of the learning phase. A parameter regulates the diversity of asset sets—randomly selected from ranked out-of-sample return predictions—across various portfolio sizes and parameters. Results show that greater diversity in asset sets increases out-of-sample portfolio diversification and risk-adjusted returns, even when the sets have lower average returns. This work links both sources of diversity (i.e., asset selection and portfolio weight optimization) to out-of-sample portfolio diversification and demonstrates that the diversification levels from both sources can be parametrically controlled prior to decision-making, offering valuable insights for practitioners.

Following are the main contributions proposed in this work: 

\begin{itemize}
    \item A consistent framework is introduced for applying structured ensemble models in multi-hypothesis prediction settings for portfolio optimization, where the ensemble combiner rule defines the portfolio target. Specifically, in the case of squared loss, this corresponds to the equal-weighted portfolio.

    \item It presents a "plug-in" and robust method enabling parametric a priori control over out-of-sample portfolio risk diversification by linking ensemble diversity, derived from the Bias-Variance-Diversity loss decomposition \citep{wood2023unified}, to portfolio risk diversification.
    \item A strategy is proposed for parametrically controlling diversity in asset or hypothesis selection, leveraging a diversity–quality trade-off in predicted returns to enhance out-of-sample portfolio diversification.
    \item The expansion of diversification limits \citep{dalio2018principles} is validated by the proposed framework, as well as through the incorporation of diversity in the asset selection stage across other methodologies, demonstrated empirically in both one-step and multi-step decision settings.
\end{itemize}

The paper is organized as follows: Section \ref{literature} provides a review of previous work on data-driven and robust optimization, Predict-then-Optimize frameworks, structured ensemble learning, and diversity. Section \ref{framework} describes the proposed framework and methodology. Section \ref{numerical} presents the experimental results and discussions. Finally, in Section \ref{concl} the final remarks and future outlook are offered.

\section{Literature Review}
\label{literature}

Portfolio optimization solutions can be broadly categorized into data-driven approaches and robust optimization methods. Data-driven methods typically rely on plug-in optimization, where the objective function is predefined by frameworks such as Modern Portfolio Theory (MPT), and data is used solely to estimate model parameters. In contrast, robust optimization modifies the objective function and relaxes certain framework assumptions to improve adaptability while tackling uncertainty. This approach was among the first to incorporate loss functions, which were later extended to predictive models and predict-then-optimize frameworks, albeit with computational challenges. Ensemble-based methods further introduce loss functions which account for diversity in predictions. This work integrates ensemble models with multiple hypothesis prediction, establishing a connection between diversity in learning and risk diversification, while also introducing computationally efficient analytical solutions.

\subsection{Data-Driven and Robust Optimization}
Reducing uncertainty in the optimization stage is critical in many applications, and two prominent approaches are data-driven optimization and robust optimization. Data-driven optimization relies on data to estimate uncertainties, using statistical models or machine learning to incorporate these estimates into decision-making \citep{articleBertsimas,bookShapiro2009}. It aims to optimize expected performance metrics such as cost or profit by leveraging historical or real-time data. On the other hand, robust optimization focuses on worst-case scenarios, ensuring decisions remain feasible and effective for all outcomes within a predefined uncertainty set, such as polyhedral or ellipsoidal bounds \citep{RobustOptimization2009}. This approach prioritizes stability and reliability, often at the expense of conservatism if the uncertainty set is poorly calibrated \citep{doi:10.1137/080734510}. Recent advances combine the strengths of both approaches, such as distributionally robust optimization (DRO), which optimizes for the worst-case distribution within plausible distributions, bridging probabilistic and robust methodologies \citep{Esfahani2015DatadrivenDR,OJMO_2022__3__A4_0}. In the context of portfolio optimization, data-driven approaches leverage historical and real-time market data to model uncertainties, employing methods such as stochastic programming, machine learning, and empirical scenario generation to predict asset returns and covariance structures \citep{Roncalli2013IntroductionTR,annurev:/content/journals/10.1146/annurev-financial-101521-104735,MA2021113973,Ozelim2023}. Robust portfolio optimization
solutions include fuzzy approaches \citep{Wu2012}, genetic algorithms \citep{Akopov2014}, multiobjective particle swarm optimization \citep{CHEN2018165}, and learning-guided multiobjective evolutionary algorithms \citep{LWIN2014757}.  Building on these, this work presents a mixed method which combines both approaches, featuring a plug-in facet but robust in that the risk diversification level can be adjusted parametrically before optimization. Therefore, this addresses the disconnection between out-of-sample diversification and input data in the optimization process.



\subsection{Predictive Modeling Integrated with Decision Optimization}
Predictive models have become essential tools for optimizing decision-making processes, particularly when the objective is to minimize decision-related costs. However, this approach overlooks cost considerations during the predictive modeling phase, leading to suboptimal decision quality \citep{articleDonti}. Recent advancements have sought to address this issue through integrated frameworks like "Predict-and-optimize" (PO), Smart Predict-then-Optimize (SPO) and End-to-End Predict-Then-Optimize (EPO) \citep{Elmachtoub2017SmartT,doi:10.1287/mnsc.2020.3922,Vanderschueren2022PredictthenoptimizeOP}. These frameworks use surrogate loss functions to manage computational challenges associated with original loss functions and link structured prediction outputs to decision variables \citep{Elmachtoub2017SmartT}. The SPO+ framework further advances this integration by defining loss functions based on the objective costs of nominal optimization problems, improving computational efficiency and decision-aware modeling \citep{doi:10.1287/mnsc.2020.3922}. However, most existing studies in structured prediction focus on classification problems \citep{Goh2016StructuredPB} and are computationally intensive, as they require backpropagation through optimization solutions, which often involves handcrafted rules, and they face challenges in handling nonconvex or discrete optimization models \citep{kotary2023predictthenoptimizeproxylearningjoint}.

\subsection{Diversity in Multi-Hypothesis Learning Ensembles for Portfolio Diversification}

 A unified theory of ensemble diversity explains that diversity is a hidden dimension in the bias-variance decomposition of the ensemble loss \citep{wood2023unified}. Moreover, studies on diversity measures in ensemble learning show their significant impact on model accuracy, generalization, and robustness \citep{pmlr-v151-ortega22a}. 
 
 Researchers link portfolio estimation to statistical decision theory by treating the difference in investor utility between the "true" optimal portfolio and the plug-in portfolio as an economic loss function \citep{Kan_Zhou_2007}. Outcomes are enhanced by incorporating the investor's utility objective directly into the statistical weight estimation process as in Bayesian decision theoretic approaches, rather than addressing estimation and utility maximization as separate issues \citep{Avramov2010BayesianPA,Kan_Zhou_2007}. Research has demonstrated the economic implications of using Mean Square Error (MSE) in portfolio optimization and financial forecasting, showing that integrating MSE into decision-making processes can enhance portfolio performance by minimizing estimation errors and optimizing risk-return trade-offs \citep{doi:10.1080/0015198X.2020.1854543,file-23SKslwIN3yLKCPK1veP3z7i,Cai2024}. This work extends this idea by incorporating diversity into the MSE loss function within an ensemble learning framework. Furthermore, for the first time, it establishes a connection between this notion of diversity and portfolio diversification, proposing parametric methods to manage diversification prior to decision-making.

\section{Framework Description}
\label{framework}






\subsection{Preliminary}
\label{prelim}
Given \( \boldsymbol{x} \in \mathbb{R}^{N\times M} \) as a vector of returns with \( M \geq 2 \) assets (each with \textit{N} time-stamps) having mean \( \boldsymbol{\mu} \) and a positive-definite covariance matrix \( \boldsymbol{\Sigma} \), then the optimal portfolio \( \boldsymbol{w} \in \mathbb{R}^M \) can be found by solving the following optimization problem \citep{https://doi.org/10.1111/j.1540-6261.1952.tb01525.x}:

\begin{equation}
\begin{aligned}
& \min_{\boldsymbol{w}} \quad \left( \boldsymbol{w}^\top \boldsymbol{\mu} - r \right)^2 + \boldsymbol{w}^\top \Sigma \boldsymbol{w} \\
& \quad \boldsymbol{w}^\top \boldsymbol{\mu} \geq r \\
& \quad \boldsymbol{w}^\top \boldsymbol{1} = 1 \\
\end{aligned}    
\label{minprob}
\end{equation}

where \(\left( \boldsymbol{w}^\top \boldsymbol{\mu} - r \right)^2\) represents the bias of the expected return of the portfolio $\boldsymbol{w}^\top \boldsymbol{\mu}$ compared to the target return $r$, and \(\boldsymbol{w}^\top \Sigma \boldsymbol{w}\) represents the variance of the portfolio. The constraint \(\boldsymbol{w}^\top \boldsymbol{\mu} \geq r\) ensures that the expected return of the portfolio is above the target return. 
The expression in (\ref{minprob}) generalizes the standard MSE 
reflecting the bias-variance trade-off \citep{Cai2024}. 

If the aforementioned portfolio optimization can be formulated as a prediction task, then the objective utility function of the mean-variance in (\ref{minprob}) can be represented as an MSE loss function \citep{Cai2024}. 
This setup allows for finding the optimal weights 
which minimize the total prediction error of the portfolio, accounting for both bias and variance, while ensuring that the expected return remains above the target value.  
In order to incorporate diversity into such an optimization, this work applies ensemble models to harness diversity in ensemble learning and link it to portfolio diversification.


In a multiple-hypothesis prediction setup in which each hypothesis is focused on a particular portfolio constituent, the portfolio can be modeled as an ensemble of hypotheses. Following that, the diversity term in the Bias-Variance-Diversity decomposition with the MSE Loss can be connected to diversification and controlled parametrically, while the centroid combiner rule is applied as the portfolio target. 
To elaborate, in a supervised setting with the data set of time series returns ($\boldsymbol{x}$, $\boldsymbol{r}$),  
the Bias-variance-diversity decomposition can be expressed in terms of the MSE loss \(\mathcal{L}(\boldsymbol{r},\boldsymbol{\bar{r}})\) as \citep{wood2023unified}:

\begin{align}
\underbrace{\frac{1}{M} \sum_{j=1}^M (\overset{\circ}{f_j}(\boldsymbol{x}_j)- \boldsymbol{r})^2}_{\text{Average Bias}} + \underbrace{\frac{1}{M} \sum_{j=1}^M \mathbb{E} \left[(f_j(\boldsymbol{x}_j) - \overset{\circ}{f_j}(\boldsymbol{x}_j))^2 \right]}_{\text{Average Variance}} -\underbrace{\mathbb{E} \left[ \frac{1}{M} \sum_{j=1}^M (f_j(\boldsymbol{x}_j) - \boldsymbol{\bar{r}})^2 \right]}_{\text{Diversity}}   
\label{biasvardivmse}
\end{align}



where \(f_j(\boldsymbol{x}_j)\) denote the temporal predictions and \(\overset{\circ}{f_j}(\boldsymbol{x}_j)=\mathbb{E}[f_j(\boldsymbol{x}_j)]\) represents the centroid prediction of the \(j\)-th model in the portfolio ensemble respectively, while $\boldsymbol{\bar{r}} = \frac{1}{M} \sum_{j=1}^M f_j(\boldsymbol{x}_j)$ represents the centroid combiner rule. 

In the next section, the Portfolio-Structured Ensemble Model (PSEM) is introduced, where each individual hypothesis predicts a specific portfolio constituent. The diversity term in equation (\ref{biasvardivmse}) is linked to both portfolio generalization performance and risk diversification. The predict-then-optimize setting with PSEM for portfolio optimization is then described, in which the target portfolio corresponds to the combiner rule. Under MSE loss, this rule becomes the arithmetic combiner, that is, the equal-weighted portfolio. In this context, the optimal ensemble parameters are equivalent to the optimal portfolio weights.




\subsection{Proposed Portfolio Structured Ensemble Model (PSEM)}
\label{psem}
The PSEM is a portfolio model formulated as a structured ensemble for prediction across multiple hypotheses.
Formally, it can be defined as a map \( \mathcal{E} : \boldsymbol{x} \in \mathbb{R}^{N \times M} \to \boldsymbol{r} \in \mathbb{R}^N \) which aggregates M predictions from multiple individual predictors or hypotheses \{\(f_{\boldsymbol{\theta}_j}(\boldsymbol{x}_j)\}_{j=1}^M\), parameterized by \(\boldsymbol{\Theta}=\{\boldsymbol{\theta}_j\}_{j=1}^M\), to produce portfolio predictions for a given asset selection time series return $\boldsymbol{x}$:

\begin{equation}
\mathcal{E}(\boldsymbol{x}) = g\left(f_{\boldsymbol{\theta}_1}(\boldsymbol{x}_1), f_{\boldsymbol{\theta}_2}(\boldsymbol{x}_2), \dots, f_{\boldsymbol{\theta}_M}(\boldsymbol{x}_M)\right)=\boldsymbol{\hat{r}}
\end{equation}

where 
\(g(\cdot)\) denotes the centroid combiner rule (e.g., averaging, voting, or weighted combination) for the ensemble \citep{wood2023unified}. 
A diversity parameter \( \varepsilon \in [0, 1] \) modulates each predictor's update at step \( i \), such that the update rule for the \( j \)-th predictor is given by:


\begin{equation}
\boldsymbol{\theta}_j = \boldsymbol{\theta}_j - \eta_j \left( \frac{\partial \mathcal{L}(f_{\boldsymbol{\theta}_j}(x_{ij}), y_{ij})}{\partial \boldsymbol{\theta}_j} + \frac{\lambda_p}{N} \boldsymbol{\theta}_j \right) \delta\left(f_{\boldsymbol{\theta}_j}(x_{ij})\right)
\end{equation}

where \(y_{ij} = x_{(i+\tau)j}\) with time lag \(\tau > 0\) is the target, \(\eta_j\) is the learning rate, and the loss function $\mathcal{L}$ 
combines a squared error term and regularization with parameter $\lambda_p$. The indicator function \(\delta(f_{\boldsymbol{\theta}_j}(x_{ij}))\), which adjusts the degree to which each predictor’s parameters are updated, is defined as:

\begin{equation}
\delta(f_{\boldsymbol{\theta}_j}(x_{ij})) = 
\begin{cases} 
1 - \varepsilon & \mathcal{L}(f_{\boldsymbol{\theta}_j}(x_{ij}), y_{ij})<\mathcal{L}(f_{\boldsymbol{\theta}_k}(x_{ik}), y_{ik})\ \ \forall k\neq j \\[10pt]
\frac{\varepsilon}{M - 1} & \text{otherwise}
\end{cases}
\end{equation}

$\delta(f_{\boldsymbol{\theta}_j}(x_{ij}))$ indicates whether the output of the \( j \)-th predictor is the top prediction for the \( i \)-th instance. A smaller \(\varepsilon\) places more emphasis on the predictors which perform the best, while a larger \(\varepsilon\) encourages updates to less accurate predictors, fostering greater diversity in the ensemble. Algorithm \ref{alg:structured_data set} provides the pseudo-code for the structured data set formation with stochastic gradient descent. 

After training, the predictions of all predictors form a structured data set:

\begin{equation}
\boldsymbol{\hat{x}}(\varepsilon) = 
\begin{bmatrix}
f_{\boldsymbol{\theta}_1}(x_{11}) & \cdots & f_{\boldsymbol{\theta}_1}(x_{N1}) \\
\vdots & \ddots & \vdots \\
f_{\boldsymbol{\theta}_M}(x_{1M}) & \cdots & f_{\boldsymbol{\theta}_M}(x_{NM})
\end{bmatrix}
\label{repsilon}
\end{equation}

\begin{algorithm}[t!]
\caption{Structured data set Formation with Gradient Descent}
\label{alg:structured_data set}
\begin{algorithmic}[1]
\Require Individual predictors $\{f_{\boldsymbol{\theta}_j}(\boldsymbol{x}_j)\}_{j=1}^M$ with parameters $\{\boldsymbol{\theta}_j\}_{j=1}^M$, input returns $\{x_{i,j}\}^{N,M}$, targets $ \{y_{i,j}\}^{N,M}$ with $y_{i,j} = x_{(i+\tau),j}$, learning rates $\eta_j$, diversity parameter $\varepsilon$, regularization parameter $\lambda_p$, loss function $\mathcal{L}$.
\Ensure Structured data set $\hat{\boldsymbol{x}}(\varepsilon)$, optimal parameters $\hat{\boldsymbol{\Theta}}$.
\State Randomly initialize $\boldsymbol{\theta}_j$ for all $j = 1$ to $M$
\For{$i = 1$ to $N$}
    \For{$j = 1$ to $M$}
        \If{$M = 1$}
            \State $\delta(f_{\boldsymbol{\theta}_j}(x_{i,j})) \gets 1$
        \Else
            \For{$k = 1$ to $M$}
                \If{$j \neq k$}
                    \If{$\mathcal{L}(f_{\boldsymbol{\theta}_j}(x_{i,j}), y_{i,j}) < \mathcal{L}(f_{\boldsymbol{\theta}_k}(x_{i,k}), y_{i,k})$}
                        \State $\delta(f_{\boldsymbol{\theta}_j}(x_{i,j})) \gets 1 - \varepsilon$
                    \Else
                        \State $\delta(f_{\boldsymbol{\theta}_j}(x_{i,j})) \gets \frac{\varepsilon}{M - 1}$
                    \EndIf
                \EndIf
            \EndFor
        \EndIf
        \State \[
        \boldsymbol{\theta}_j \gets \boldsymbol{\theta}_j - \eta_j \left( \frac{\partial \mathcal{L}(f_{\boldsymbol{\theta}_j}(x_{i,j}), y_{i,j})}{\partial \boldsymbol{\theta}_j} + \frac{\lambda_p}{N} \boldsymbol{\theta}_j \right) \delta(f_{\boldsymbol{\theta}_j}(x_{i,j}))
        \]
        \State $\hat{x}_{i,j}(\varepsilon) \gets f_{\boldsymbol{\theta}_j}(x_{i,j})$
    \EndFor
\EndFor
\State \Return $\hat{\boldsymbol{x}}(\varepsilon)$ and $\hat{\boldsymbol{\Theta}}$
\end{algorithmic}
\end{algorithm}

which serves as input to a structured ensemble model. The PSEM predictions are given by:

\begin{equation}
\boldsymbol{\hat{r}} = g(\boldsymbol{\hat{x}}(\varepsilon)^T)= \boldsymbol{\hat{x}}(\varepsilon)^T\boldsymbol{w}
\label{ensemble}
\end{equation}

with ensemble parameters $\boldsymbol{w}$, which, when properly scaled, are equivalent to portfolio weights. 
By incorporating the diversity parameter \(\varepsilon\), the ensemble achieves a balance between prediction accuracy and diversity, enhancing its generalization capabilities in portfolio forecasting tasks. Interestingly, this diversity aligns with the concept of portfolio diversification, as originally proposed in MPT \citep{https://doi.org/10.1111/j.1540-6261.1952.tb01525.x}.

\subsection{Predict-then-optimize Method with PSEM}

In the predict-then-optimize setting with PSEM, the optimal portfolio weights \( \boldsymbol{w} \) are obtained by minimizing the loss function \( \mathcal{L}(\boldsymbol{\hat{r}}, \boldsymbol{\bar{r}}) \), formulated as a supervised learning problem between the PSEM prediction \( \boldsymbol{\hat{r}} \), as defined in Section \ref{psem}, and a portfolio target \( \boldsymbol{\bar{r}} \) whose weights reflect the ensemble combiner rule in (\ref{biasvardivmse}). This combiner rule depends on the choice of loss function, following the bias-variance-diversity decomposition framework proposed by \citep{wood2023unified}. For the squared loss, the combiner rule reduces to the arithmetic combiner. When the target portfolio time series is defined using the arithmetic combiner, it is equivalent to the equal-weighted portfolio representing the naive diversification strategy \citep{DeMiguel2009OptimalVN}. The target portfolio returns are given by \( \boldsymbol{\bar{r}} = \boldsymbol{x}^T \boldsymbol{w}_{\text{target}} = \boldsymbol{x}^T \frac{1}{\| \boldsymbol{x}^T \|} \), and the optimization is formulated as:

\begin{equation}
\min_{\boldsymbol{w}} \left\| \boldsymbol{\bar{r}} - \boldsymbol{\hat{x}}(\varepsilon)^T \boldsymbol{w} \right\|^2 + \lambda_s \| \boldsymbol{w} \|_p
\label{optweights}
\end{equation}

Here, \( \lambda_s \) denotes the PSEM regularization parameter. The predict-then-optimize approach can be implemented in a "plug-in" format, where the structured data set \( \boldsymbol{\hat{x}}(\varepsilon)^T \) is constructed independently prior to training the structured ensemble and deriving the portfolio weights. In the experimental section, portfolio weight constraints are applied post-optimization, including non-negativity constraints, \( w_i \geq 0, \ \forall i \in \{1, \ldots, M\} \), and normalization, \( \sum_{i=1}^{M} w_i = 1 \). However, the method can also be extended to a fully constrained plug-in optimization framework using quadratic solvers, which is left for future work.

Although the ensemble prediction \( \bar{\boldsymbol{r}} = \frac{1}{M} \sum_{j=1}^M \hat{\boldsymbol{x}}_j(\varepsilon) \) is computed as an equally weighted average of base learner outputs, the optimal ensemble weights \( \boldsymbol{w} \) derived from (\ref{optweights}) do not generally coincide with uniform or equal-weighting. This discrepancy arises due to induced diversity via the parameter \( \varepsilon \), variation in base learner initialization and regularization, and inherent dependencies among predictors. These factors collectively break the symmetry of the hypothesis space, resulting in non-uniform projections of the ensemble mean onto the span of predictions. As a result, the final ensemble weights adapt to capture these structural and informational asymmetries, ultimately improving loss alignment while diverging from naive equal weighting.

\begin{figure}[t!]
    \centering
    \includegraphics[width=\linewidth]{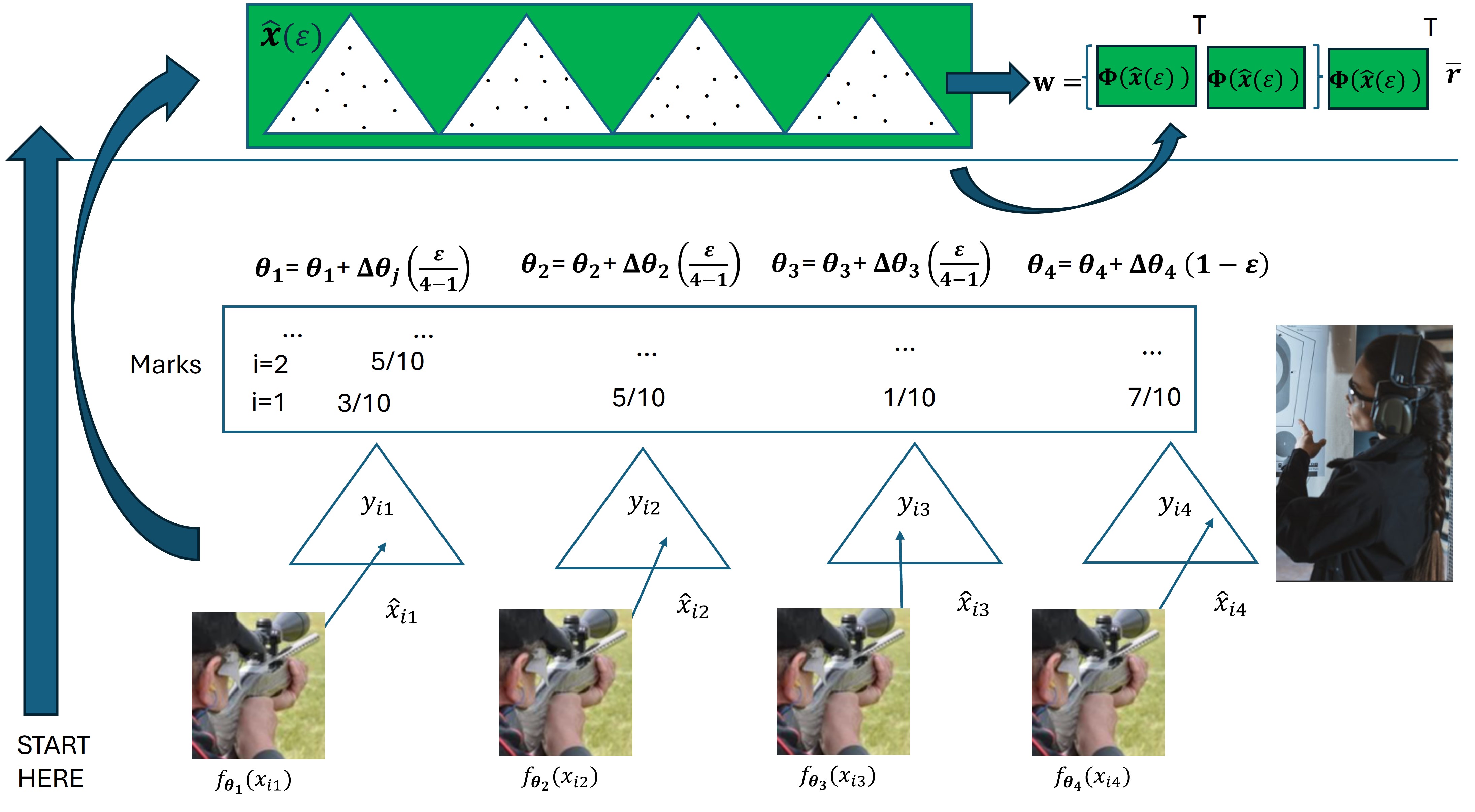} 
    \caption{Structured data set formation, including diversity ($\varepsilon$), and analytical ensemble training of the s-RBFN model, described in Section \ref{sRBFNmodel}. $f_{\boldsymbol{\theta}_j}(x_{ij})$ is the $j^{th}$ shooter (predictor) with input data $x_{ij}$ (for example wind, distance, etc.) and $\boldsymbol{\theta}_j$ her shooting skills, $\hat{x}_{ij}$ is the $i^{th}$ shoot from $j^{th}$ shooter and $\boldsymbol{\hat{x}}=\boldsymbol{\hat{x}}(\varepsilon)$ is the full set of shoots after training, $\boldsymbol{\bar{r}}$ is the portfolio target for the training set (equal-weighted portfolio), and $\boldsymbol{w}$ are the s-RBFN model parameters optimized by least-squares.}
    \label{fig:example_shoot} 
\end{figure}

Figure \ref{fig:example_shoot} shows a representation of the PSEM optimization, illustrated as a shooting gallery where the shooters represent individual predictors. Each shooter receives inputs such as wind conditions, distance, and target position, aiming to hit their respective targets. In each iteration, the trainer penalizes the best shooter by limiting how much their parameters update during gradient descent, while the remaining shooters update equally. This process introduces diversity in learning through $\varepsilon$. Once the training of the individual predictors is completed, all shots are gathered into a diversified data set, which can be used as input for structured prediction models, like the Structured Radial Basis Function Network (s-RBFN) \citep{10.1007/978-3-031-77915-2_7}, shown in this figure and trained using least-squares.

In the next section, the PSEM optimization is detailed for the case of the s-RBFN model. 

\subsection{Example with Structured Radial Basis Function Network (s-RBFN)}
\label{sRBFNmodel}
In this case, the structured data set is used as input to a radial basis function network, where each \(j\)-th predictor \(f_{\boldsymbol{\theta}_j}(\boldsymbol{x}_j)\) is associated with a specific basis function \(\phi_j(\hat{x}_{ij}(\varepsilon), \mu_j, \sigma_j)\). The basis function \(\phi_j\) is applied element-wise to the entries of the \(j\)-th input vector of predictions \(\boldsymbol{\hat{x}}_j(\varepsilon) \in \mathbb{R}^N\), with \(\mu_j\) and \(\sigma_j\) representing the center and scale parameters specific to the \(j\)-th predictor. This defines a transformation map $\boldsymbol{\Phi}(\boldsymbol{\hat{x}}(\varepsilon)) : \mathbb{R}^{N \times M} \to \mathbb{R}^{N \times M}$, where \(\boldsymbol{\Phi}(\boldsymbol{\hat{x}}(\varepsilon))_{ij} = \phi_j(\hat{x}_{ij}(\varepsilon), \mu_j, \sigma_j)\). 
For example, the Gaussian basis function $\phi_j\left(\cdot\right)=\exp{\left(\frac{-1}{2\sigma_{j}^2}\left|{\hat{x}}_{ij}(\varepsilon)-\mu_{j}\right|^2\right)}$ can be used where the parameters of the centers $\mu_{j}$ and the scales $\sigma_{j}$ for the basis functions are calculated from each column of $\boldsymbol{\hat{x}}(\varepsilon)$ as the mean and standard deviation, respectively. The s-RBFN formulation can now be expressed in matrix form as follows:

\begin{align}
 \label{matrixform}
&\boldsymbol{\Phi}\left(\boldsymbol{\hat{x}}(\varepsilon)\right)^T\boldsymbol{w}
=\left[\begin{matrix}\phi\left(\hat{x}_{11}(\varepsilon),\mu_1, \sigma_1\right)&\ldots&\phi\left(\hat{x}_{1M}(\varepsilon), \mu_M, \sigma_M\right)\\\vdots&\ddots&\vdots\\\phi\left(\hat{x}_{N1}(\varepsilon),\mu_1, \sigma_1\right)&\ldots&\phi\left(\hat{x}_{NM}(\varepsilon), \mu_M, \sigma_M\right)\\\end{matrix}\right]\left[\begin{matrix}w_1\\\vdots\\w_M\\\end{matrix}\right]   
\end{align}

The optimal s-RBFN parameters are obtained by least-squares with the regularization parameter $\lambda_s$ and target portfolio, the arithmetic combiner or equal-weighted portfolio, $\boldsymbol{\bar{r}}$:

\begin{equation}    \boldsymbol{w}=\left(\boldsymbol{\Phi}\left(\boldsymbol{\hat{x}}(\varepsilon)\right)^{\rm T}\boldsymbol{\Phi}\left(\boldsymbol{\hat{x}}(\varepsilon)\right)+\lambda_s\ast \boldsymbol{I}_{(mxm)}\right)^{-1}\boldsymbol{\Phi}\left(\boldsymbol{\hat{x}}(\varepsilon)\right)^{\rm T}\boldsymbol{\bar{r}}
    \label{weights}
\end{equation}
The optimal portfolio weights are equivalent to the s-RBFN optimal parameters \( \boldsymbol{w} \), after applying weight normalization and other relevant constraints.

\begin{figure}[t!]
    \centering
    \includegraphics[width=\linewidth]{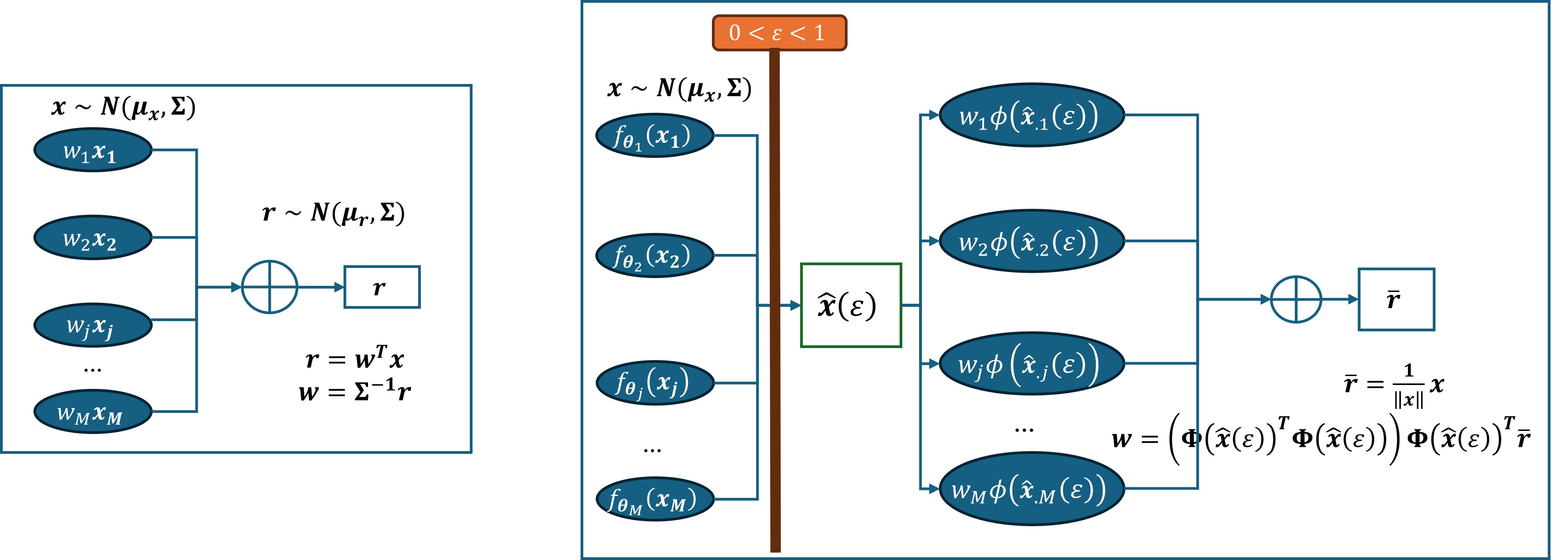} 
    \caption{Systems representation comparison between the Mean-Variance framework for portfolio optimization from Modern Portfolio Theory \citep{https://doi.org/10.1111/j.1540-6261.1952.tb01525.x} (left blue box) and the predict-then-optimize PSEM framework (right blue box). In the Mean-Variance case, the portfolio constituents' returns, \( \boldsymbol{x} \), and the target portfolio returns, \( \boldsymbol{r} \), are assumed to be normally distributed. In contrast, in the PSEM framework, the target is the equal-weighted portfolio \( \boldsymbol{\bar{r}} \). Additional parameters and variables include the portfolio weights \( \boldsymbol{w} \), the diversity parameter \( 0 < \varepsilon < 1 \) used for forming the structured data set \( \boldsymbol{\hat{x}}(\varepsilon) \), the individual predictors \( \{f_{\boldsymbol{\theta}_j}(\boldsymbol{x}_j)\}_{j=1}^M \), and the s-RBFN model basis functions represented in matrix form by \( \boldsymbol{\Phi} \).}
    \label{fig:example_system} 
\end{figure}

In Figure \ref{fig:example_system}, a comparative analysis between two portfolio optimization frameworks is presented: the classical mean-variance approach from Modern Portfolio Theory (MPT) \citep{https://doi.org/10.1111/j.1540-6261.1952.tb01525.x}, shown in the blue box on the left, and the s-RBFN model trained via the least-squares algorithm \citep{10.1007/978-3-031-77915-2_7}, as the PSEM model, shown in the blue box on the right. Despite their structural differences, both methodologies exhibit notable conceptual and computational parallels.

In the mean-variance framework, the key plug-in component is the covariance matrix, which is based exclusively on the input data set $\boldsymbol{x}$ and does not provide a guarantee of achieving portfolio diversification outside the sample. In contrast, the PSEM framework incorporates portfolio diversification directly into the modeling process by treating assets as hypotheses and the portfolio as an ensemble predictor, leveraging the diversity among individual predictors during training. 
The structured data set $\boldsymbol{\hat{x}}(\varepsilon)$, constructed from predictions generated by models using the same input data as the mean-variance model, enables the use of a continuous diversification parameter $0 < \varepsilon < 1$. This allows for the creation of multiple diversified plug-in data sets from the same input and for portfolio diversification to be adjusted prior to weight optimization.

Both approaches allow closed-form solutions via the least-squares algorithm, rendering them computationally equivalent in terms of efficiency. In the PSEM framework, the input data \( \boldsymbol{x} \) is replaced with the diversified structured data set of predictions \( \boldsymbol{\hat{x}}(\varepsilon) \), which is then augmented using basis functions within the s-RBFN model. These basis functions—such as radial, Gaussian, or other types—introduce additional flexibility to the s-RBFN model, enabling it to capture more complex non-linear relationships within the data.

The focus of this work is on portfolio allocation, which encompasses both asset selection and portfolio optimization. The latter has been addressed in the previous sections. The former is introduced in the next section through a novel method which leverages the diversity of asset return predictions as an asset selection tool. This approach seeks to uncover connections between diversity in asset selection and out-of-sample portfolio diversification, identify new limits on portfolio diversification introduced by this additional source of variability, and evaluate the ability to control these effects parametrically prior to weight optimization.


\subsection{Parametric Control of Out-of-Sample Diversification at the Asset or Hypothesis Selection Stage}
\label{assetselection}

The portfolio allocation is structured into two stages \citep{https://doi.org/10.1111/j.1540-6261.1952.tb01525.x}: asset selection and portfolio optimization. In the asset selection, the cumulative 1-month forecasts for the $m$ members of a hypothetical market are computed and ranked based on their predicted cumulative returns. A threshold \( T \) is applied to the ranking, and the top-performing assets exceeding this threshold are selected based on rational investor behavior, prioritizing those with the highest expected or predicted returns \citep{https://doi.org/10.1111/j.1540-6261.1952.tb01525.x}. To incorporate diversity into the asset selection process, a parameter $\gamma$ is introduced. This parameter acts as a multiplier of the number of portfolio constituents \(M\), generating a sample of size \(m = \gamma \times M\) consisting of assets with predictions exceeding the threshold \(T\), from which a set of \(M\) portfolio constituents is randomly selected.

\begin{figure}[t]
    \centering
    \includegraphics[width=\linewidth]{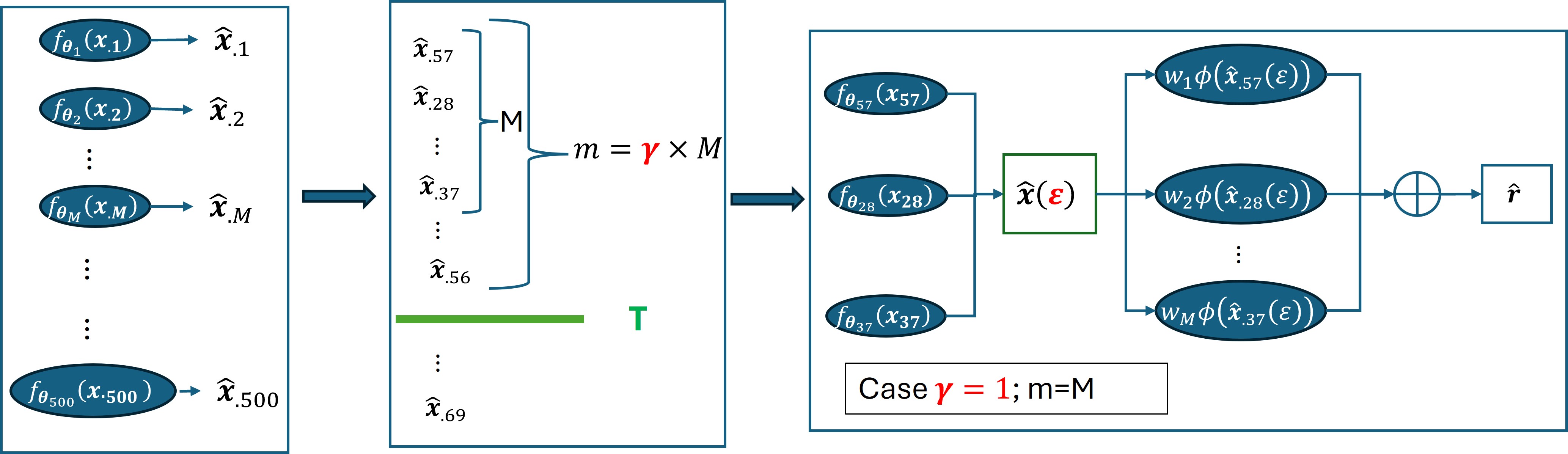} 
    \caption{Assets or hypotheses selection including forecasting (see left block). A 1-month cumulative return ranking with threshold $T$, random selection of $M$ constituents from a sample of $m$ candidates ($m=\gamma\times M\geq M$) and $\gamma$ the diversity parameter in the hypothesis selection stage (see middle block). Model architecture for s-RBFN for $m=M$ and $\gamma=1$, and $\varepsilon$ the diversity parameter for the learning stage (see right block).}
    \label{fig:example_gamma_varepsilon} 
\end{figure}

This can be seen in Figure \ref{fig:example_gamma_varepsilon}, where the forecast is first computed (see left block). Then, depending on whether \(\gamma = 1\), no diversity is added; if \(\gamma > 1\), diversity is introduced into the asset selection, provided the sample size is such that all predictions in the sample exceed \(T\). From this sample, \(M\) portfolio constituents are randomly selected, with the parameter \(\gamma\) controlling the degree of diversity in asset or hypothesis selection (see central block). The right block refers to the structured prediction model with multiple hypotheses, where the hypotheses have already been selected. At this stage, the parameter \(\varepsilon\) regulates the diversity in the learning process of individual predictors and the ensemble. Once the prediction data set with the selected diversity is obtained, the portfolio weights are optimized as described in Section \ref{sRBFNmodel} (see right block in Figure \ref{fig:example_gamma_varepsilon}).

This setup allows the testing of the hypothesis that there is a trade-off between the diversity of expected or predicted future returns and their average value as a group. A more diverse group of predictions is preferred, even if it includes lower returns which reduce the group average, highlighting a balance between diversity and the quality of return predictions for asset or hypothesis selection. This trade-off is verified by the experiments.

In conclusion, modeling the portfolio allocation problem as a structured prediction task over multiple hypotheses allows the separation of the process into two key components: hypotheses selection and weight optimization. This approach provides parametric methodologies to induce diversity in hypothesis selection through the parameter \(\gamma\) and to regulate diversity in the learning process of individual predictors based on the selected hypotheses through the parameter \(\varepsilon\). These two sources of diversity are directly connected to out-of-sample portfolio risk diversification and can be controlled and adjusted prior to the optimization and decision-making process.

\section{Numerical Results}
\label{numerical}

\subsection{Data and Experimental Setup}
The time series of daily prices for the S\&P 500 index members \citep{wikiSP500} (excluding weekends) was obtained from Bloomberg \citep{Bloomberg}. Returns were calculated as the percentage change in daily prices. 

To empirically demonstrate the connection between diversity in the structured prediction setting with multiple hypotheses and out-of-sample portfolio diversification, numerous out-of-sample Sharpe ratios are computed using the predictive setting. The Sharpe ratio is the quotient between return and volatility (risk) for a portfolio of \(M\) stocks over a time period \(\Delta t\). The Sharpe ratio is computed for different values of portfolio size \(M\) and diversity parameter \(\varepsilon\), both for the case without diversity in asset selection and for the scenario where the portfolio allocation decision is made once, performing 100 simulations for each trial, as in Modern Portfolio Theory (MPT) \citep{https://doi.org/10.1111/j.1540-6261.1952.tb01525.x}. This allows comparison with traditional methods and reveals the pattern of portfolio diversification behavior in relation to the parameters \(M\) and \(\varepsilon\).

Next, a more realistic case not considered in MPT is included, where the problem becomes sequential over 10-months, making it more practical and useful for practitioners (in this case, the Sharpe ratio over the 10-month period is reported). For these cases, the asset selection is performed by ranking 1-month cumulative return predictions and choosing the top \(M\) assets based on different threshold values \(T\). In the one-step decision-making cases, the same assets are usually selected across simulations, whereas in the sequential cases, the non-stationarity in the data may alter prediction ranking and asset selection. Finally, a second source of diversity is introduced through asset or hypothesis selection, controlled by the parameter \(\gamma\). The experiments are performed for both sources of diversity in one-step as well as in the multi-step decision-making cases. 

The structured ensemble models used are the s-RBFN optimized by least-squares and two-layer networks as individual predictors, with the same set of hyperparameter configurations for all cases as described in \citep{10.1007/978-3-031-77915-2_7}. The equal-weighted model and the MSE-weighted model are also included, where the weights are either equally distributed across all constituents or inversely proportional to their generalization prediction error. For asset selection, the ranking is constructed by calculating the 1-month recurrent prediction of the daily time series and then ranking them based on the accumulated monthly return. The prediction is performed using the same model as the individual predictors with identical hyperparameters, but each of the 500 index members has its own network.

To evaluate the generalization of the proposed framework across different market regimes and asset classes, additional experiments are presented in \ref{appendixD}. These include 1,336 global bond time series from 2014–2018, with the distribution by sector, seniority, rating, tenor (curve), and country (ISO) shown in Figure~\ref{fig:bond_pie_subplots}, as well as all S\&P 500 constituents during the Global Financial Crisis (2007–2009) and the U.S. equity rally (2009–2016).

\subsection{Model Evaluation and Performance Metrics}
One of the most commonly used portfolio performance metrics is the Sharpe Ratio \citep{Sharpe1994TheSR}, defined as the ratio of the average daily return over a period to its standard deviation. In this work, return predictions are evaluated using 1-month cumulative returns. Accordingly, a slight variation of the Sharpe Ratio is considered, where the cumulative return over one month is divided by the standard deviation of daily returns during that same period. This modified Sharpe Ratio is used to assess out-of-sample portfolio performance in the one-step decision scenario. For the multi-step decision setting, the modified Sharpe Ratio is computed over a 10-month sequential investment horizon.

For the individual predictors, a two-layer multi-layer perceptron (MLP) is used with hyperbolic tangent activation functions. The number of neurons in each layer is denoted by $\kappa$, the learning rate by $\eta$, and a multiplicative factor for the random initial weights during gradient descent by $\chi$, which affects model performance. Regularization parameters are represented by $\lambda_p$. These individual predictors are applied during the asset selection stage, where RMSE is used as the metric for time-series forecasting performance. Assets are selected from a pool of diversified assets based on their predicted values, with diversity scaled by the parameter $\gamma$.

The s-RBFN is employed as the PSEM model in the experiments, where the number of predictors or hypotheses is denoted by $M$, and the regularization parameter is represented by $\lambda_s$. Once the optimal portfolio weights are obtained—by normalizing the ensemble parameters, they are used to evaluate out-of-sample portfolio performance using the modified Sharpe Ratio: 1-month for the one-step decision case and 10-month for the multi-step decision case. Portfolio performance serves as the criterion for selecting the optimal network configuration for both the individual predictors and the s-RBFN. Multiple hyperparameter combinations are tested in the experiments, and all hyperparameter values used are listed in Table~\ref{hyper}.

\begin{table}[t]
\caption{Sets of values for the individual predictors and s-RBFN hyperparameters, \citep{10.1007/978-3-031-77915-2_7}.}
\label{Table1values}
\centering
\begin{subtable}[t]{0.9\textwidth}
\centering
\caption{$M$ number of hypotheses, $\kappa$ number of neurons per layer, $\eta$ learning rates for the predictors, $\chi$ is a multiplicative factor for random initial predictors' parameters $\Theta$.}
\begin{tabular}{|c| c |c| c |}\hline 
$M$     & $\kappa$   &   $\eta$   &    $\chi$  \\
\hline  
$[2, 5, 10, 20, 35]$     & $[20, 200, 2000]$    &   $[0.03, 0.3]$    &    $[0.0001, 0.01, 0.1, 1]$  \\  \hline
\end{tabular}
\end{subtable}

\vspace{0.5cm}

\begin{subtable}[t]{0.9\textwidth}
\centering
\caption{$\varepsilon$ is the diversity parameter in the optimization and $\gamma$ in the asset selection, $\lambda_p$ is the regularization parameter for the predictors and $\lambda_s$ for the s-RBFN.}
\begin{tabular}{|c| c |c| c |}\hline 
$\varepsilon$  &   $\lambda_p$  &  $\lambda_s$ & $\gamma$\\
\hline  
$[0, 0.1, 0.35, 0.5]$   &   $[0, 0.0001, 0.01, 0.07]$   &  $[0, 3, 5]$ &  $[1, 2, 3,5]$\\  \hline
\end{tabular}
\end{subtable}
\label{hyper}
\end{table}

Additional evaluation metrics—Maximum Drawdown, Sortino Ratio, and Omega Ratio, alongside Sharpe Ratio—are reported in Appendix~\ref{appendixD}. These are applied to both the regime-based analyses and the global bond data set, supporting the sensitivity study in Section~\ref{Sensisection} and the comparative performance analysis in Section~\ref{compdifreg}.

\subsection{Parametric Portfolio Risk Diversification}
Experiments validate the hypothesis that out-of-sample portfolio risk diversification can be controlled prior to the decision making process if the portfolio is modeled as a structured ensemble in a multiple-hypothesis prediction setting. 

\subsubsection{Diversity in the Ensemble Learning Stage}

In this section, asset selection is performed for each portfolio allocation as described previously, taking portfolio constituents as the top $M$ of the ranking based on the 1-month accumulated return that are also higher than the threshold $T$. Two types of experiments are carried out: one-step decision-making (1-month and 100 simulations per trial) and sequential multi-step decision-making (10-months and 1 simulation per trial). The structured prediction model s-RBFN with radial and Gaussian basis functions is implemented, both with and without regularization. The experiments are repeated for different threshold values $T = -1\%, -0.5\%, 0.0\%, 0.5\%, 1\%$ and diversity parameter $\varepsilon=0,0.1,0.35$.

\begin{figure}[H]
    \centering
    \includegraphics[width=0.8\linewidth]{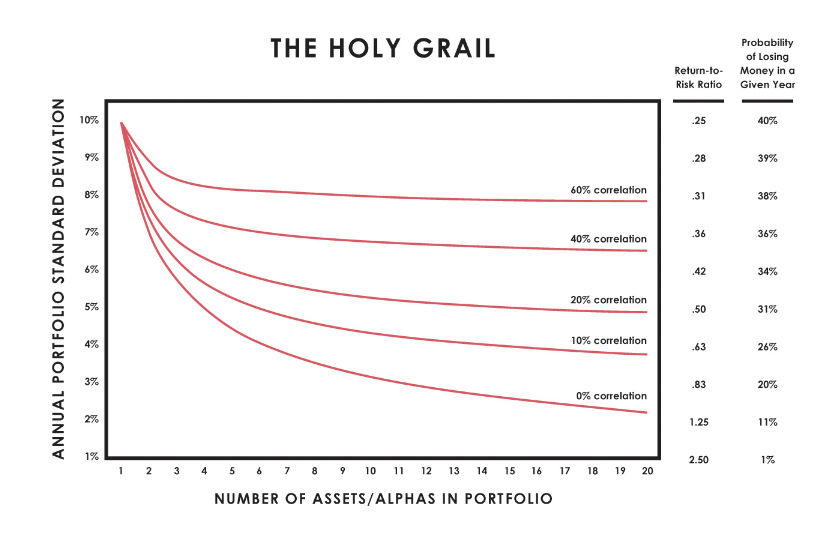}
    \caption{The limits of diversification, \citep{dalio2018principles}. Sharpe and Return-to-Risk ratios are the same (inverted axis). The standard deviation is reduced up to a diversification limit.}
    \label{DiverRayd}
\end{figure}

\begin{figure}[H]
    \centering
    \begin{subfigure}{0.45\textwidth}
        \centering
        \includegraphics[width=\textwidth]{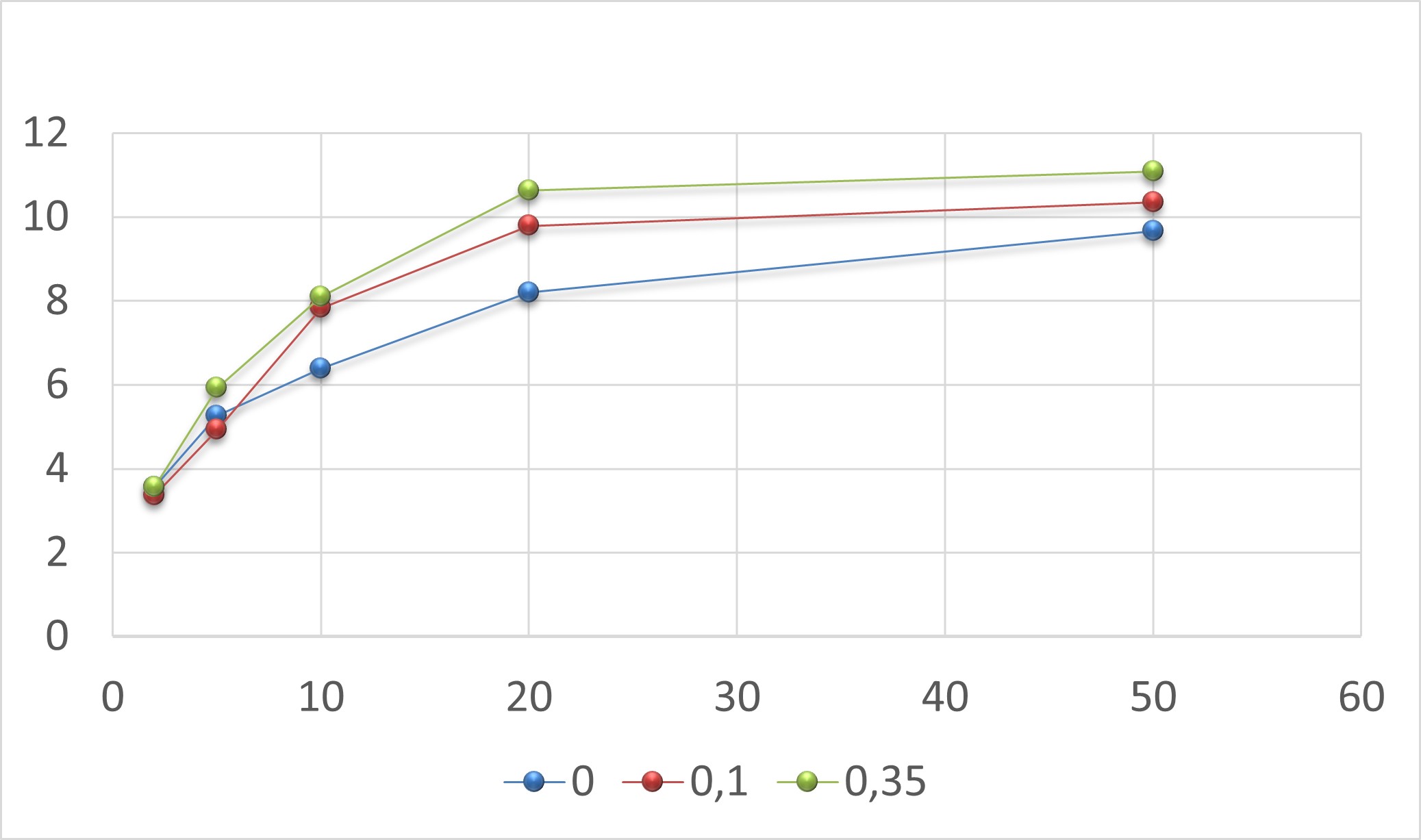}
        \caption{$\lambda_s=0$; $T=1\%$}
        \label{sub5G1}
    \end{subfigure}
    \hfill
    \begin{subfigure}{0.45\textwidth}
        \centering
        \includegraphics[width=\textwidth]{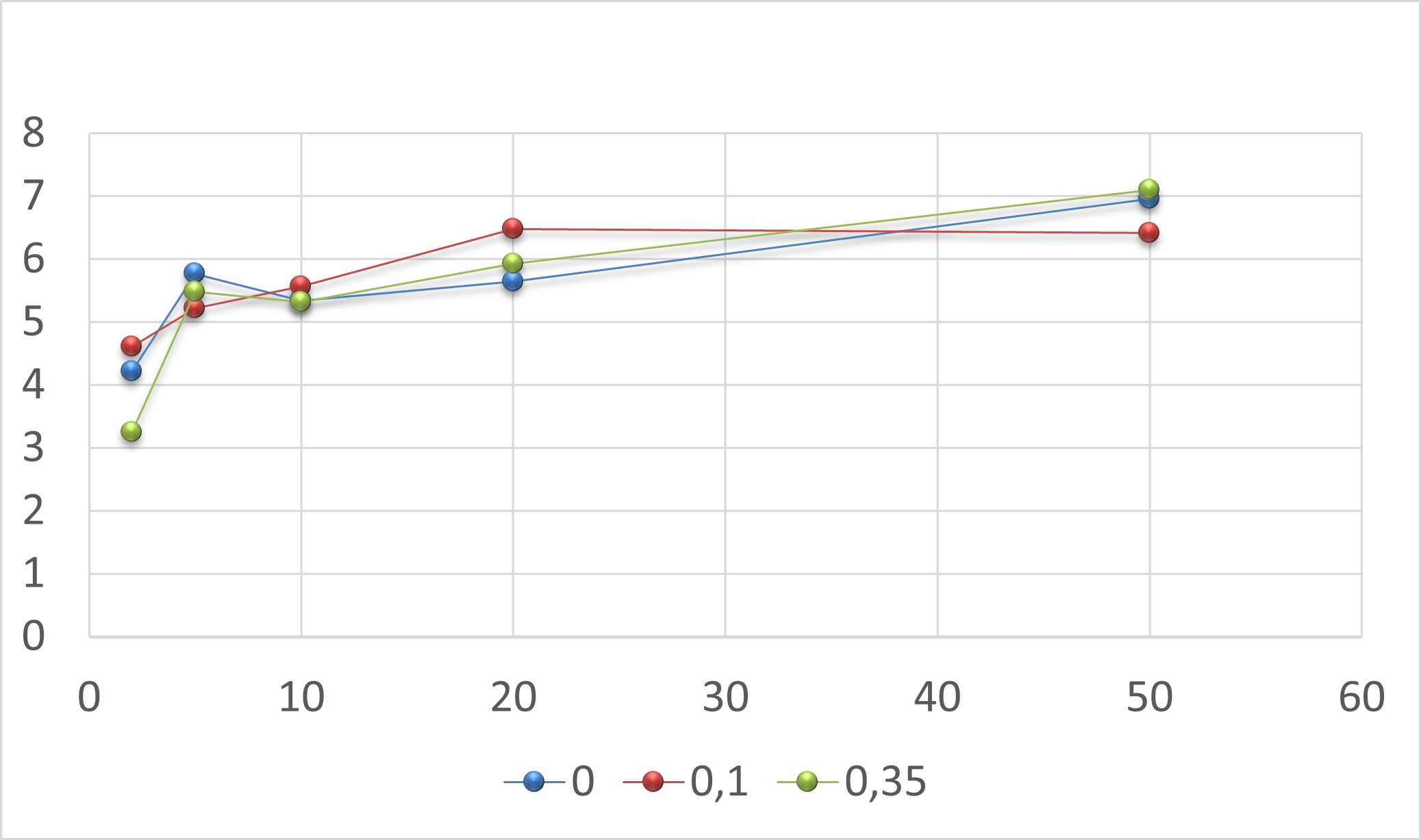}
        \caption{$\lambda_s=10$; $T=1\%$}
        \label{sub6G1}
    \end{subfigure}
    \vspace{0.5cm}    
    \begin{subfigure}{0.45\textwidth}
        \centering
        \includegraphics[width=\textwidth]{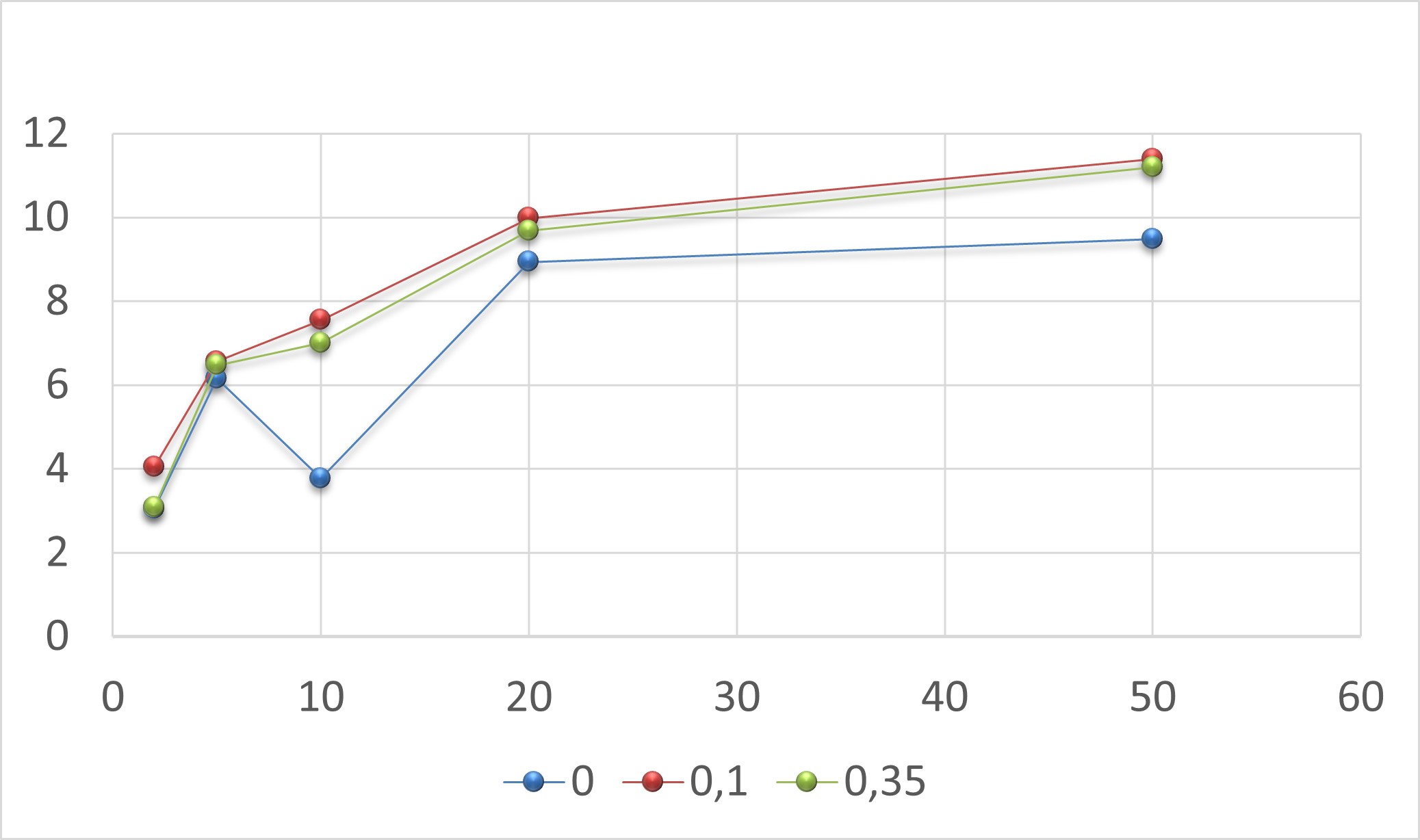}
        \caption{$\lambda_s=0$; $T=0.5\%$}
        \label{sub7G1}
    \end{subfigure}
    \hfill
    \begin{subfigure}{0.45\textwidth}
        \centering
        \includegraphics[width=\textwidth]{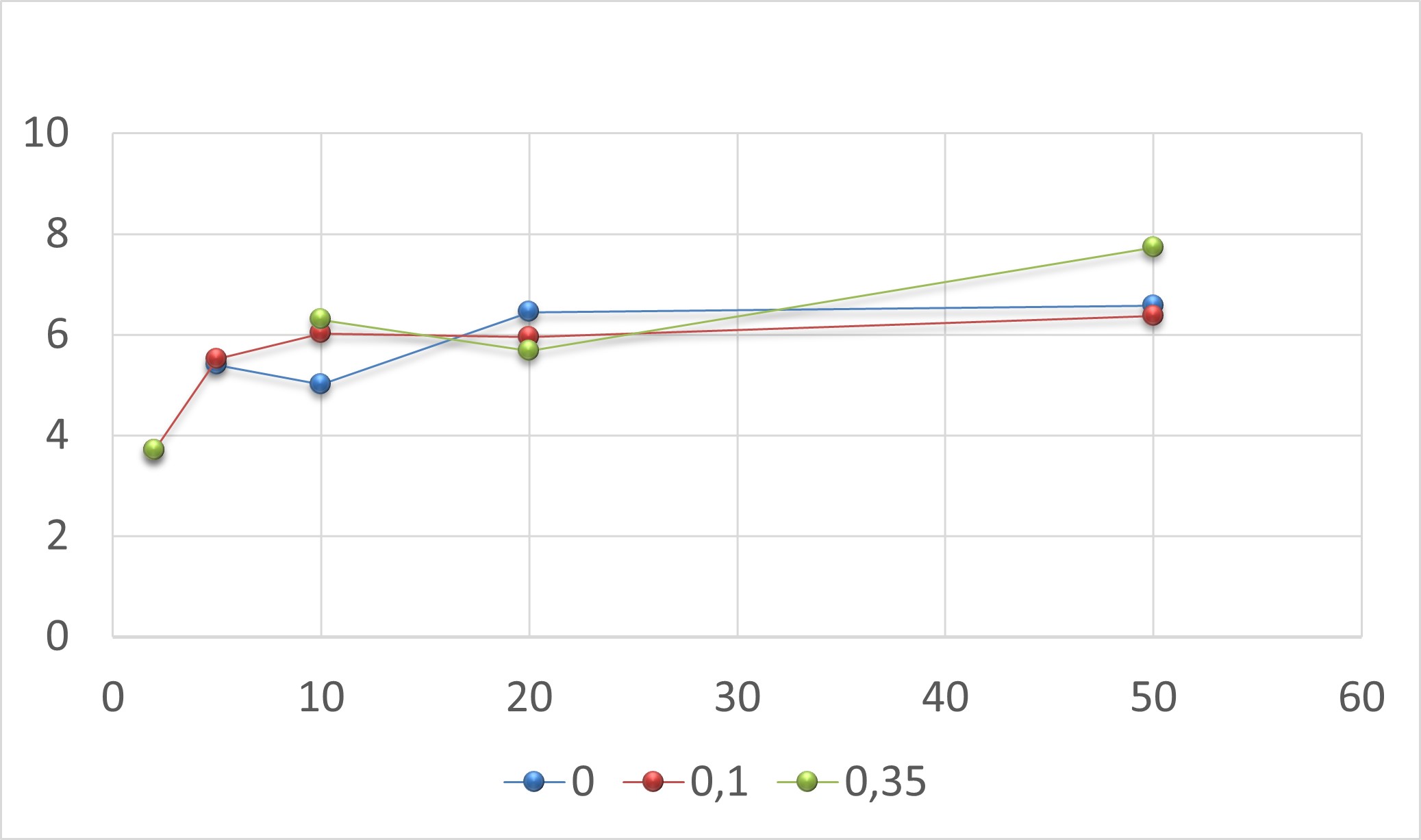}
        \caption{$\lambda_s=10$; $T=0.5\%$}
        \label{sub8G1}
    \end{subfigure}
    \vspace{0.5cm} 
    \begin{subfigure}{0.45\textwidth}
        \centering
        \includegraphics[width=\textwidth]{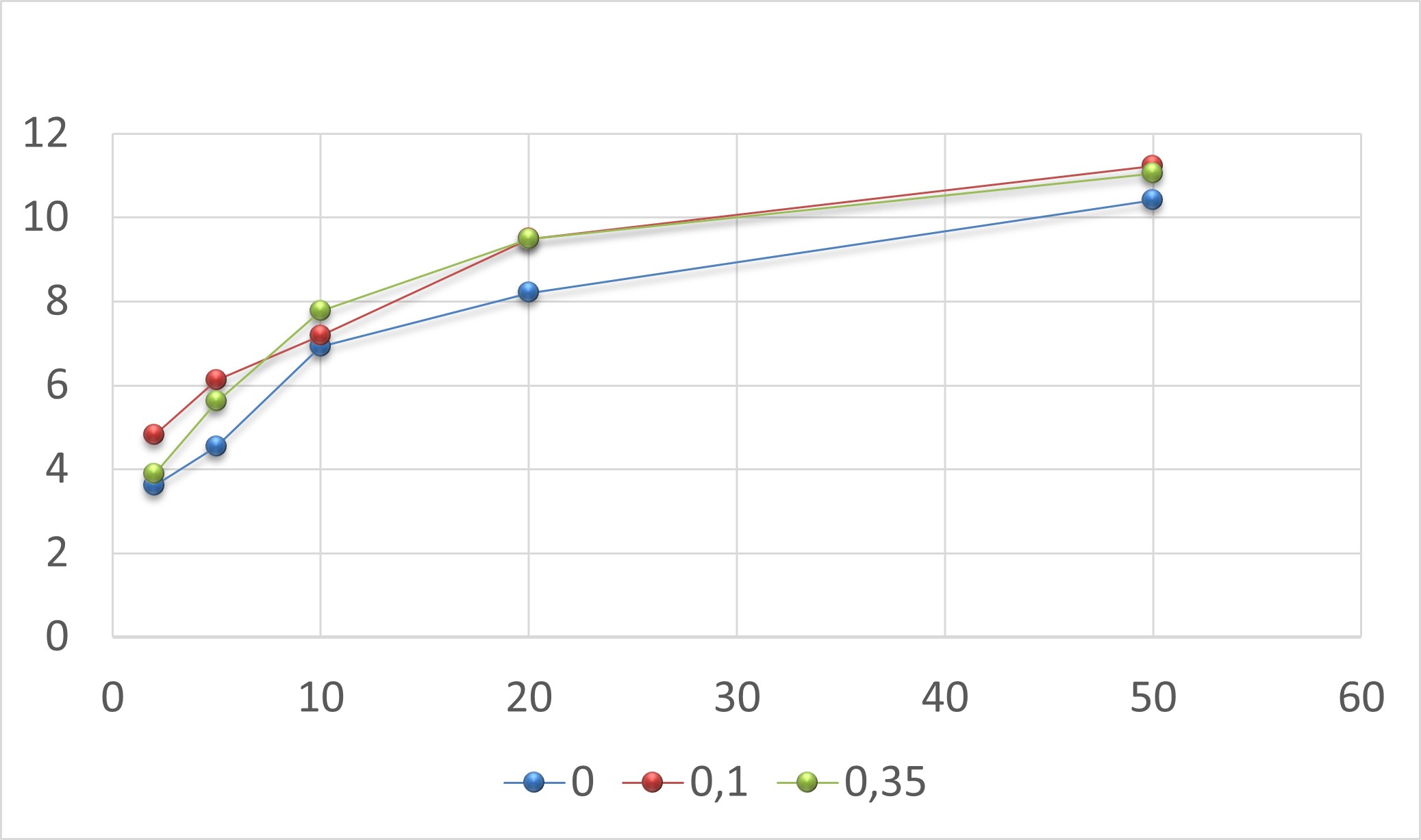}
        \caption{$\lambda_s=0$; $T=-0.5\%$}
        \label{sub9G1}
    \end{subfigure}
    \hfill
    \begin{subfigure}{0.45\textwidth}
        \centering
        \includegraphics[width=\textwidth]{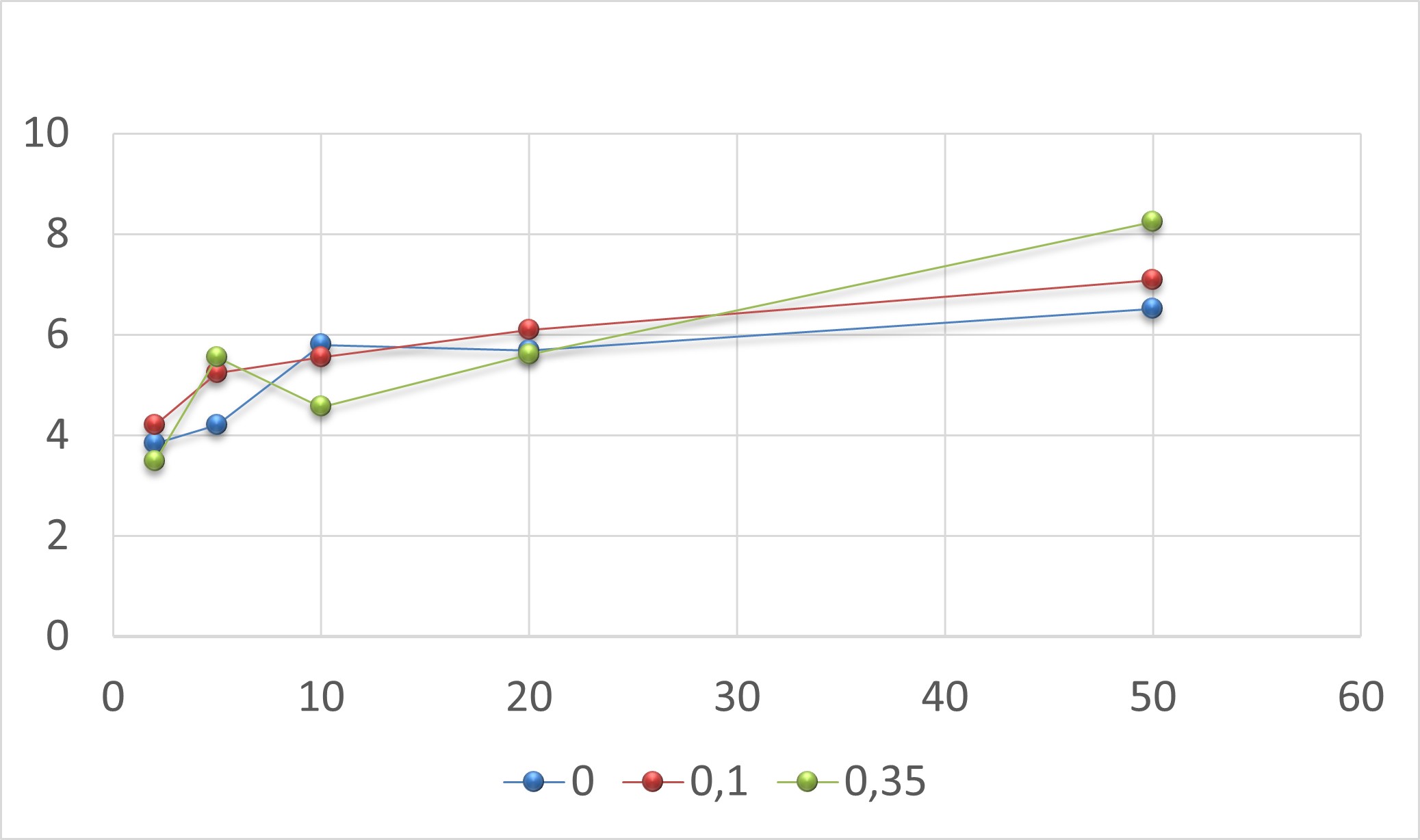}
        \caption{$\lambda_s=10$; $T=-0.5\%$}
        \label{sub10G1}
    \end{subfigure}
    \vspace{0.5cm}
    \begin{subfigure}{0.45\textwidth}
        \centering
        \includegraphics[width=\textwidth]{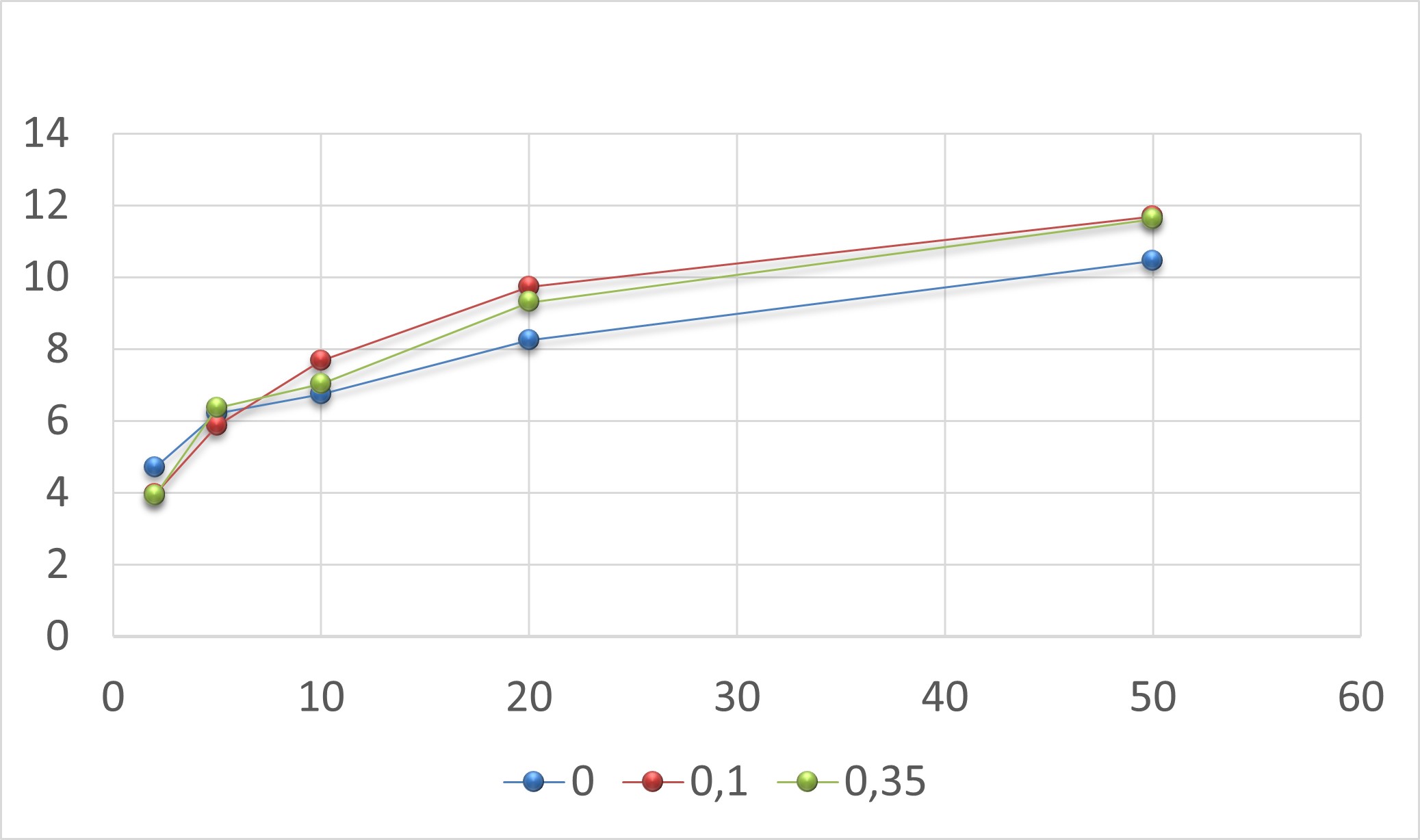}
        \caption{$\lambda_s=0$; $T=-1\%$}
        \label{sub11G1}
    \end{subfigure}
    \hfill
    \begin{subfigure}{0.45\textwidth}
        \centering
        \includegraphics[width=\textwidth]{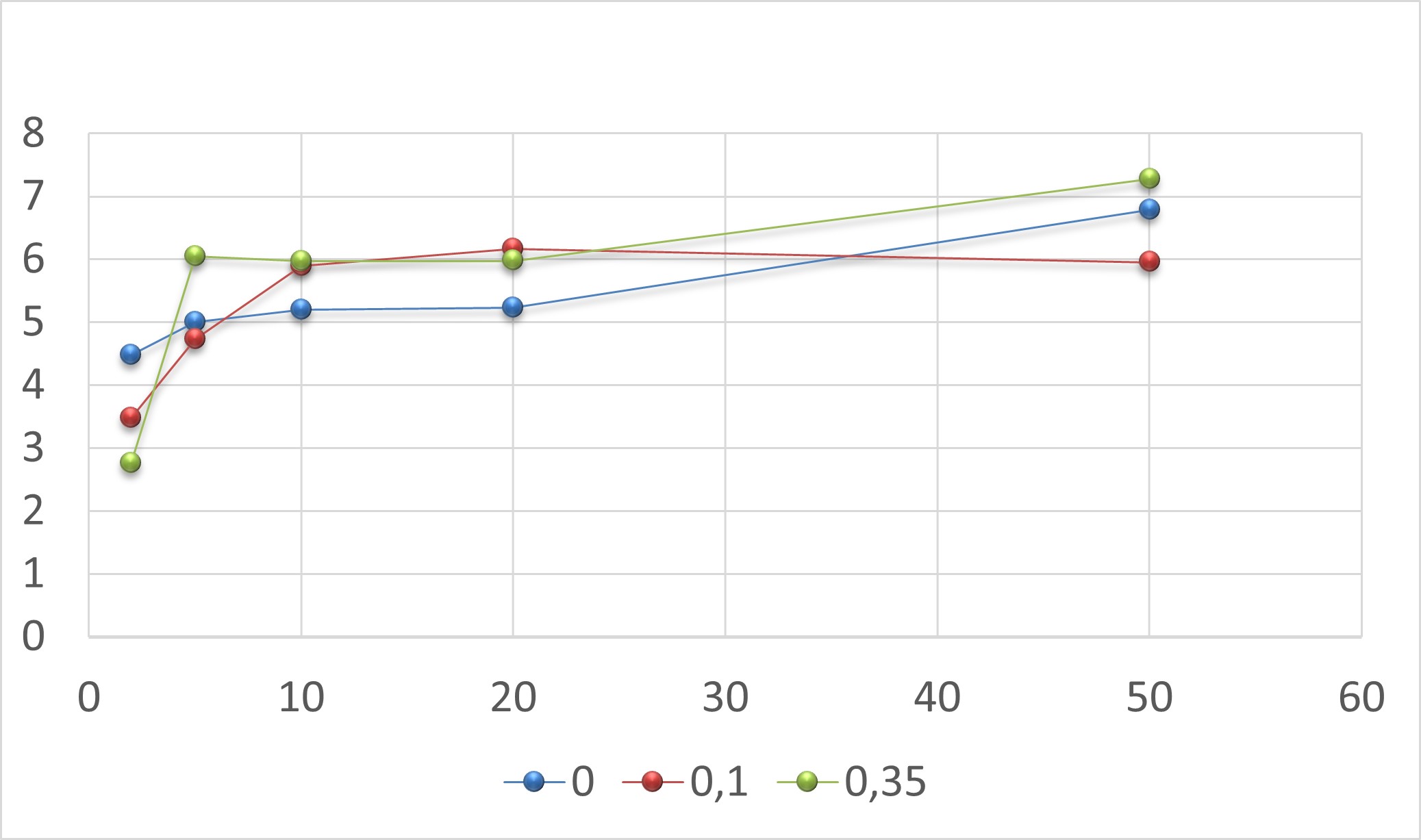}
        \caption{$\lambda_s=10$; $T=-1\%$}
        \label{sub12G1}
    \end{subfigure}
    \caption{One-step (1-month) 100 simulations Sharpe ratios (Y-axis) for different number of predictors $M$ (X-axis), diversity parameter $\varepsilon=0,0.1,0.35$ (colours), s-RBFN regularization parameter $\lambda_s$, and $-1\%\leq T\leq 1\%$. s-RBFN with Gaussian basis functions.}
    \label{FigureSR_Gaus1-Simp}
\end{figure}

The result for the one-step decision-making case can be seen in Figure \ref{FigureSR_Gaus1-Simp}, showing a strong similarity with the results from the MPT setup \citep{https://doi.org/10.1111/j.1540-6261.1952.tb01525.x}, which is illustrated in Figure \ref{DiverRayd}. Figure \ref{DiverRayd} shows the diversification limits according to MPT, where the Sharpe ratios are presented with an inverted axis compared to Figure \ref{FigureSR_Gaus1-Simp}, with the same horizontal axes referring to the number of assets or stocks $M$. In the case of MPT, the Sharpe ratios are displayed for different values of the correlation between assets, and it is observed that the negative growth of the correlation has the same effect as the positive growth of the diversity parameter in ensemble learning in the multi-hypothesis prediction setting, as shown in Figure \ref{FigureSR_Gaus1-Simp}. Thus, portfolio diversification, which depends on correlation (the lower, the better) and the number of portfolio constituents, is to some extent equivalent to diversity in the learning of structured ensembles in a setting like the one discussed (regulated by $\varepsilon$) and the number of predictors. The experiments empirically validate the equivalence between out-of-sample portfolio diversification and diversity in the learning of a structured ensemble in a multi-hypothesis prediction setting when used to model a portfolio, according to Section \ref{psem}.

\begin{figure}[H]
    \centering
    \begin{subfigure}{0.45\textwidth}
        \centering
        \includegraphics[width=\textwidth]{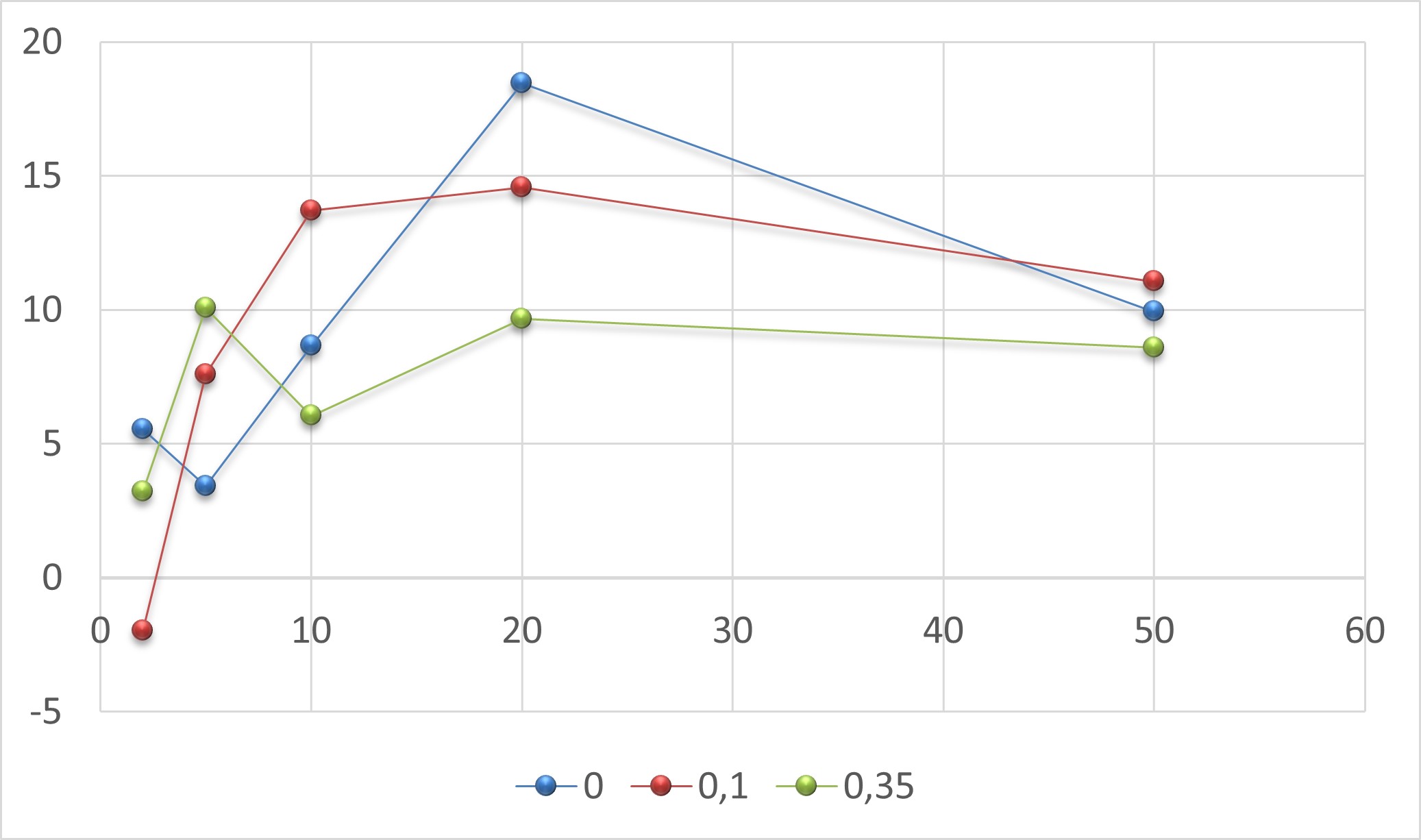}
        \caption{$\lambda_s=0$; $T=1\%$}
        \label{sub5G10}
    \end{subfigure}
    \hfill
    \begin{subfigure}{0.45\textwidth}
        \centering
        \includegraphics[width=\textwidth]{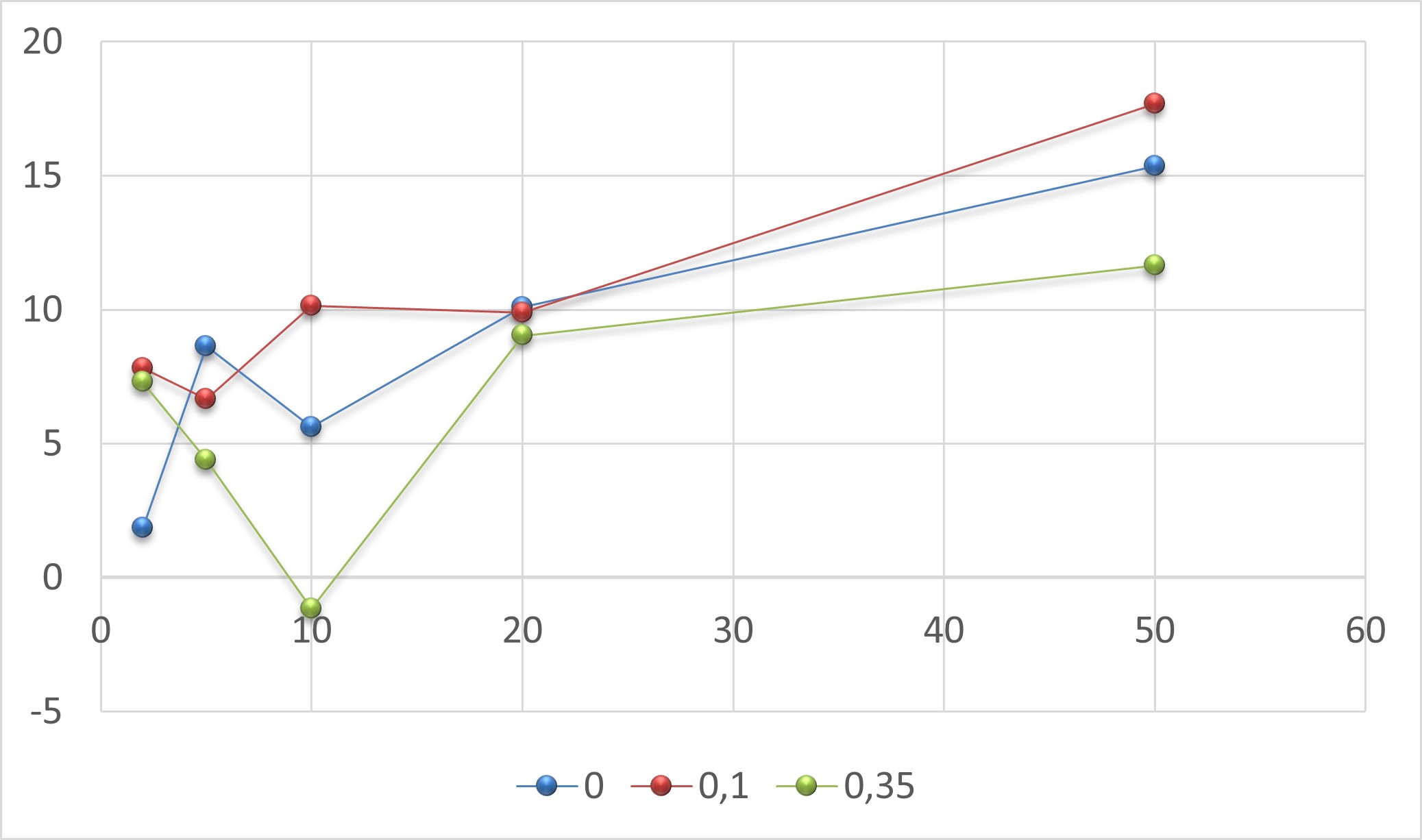}
        \caption{$\lambda_s=10$; $T=1\%$}
        \label{sub6G10}
    \end{subfigure}
    \vspace{0.5cm}    
    \begin{subfigure}{0.45\textwidth}
        \centering
        \includegraphics[width=\textwidth]{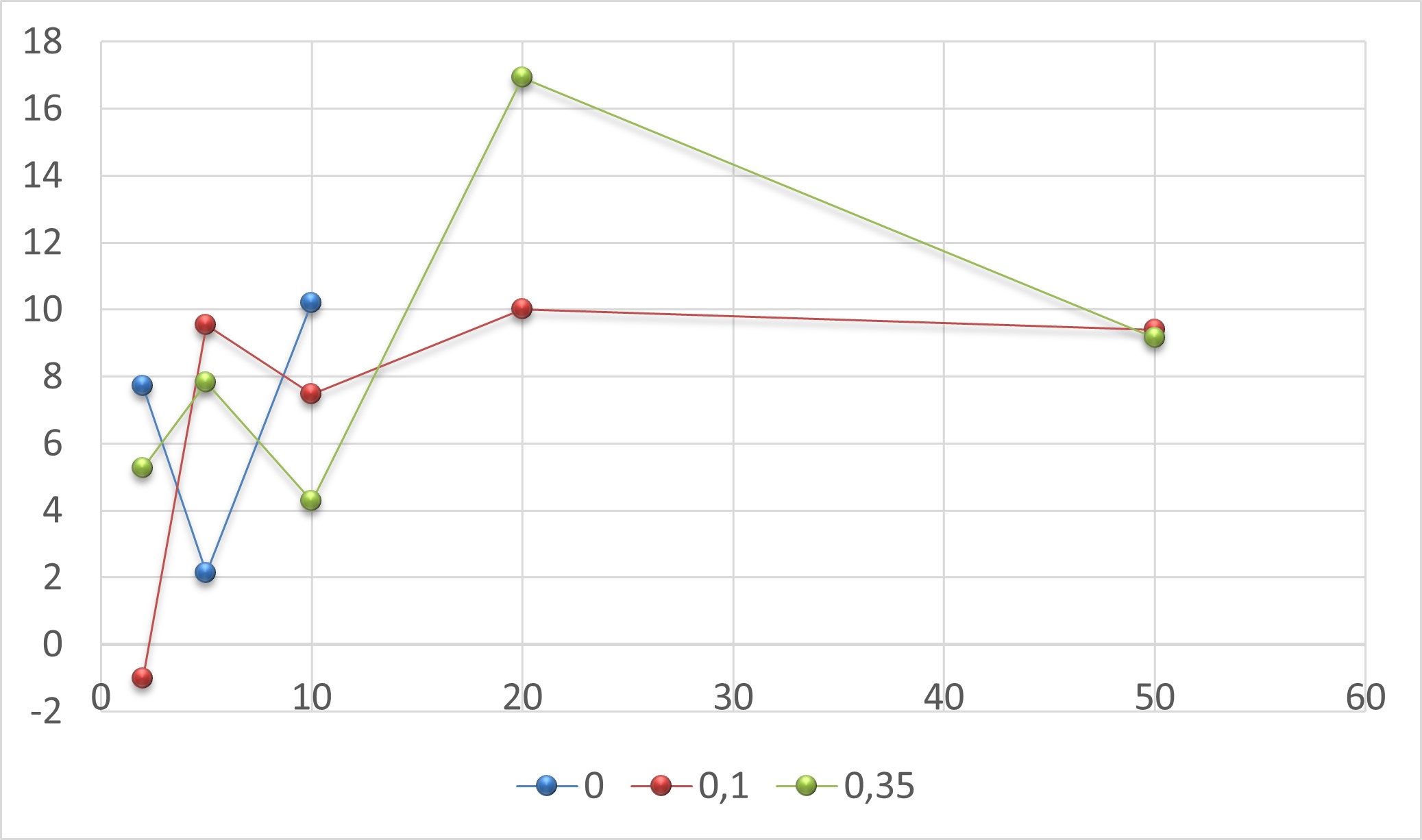}
        \caption{$\lambda_s=0$; $T=0.5\%$}
        \label{sub7G10}
    \end{subfigure}
    \hfill
    \begin{subfigure}{0.45\textwidth}
        \centering
        \includegraphics[width=\textwidth]{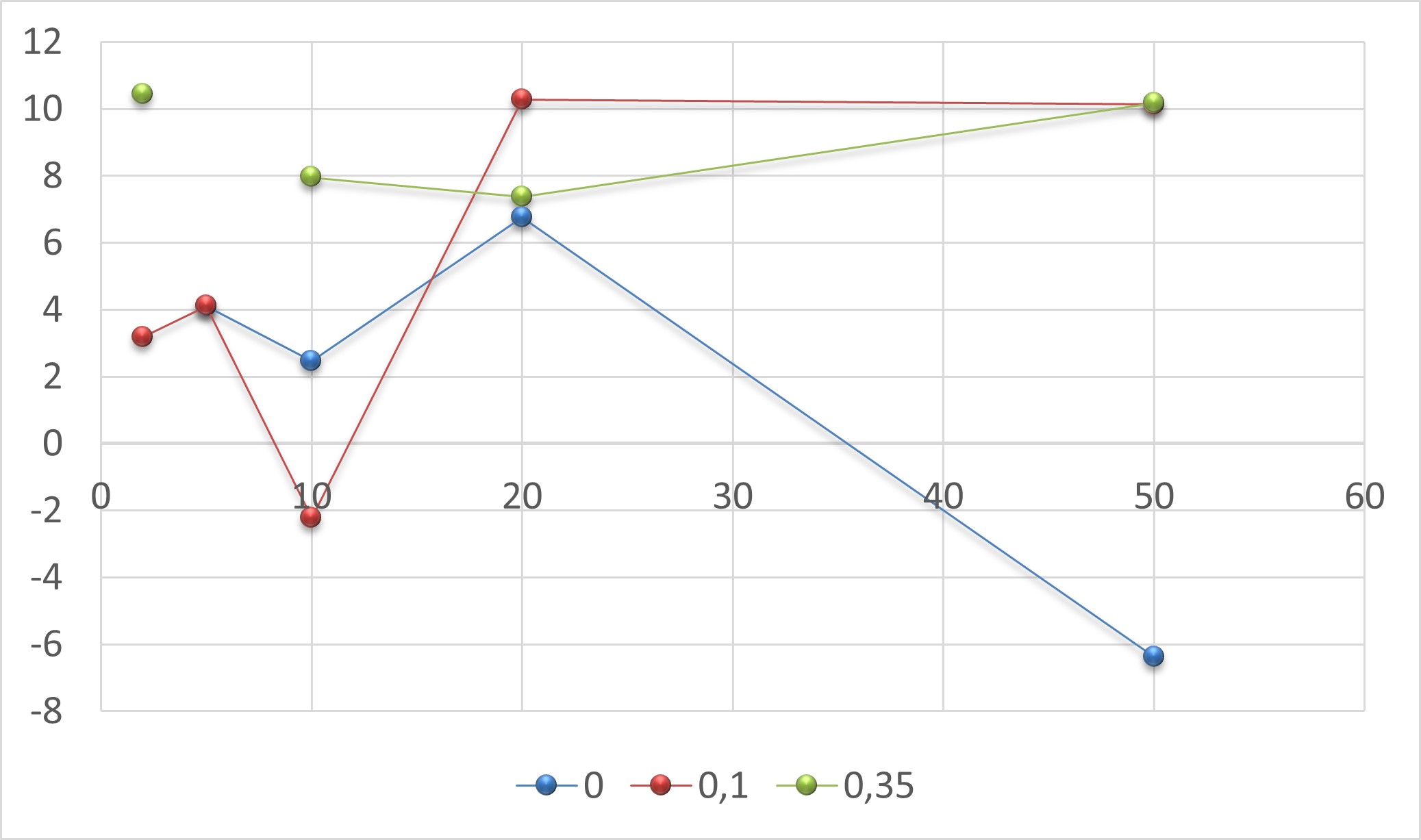}
        \caption{$\lambda_s=10$; $T=0.5\%$}
        \label{sub8G10}
    \end{subfigure}
    \vspace{0.5cm} 
    \begin{subfigure}{0.45\textwidth}
        \centering
        \includegraphics[width=\textwidth]{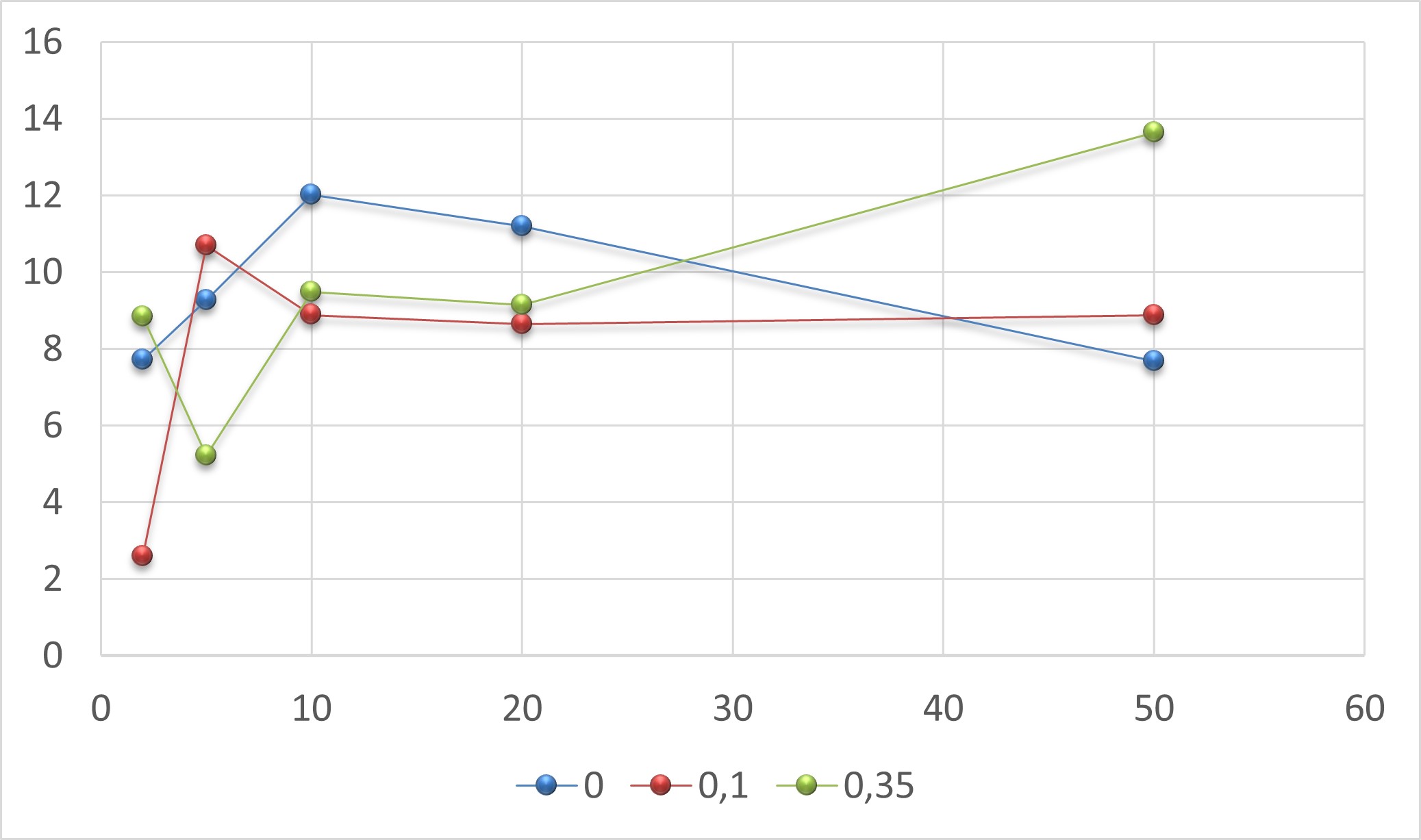}
        \caption{$\lambda_s=0$; $T=-0.5\%$}
        \label{sub9G10}
    \end{subfigure}
    \hfill
    \begin{subfigure}{0.45\textwidth}
        \centering
        \includegraphics[width=\textwidth]{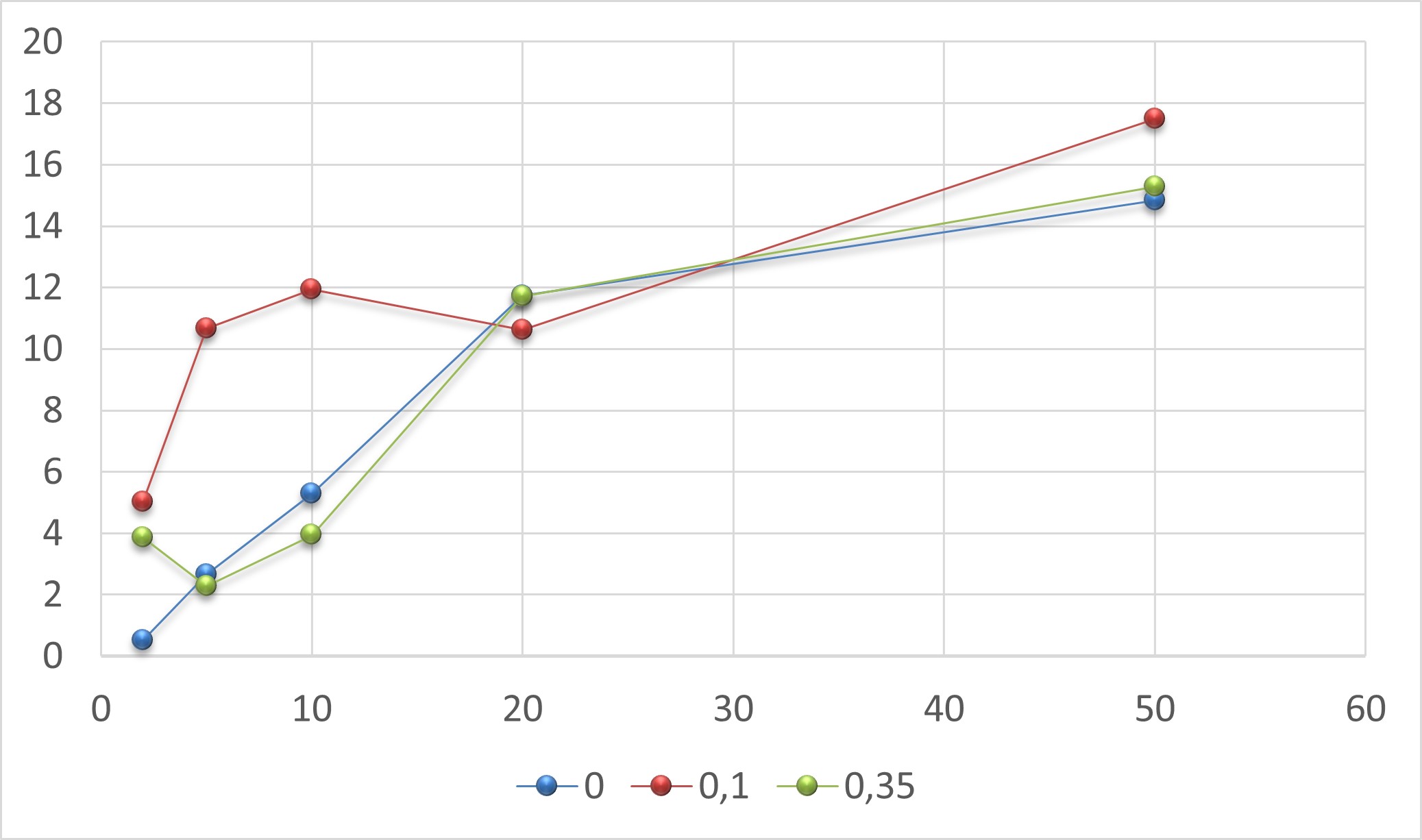}
        \caption{$\lambda_s=10$; $T=-0.5\%$}
        \label{sub10G10}
    \end{subfigure}
    \vspace{0.5cm}
    \begin{subfigure}{0.45\textwidth}
        \centering
        \includegraphics[width=\textwidth]{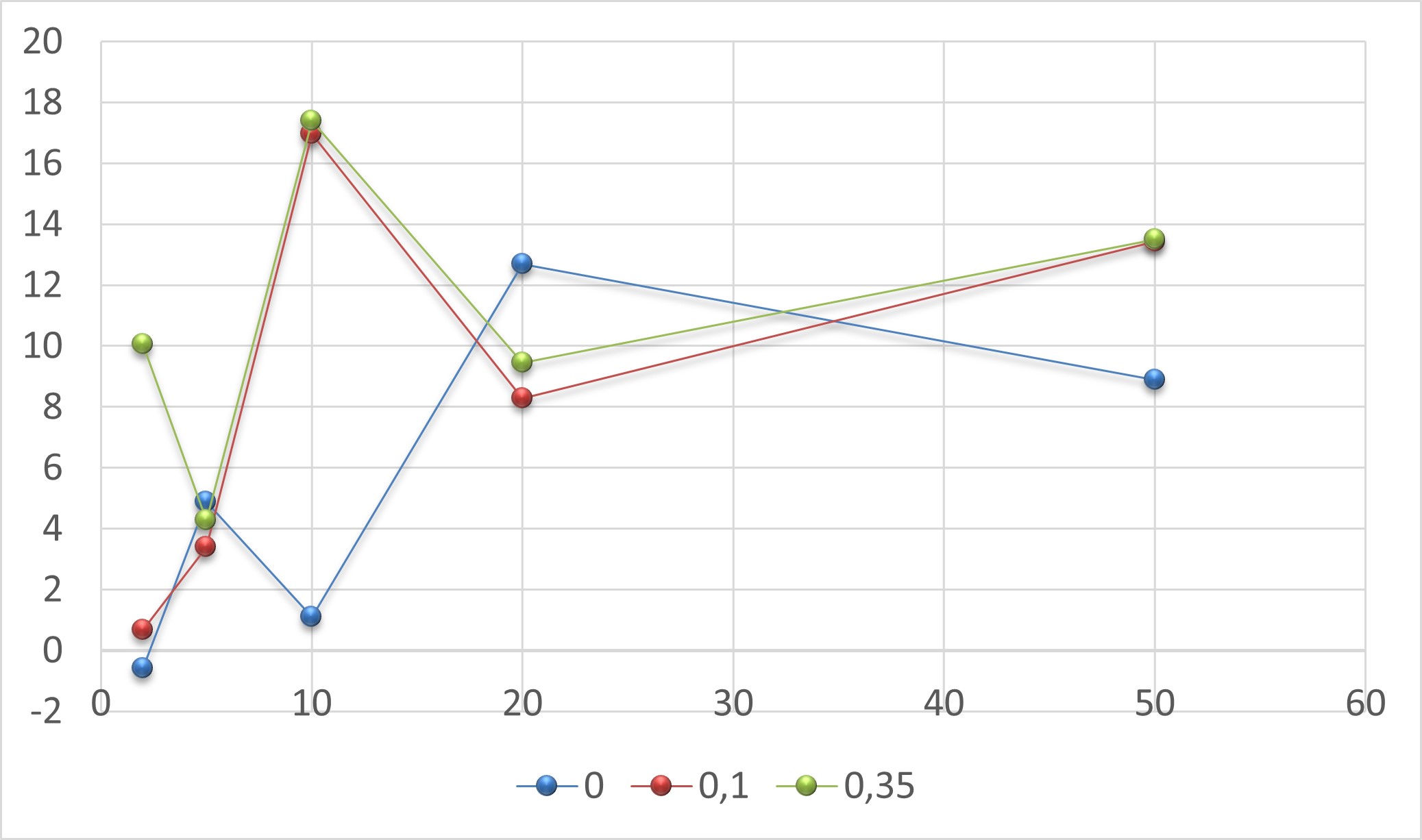}
        \caption{$\lambda_s=0$; $T=-1\%$}
        \label{sub11G10}
    \end{subfigure}
    \hfill
    \begin{subfigure}{0.45\textwidth}
        \centering
        \includegraphics[width=\textwidth]{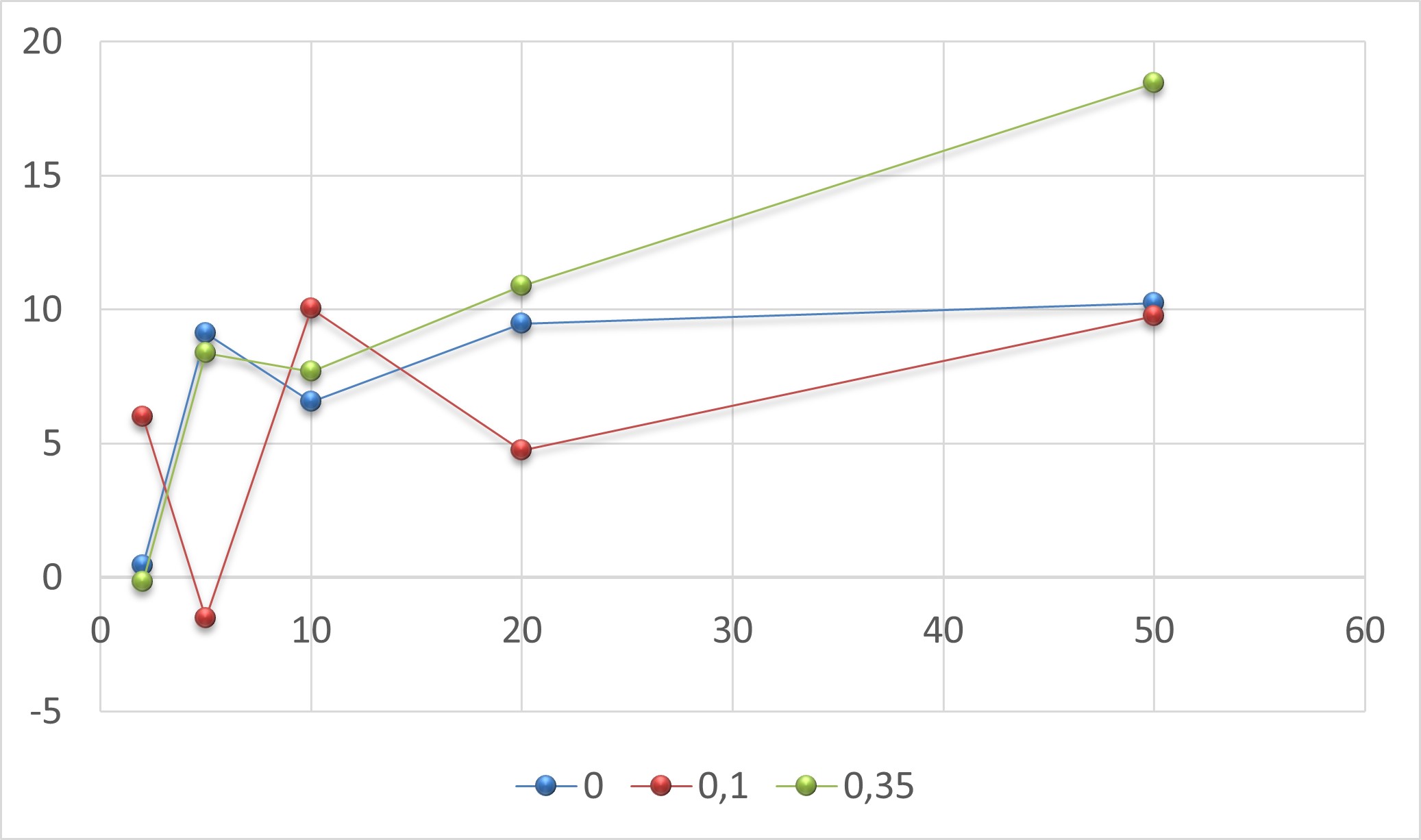}
        \caption{$\lambda_s=10$; $T=-1\%$}
        \label{sub12G10}
    \end{subfigure}
    \caption{Multi-step (10-month)  Sharpe ratios (Y-axis) for different number of predictors $M$ (X-axis), diversity parameter $\varepsilon=0,0.1,0.35$ (colours), s-RBFN regularization parameter $\lambda_s$, and $-1\%\leq T\leq 1\%$. s-RBFN with Gaussian basis functions.}
    \label{FigureSR_Gaus10-Simp}
\end{figure}

In Figure \ref{FigureSR_Gaus10-Simp}, the same set of experiments is carried out, but for the sequential multi-step decision-making case is presented. In this case, the resemblance is different from that shown in MPT and Figure \ref{DiverRayd}, which makes sense since those setups are one-step, whereas the results here are sequential multi-step cases. Nevertheless, the conclusions remain unchanged, and the same patterns can be observed regarding the relationship between diversity in ensemble learning and out-of-sample portfolio diversification. A consistent increase in Sharpe ratios can be seen as the parameter $\varepsilon$ and the number of predictors, hypotheses, or assets $M$ increase. On the other hand, the regularization parameter has a greater impact in this case than in the one-step scenario shown in Figure \ref{FigureSR_Gaus1-Simp}. It can be observed that the out-of-sample risk-adjusted portfolio performance improves even in the case where the diversity parameter $\varepsilon$ is not applied. This may indicate that performance in multi-step settings depends not only on diversification but also on the complexity of the data. The same experiments can be seen for the case of radial basis functions with the same conclusions in the Appendix in Figures \ref{FigureSR_rad1-Simp} and \ref{FigureSR_rad10-Simp}.

\subsubsection{Diversity-quality Trade-off of Returns Predictions: Including Diversity in the Asset Selection Stage}
\label{gammasection}
In this section, the same experiments are performed, but here the $M$ portfolio constituents or hypotheses are randomly selected from the top $m = M \times \gamma$ stocks based on the 1-month cumulative forecast ranking. The parameter of asset selection diversity $\gamma$ is a multiplier which scales the candidate pool relative to the final selection size $M$. This approach seeks to test the hypothesis that out-of-sample diversification in portfolio allocation is achieved not only during the optimization process—whether through the inclusion of the parameter $\varepsilon$ in this structured prediction model framework or through other plug-in methods without such aid—but also at the stage of hypothesis selection, when choosing assets or stocks prior to optimizing portfolio weights. For this purpose, the experiments are repeated for different values of the multiplier, including in the no-diversity case, which is equivalent to the results in the previous section, with \(\gamma = 1, 2, 3, 5\) for both one-step-ahead and multi-step-ahead cases, using the s-RBFN model with radial basis functions.
\par

\begin{figure}[H]
    \centering
    \begin{subfigure}{0.45\textwidth}
        \includegraphics[width=\linewidth]{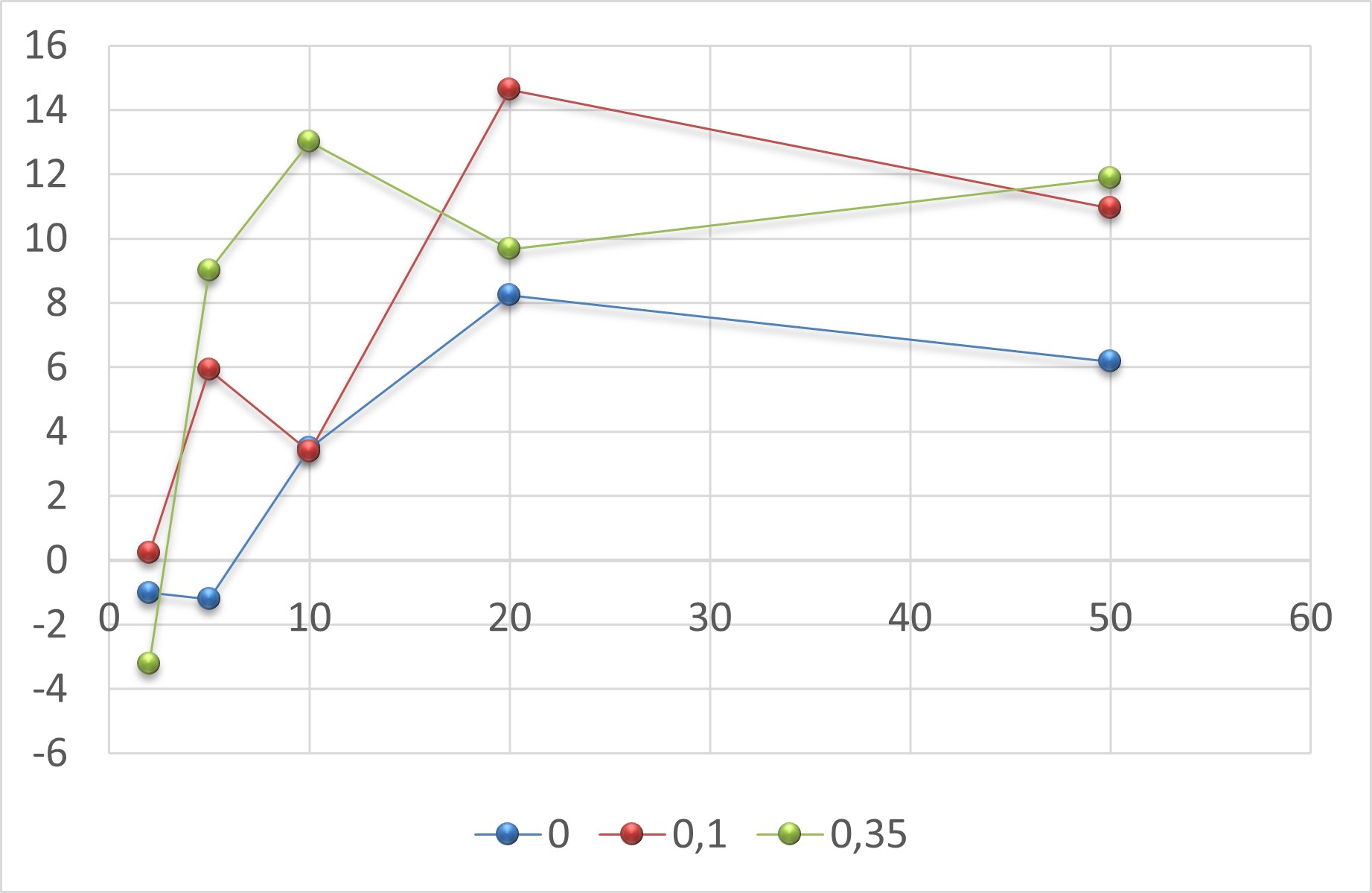} 
        \caption{$\gamma=1;0.5\%$}
        \label{fig:image110m}
    \end{subfigure}
    \begin{subfigure}{0.45\textwidth}
        \includegraphics[width=\linewidth]{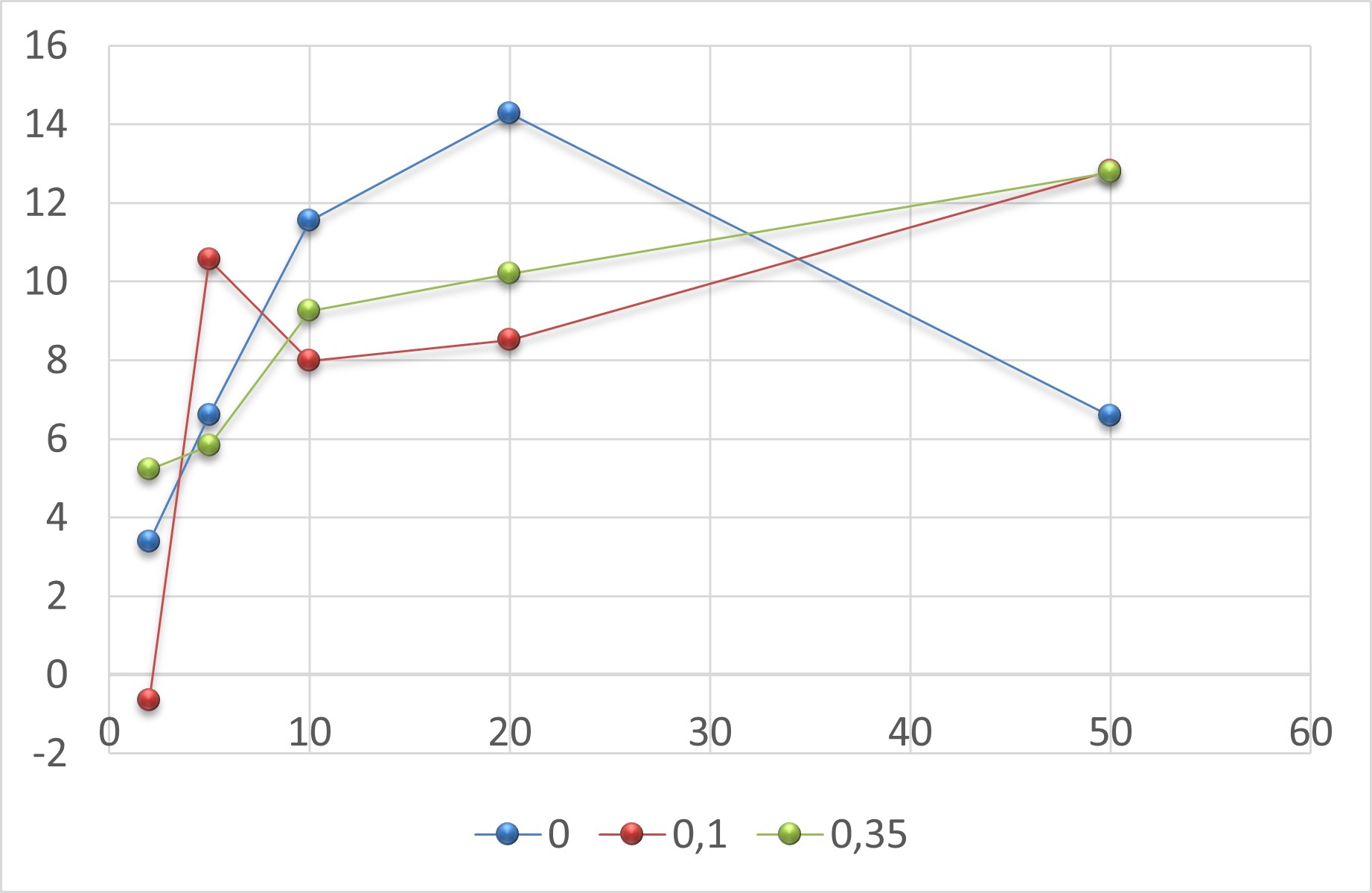} 
        \caption{$\gamma=2;0.5\%$}
        \label{fig:image210m}
    \end{subfigure}

    \begin{subfigure}{0.45\textwidth}
        \includegraphics[width=\linewidth]{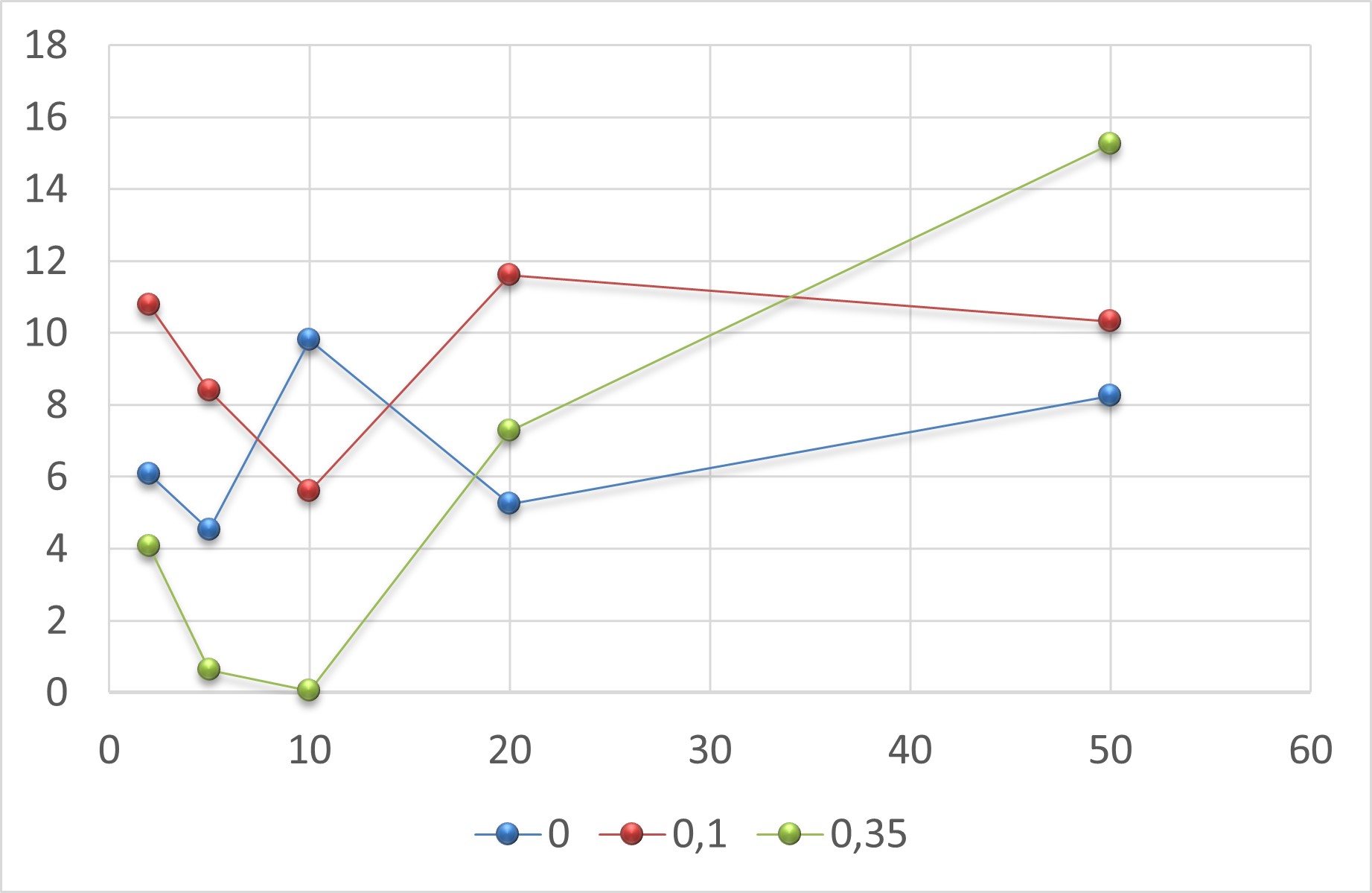} 
        \caption{$\gamma=3;0.5\%$}
        \label{fig:image310m}
    \end{subfigure}
    \begin{subfigure}{0.45\textwidth}
        \includegraphics[width=\linewidth]{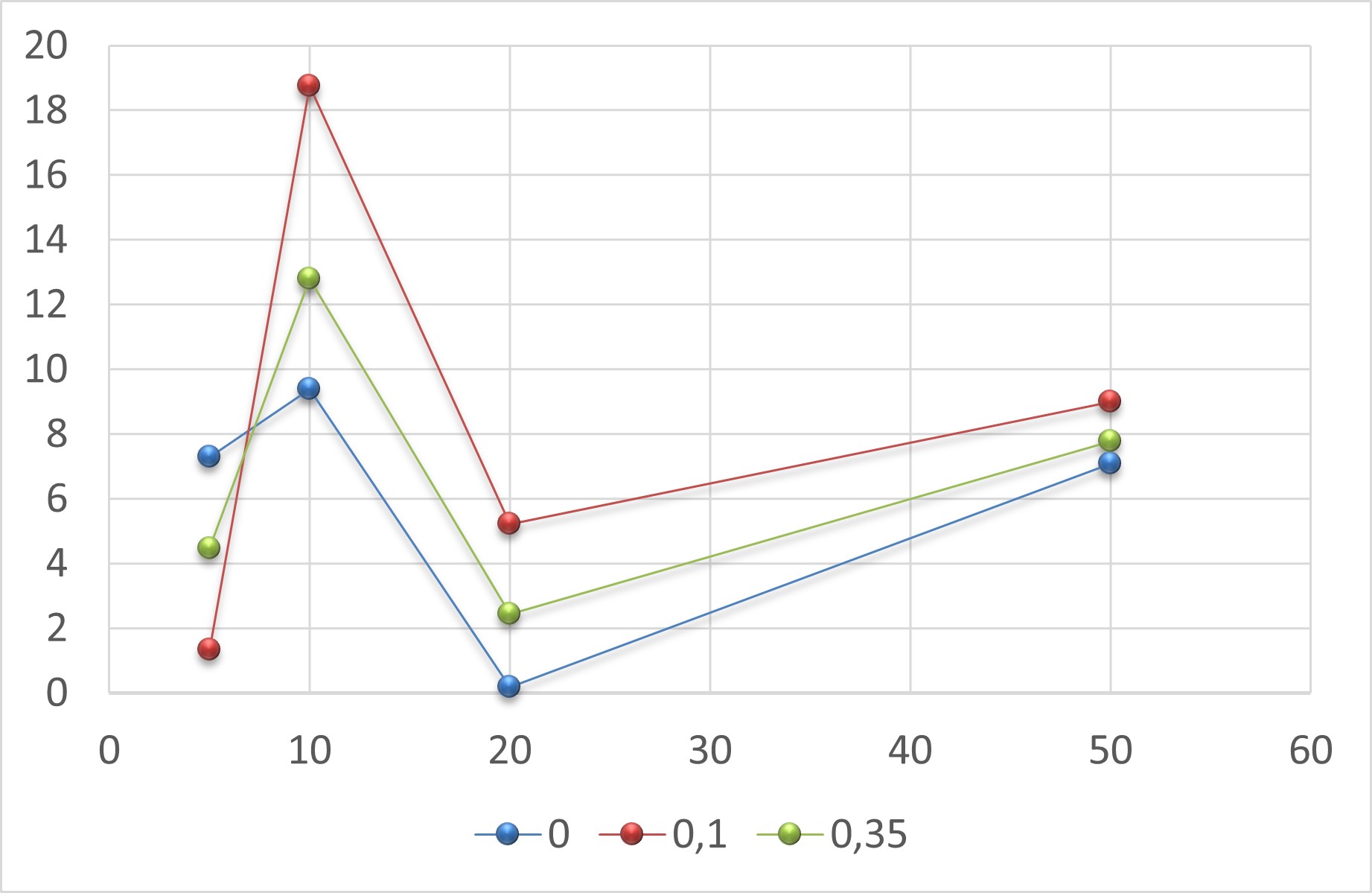} 
        \caption{$\gamma=5;0.5\%$}
        \label{fig:image410m}
    \end{subfigure}

    \begin{subfigure}{0.45\textwidth}
        \includegraphics[width=\linewidth]{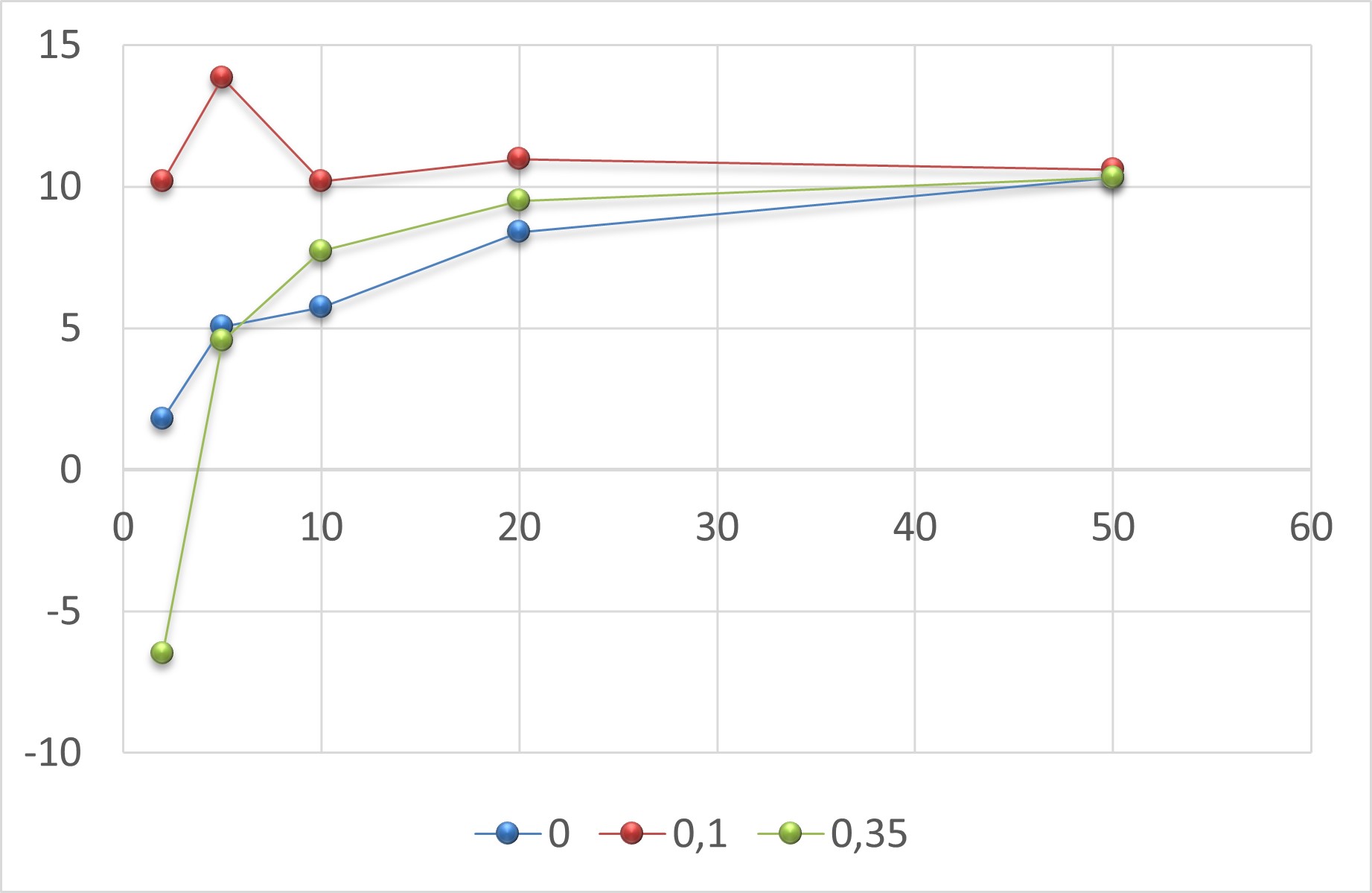} 
        \caption{$\gamma=1;0.0\%$}
        \label{fig:image510m}
    \end{subfigure}
    \begin{subfigure}{0.45\textwidth}
        \includegraphics[width=\linewidth]{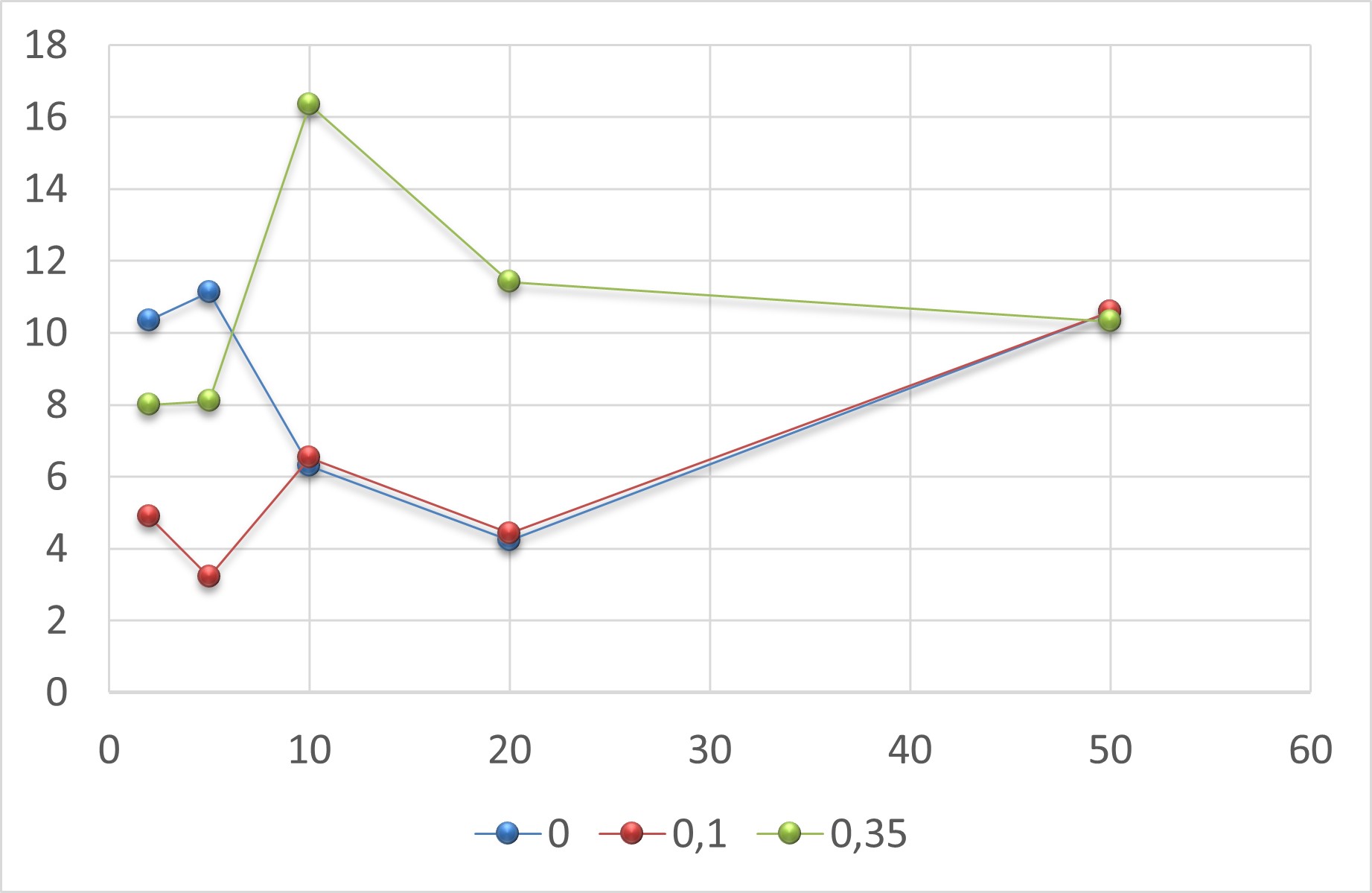} 
        \caption{$\gamma=2;0.0\%$}
        \label{fig:image610m}
    \end{subfigure}

    \begin{subfigure}{0.45\textwidth}
        \includegraphics[width=\linewidth]{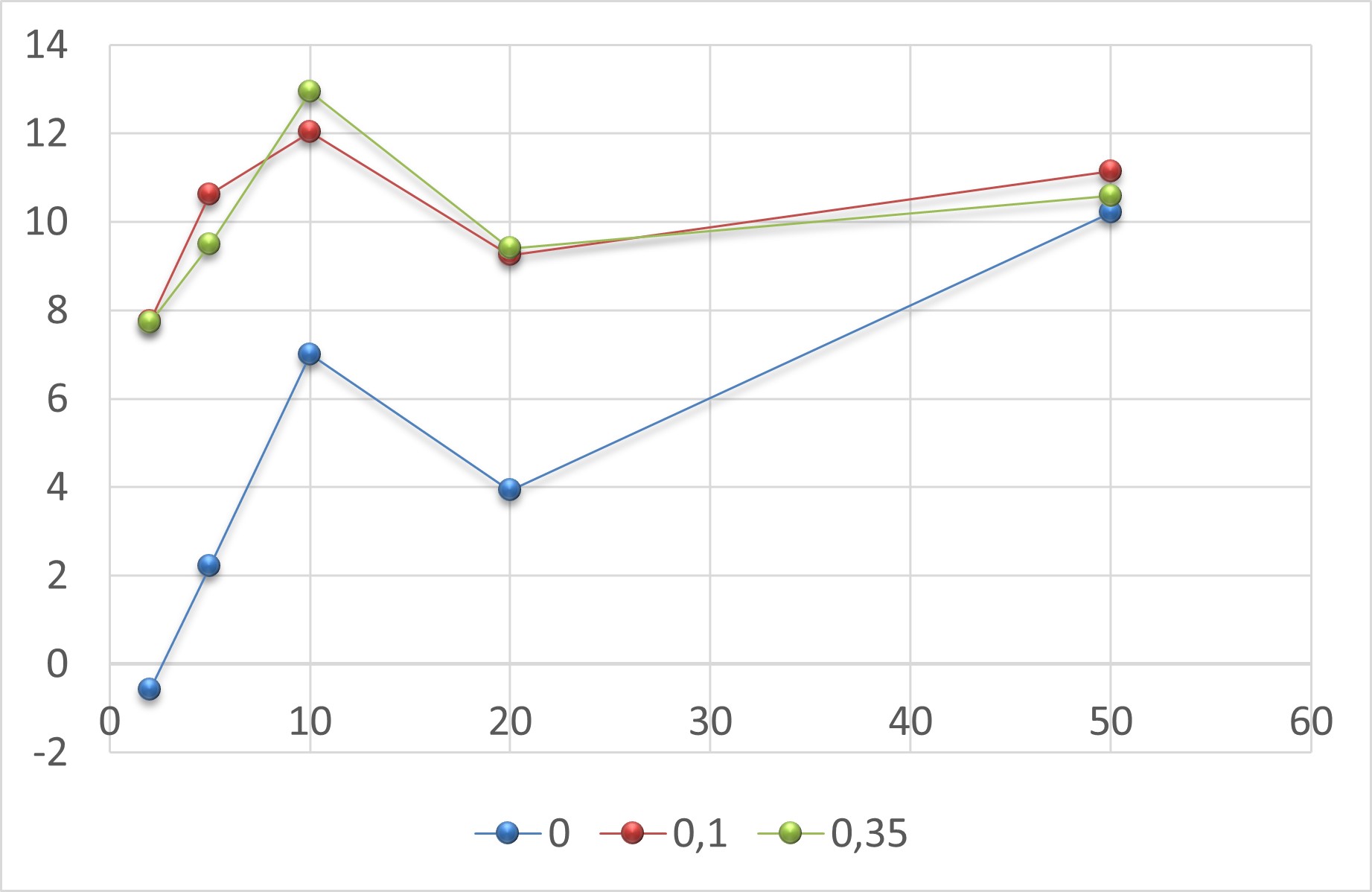} 
        \caption{$\gamma=3;0.0\%$}
        \label{fig:image710m}
    \end{subfigure}
    \begin{subfigure}{0.45\textwidth}
        \includegraphics[width=\linewidth]{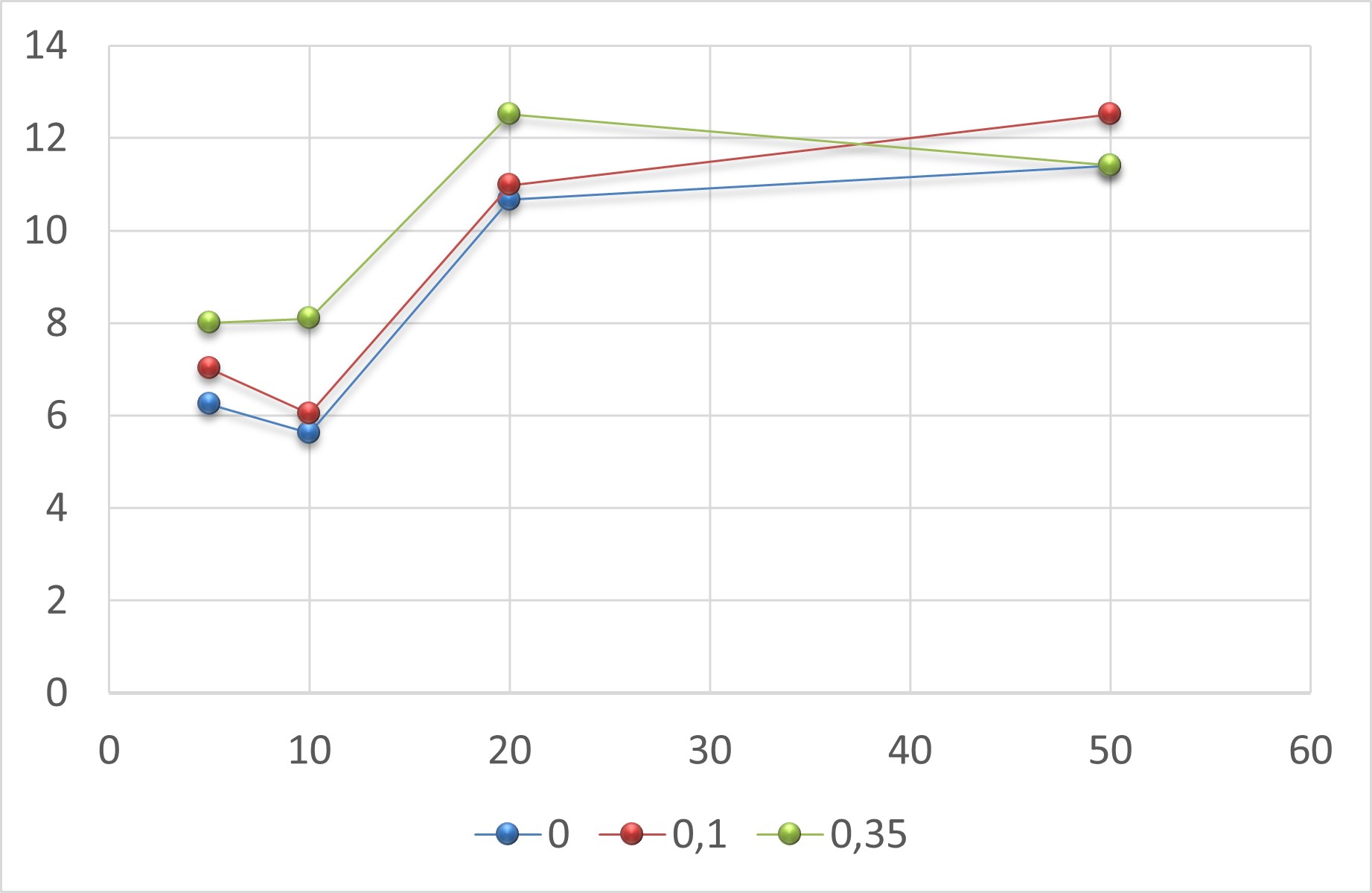} 
        \caption{$\gamma=5;0.0\%$}
        \label{fig:image810m}
    \end{subfigure}

    \caption{Multi-step (10-month)  Sharpe ratios (Y-axis) for different number of predictors $M$ (X-axis), diversity parameter $\varepsilon=0,0.1,0.35$ (colours), multiplicative factor $\gamma$, and $0\%\leq T\leq 0.5\%$. s-RBFN with radial basis functions.}
    \label{fig:large_figure10m}
\end{figure}

\begin{figure}[H]
    \centering
    \begin{subfigure}{0.45\textwidth}
        \includegraphics[width=\linewidth]{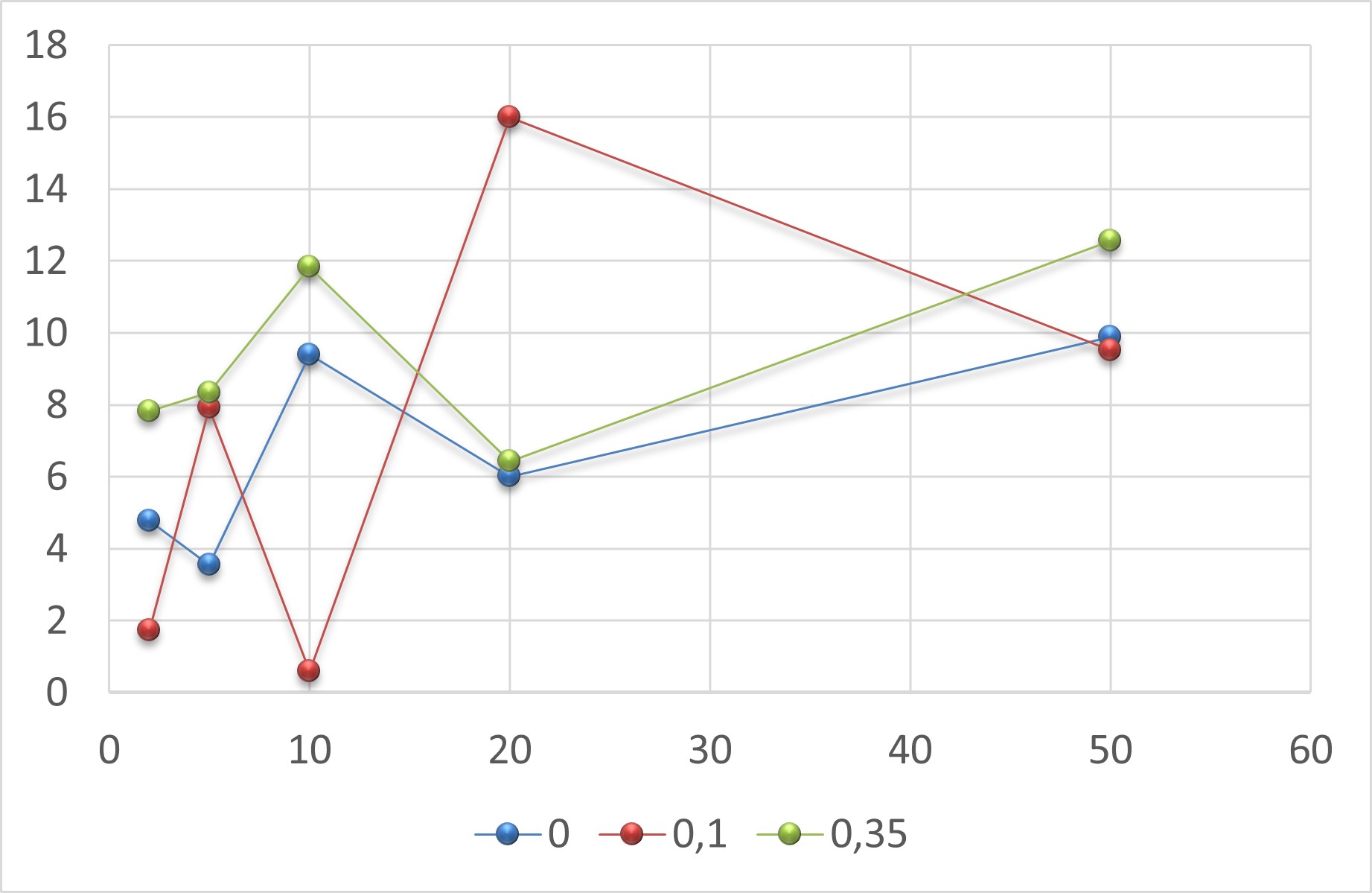} 
        \caption{$\gamma=1;-0.5\%$}
        \label{fig:image110min}
    \end{subfigure}
    \begin{subfigure}{0.45\textwidth}
        \includegraphics[width=\linewidth]{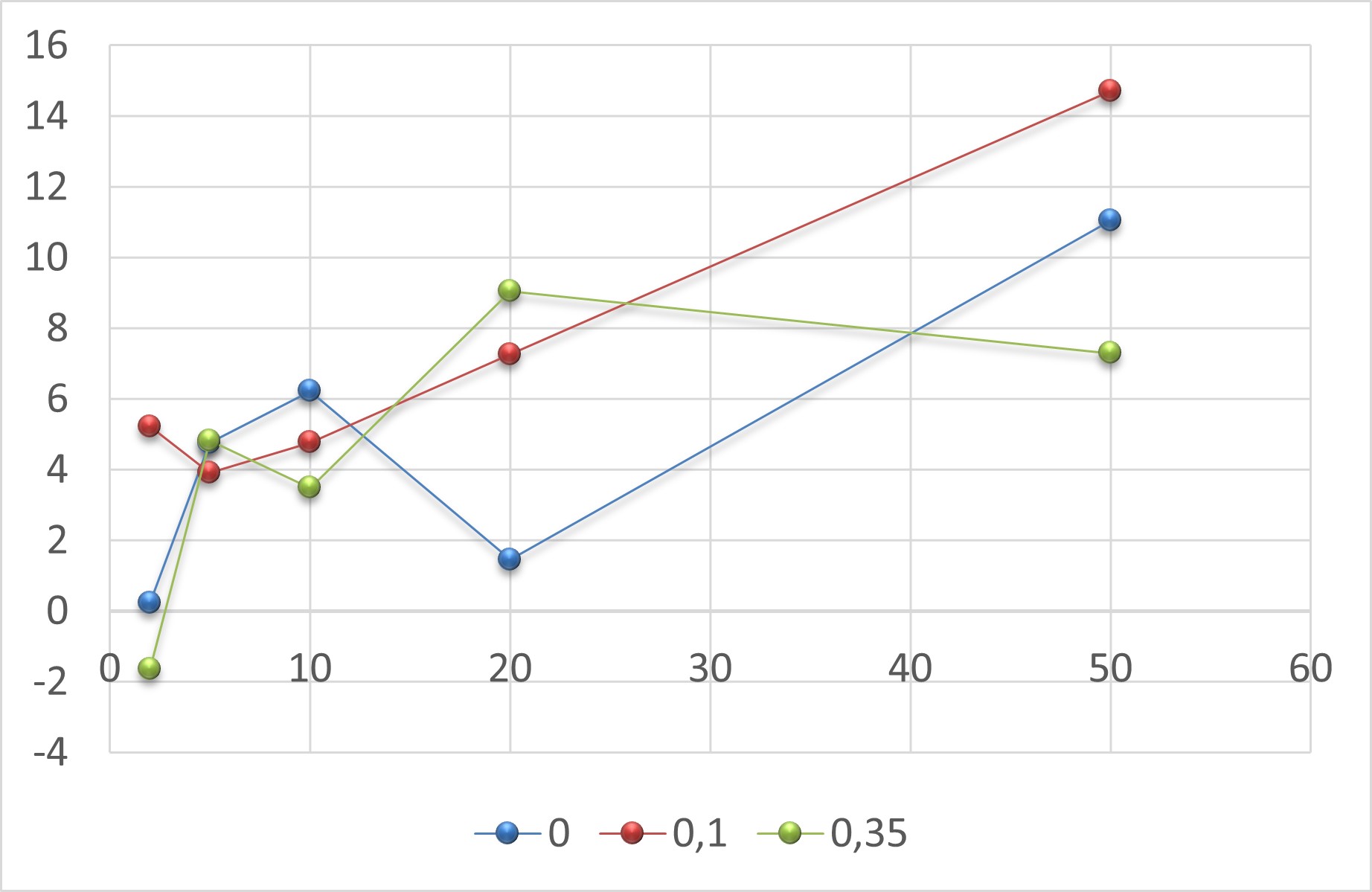} 
        \caption{$\gamma=2;-0.5\%$}
        \label{fig:image210min}
    \end{subfigure}

    \begin{subfigure}{0.45\textwidth}
        \includegraphics[width=\linewidth]{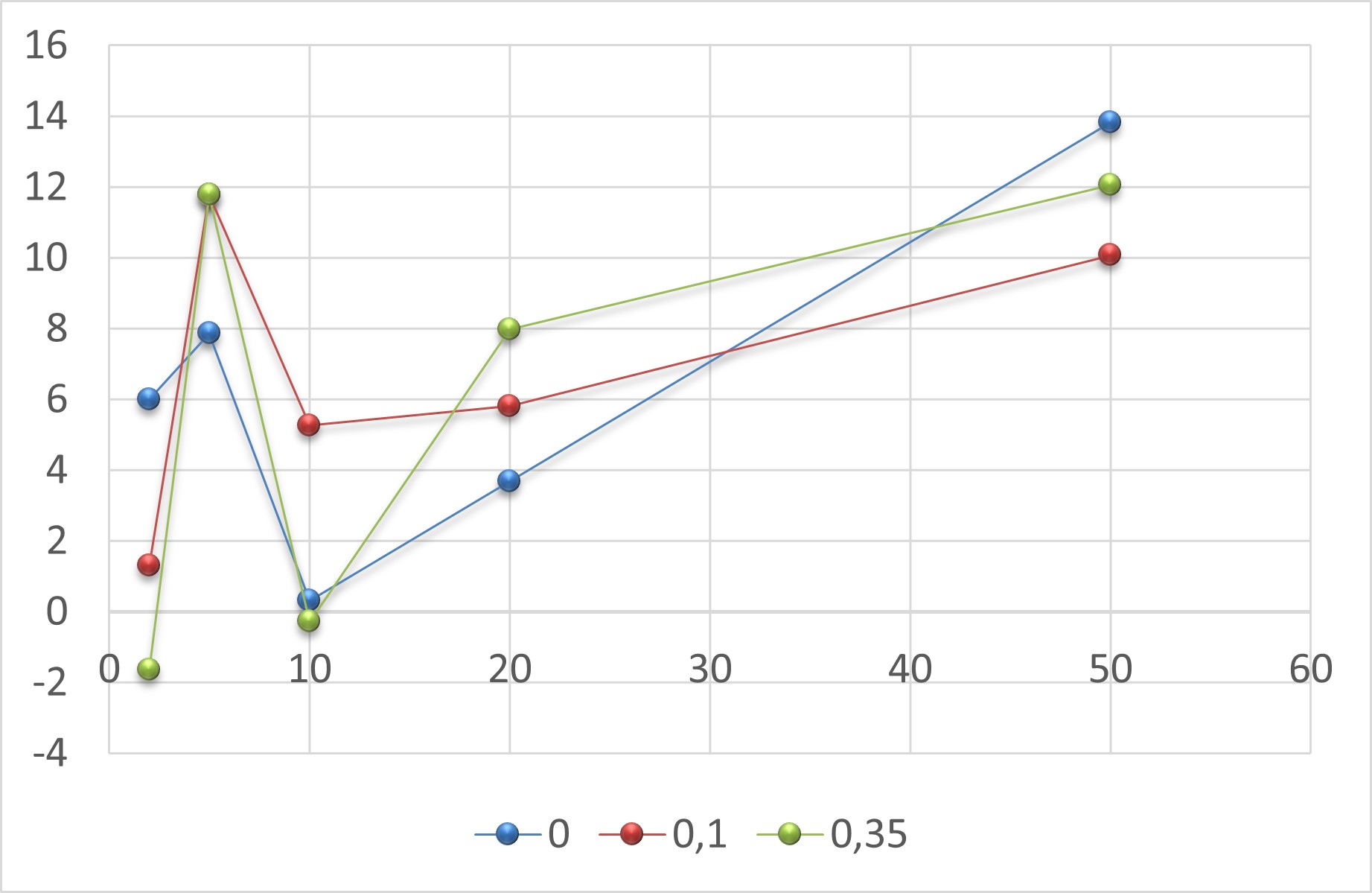} 
        \caption{$\gamma=3;-0.5\%$}
        \label{fig:image310min}
    \end{subfigure}
    \begin{subfigure}{0.45\textwidth}
        \includegraphics[width=\linewidth]{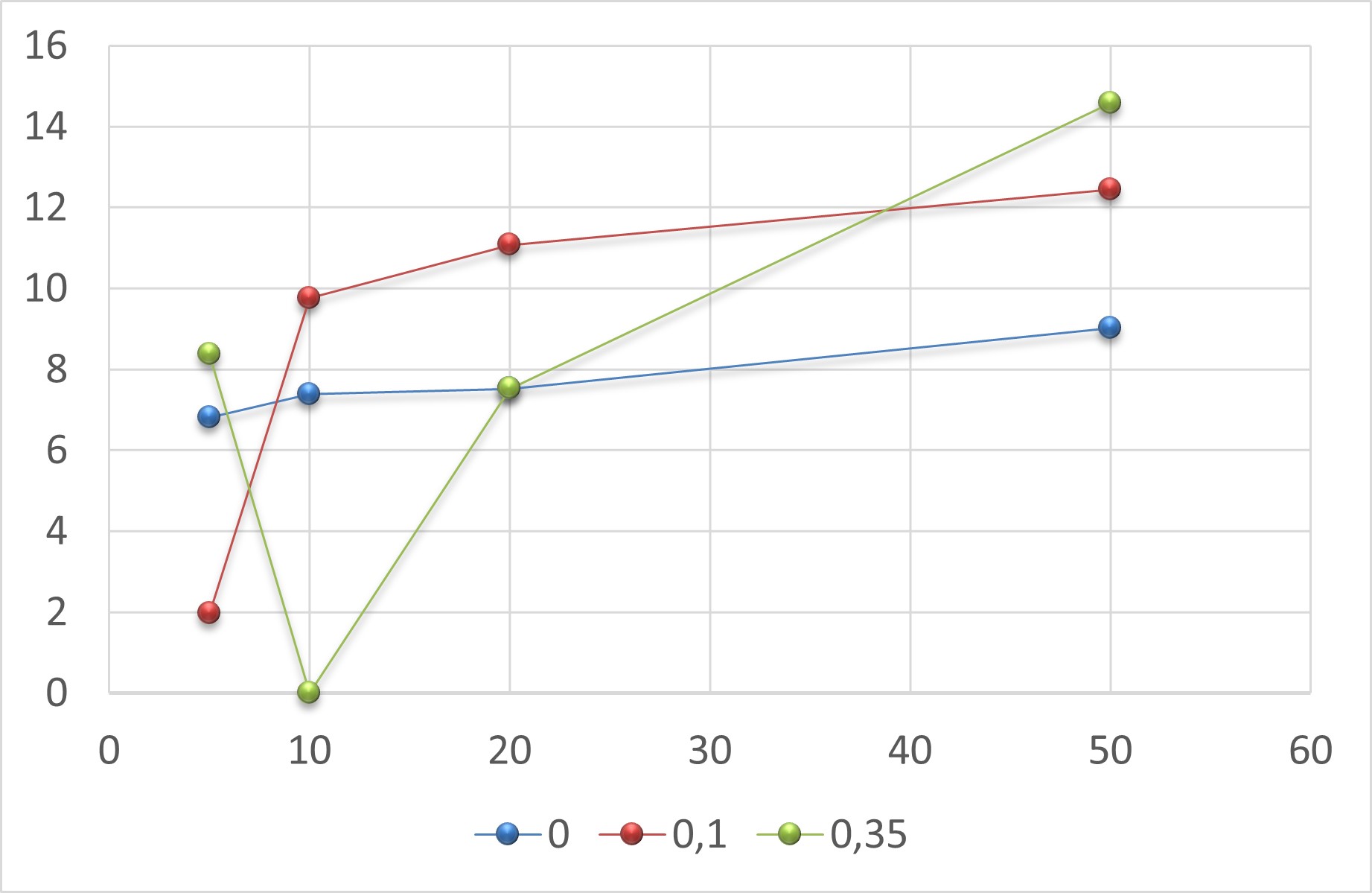} 
        \caption{$\gamma=5;-0.5\%$}
        \label{fig:image410min}
    \end{subfigure}

    \begin{subfigure}{0.45\textwidth}
        \includegraphics[width=\linewidth]{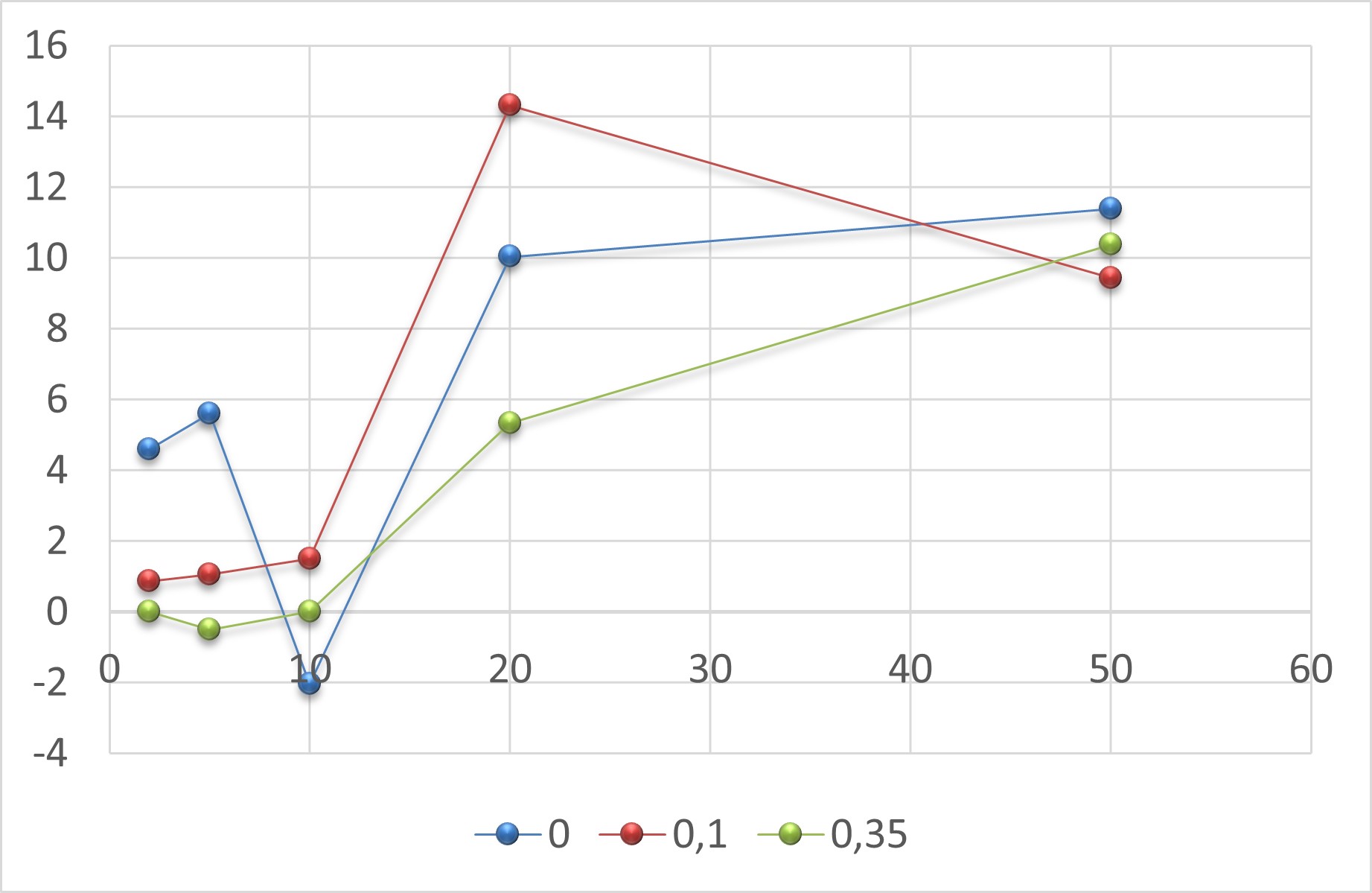} 
        \caption{$\gamma=1;-1\%$}
        \label{fig:image510min}
    \end{subfigure}
    \begin{subfigure}{0.45\textwidth}
        \includegraphics[width=\linewidth]{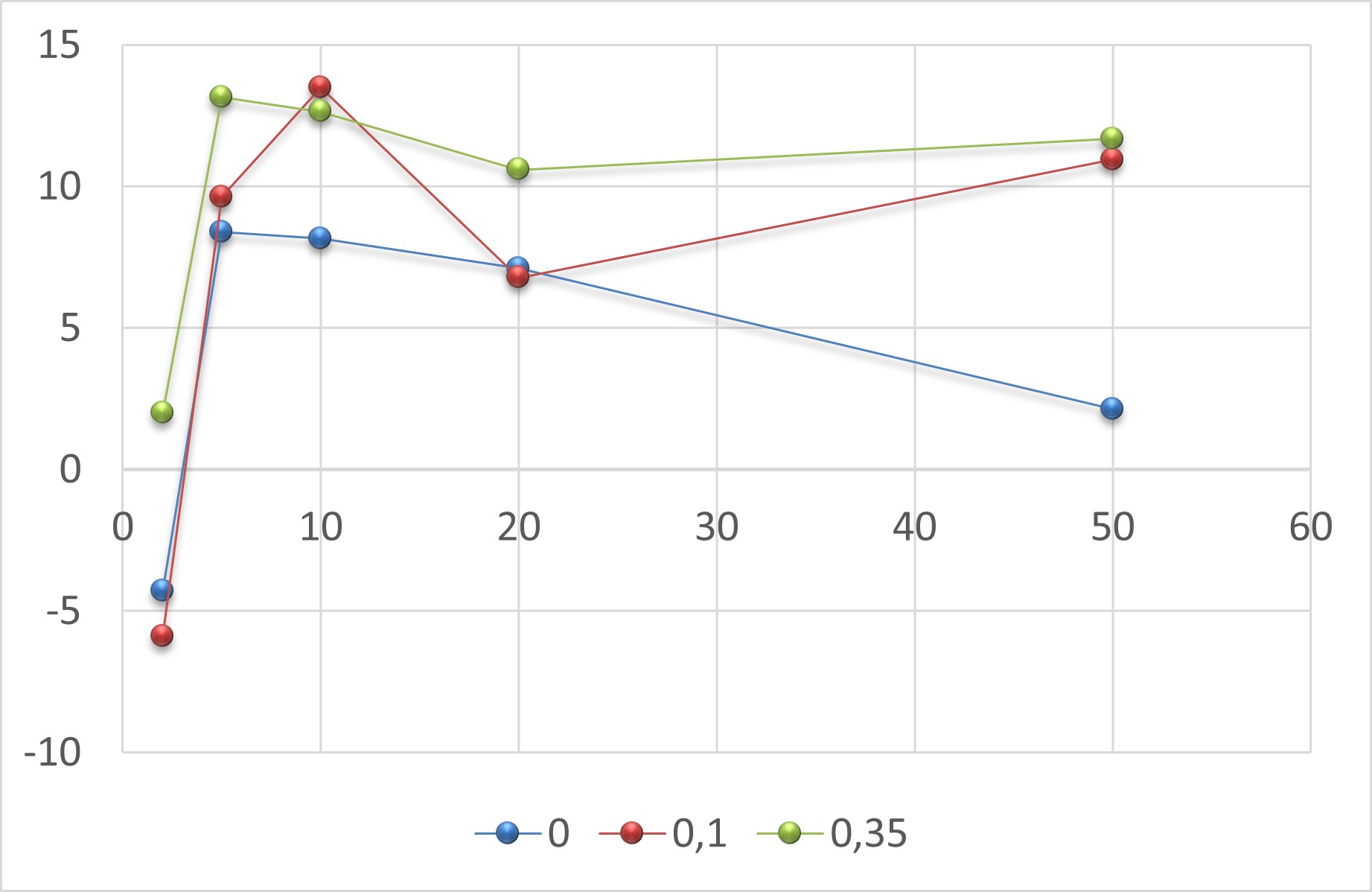} 
        \caption{$\gamma=2;-1\%$}
        \label{fig:image610min}
    \end{subfigure}

    \begin{subfigure}{0.45\textwidth}
        \includegraphics[width=\linewidth]{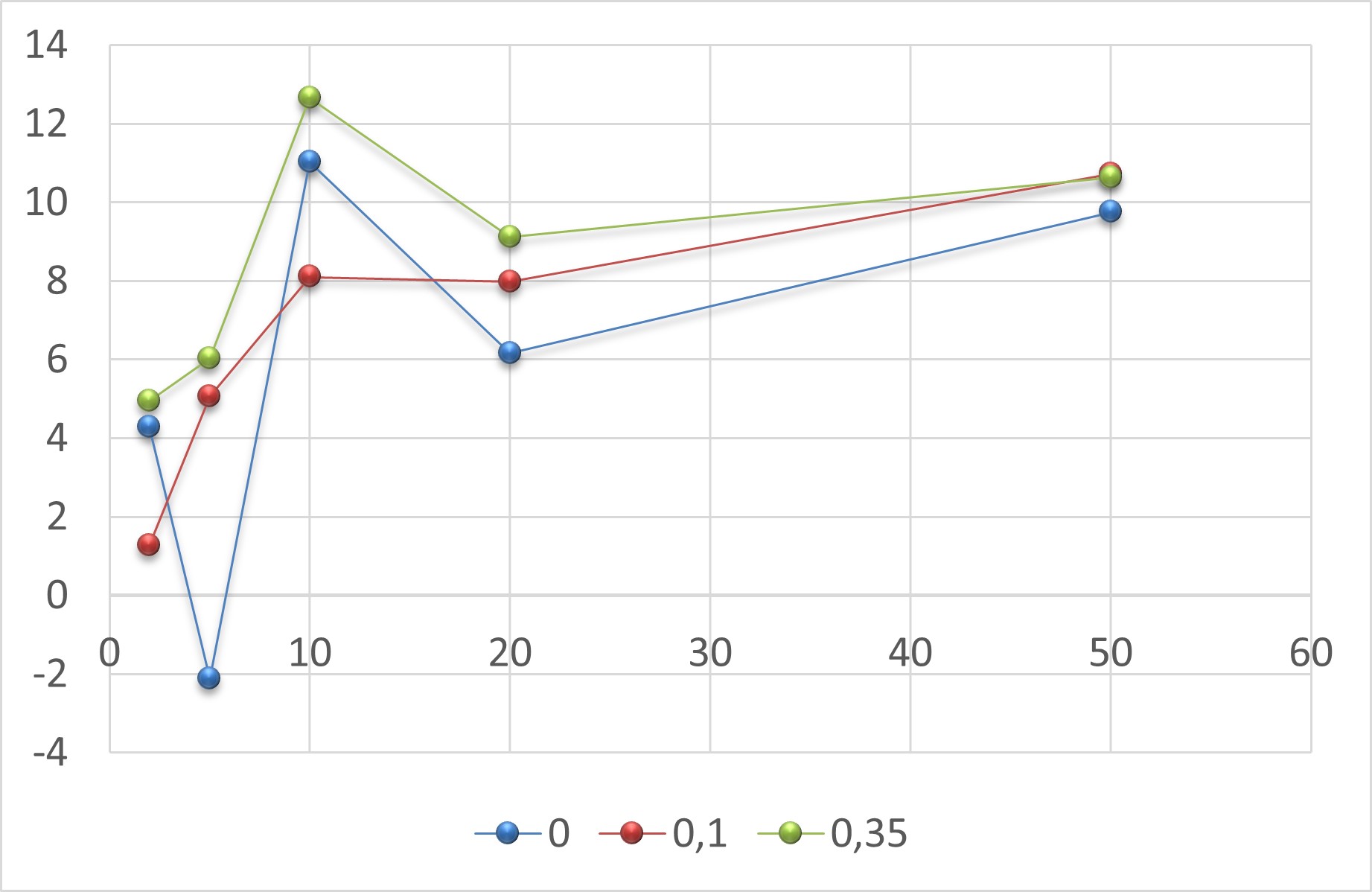} 
        \caption{$\gamma=3;-1\%$}
        \label{fig:image710min}
    \end{subfigure}
    \begin{subfigure}{0.45\textwidth}
        \includegraphics[width=\linewidth]{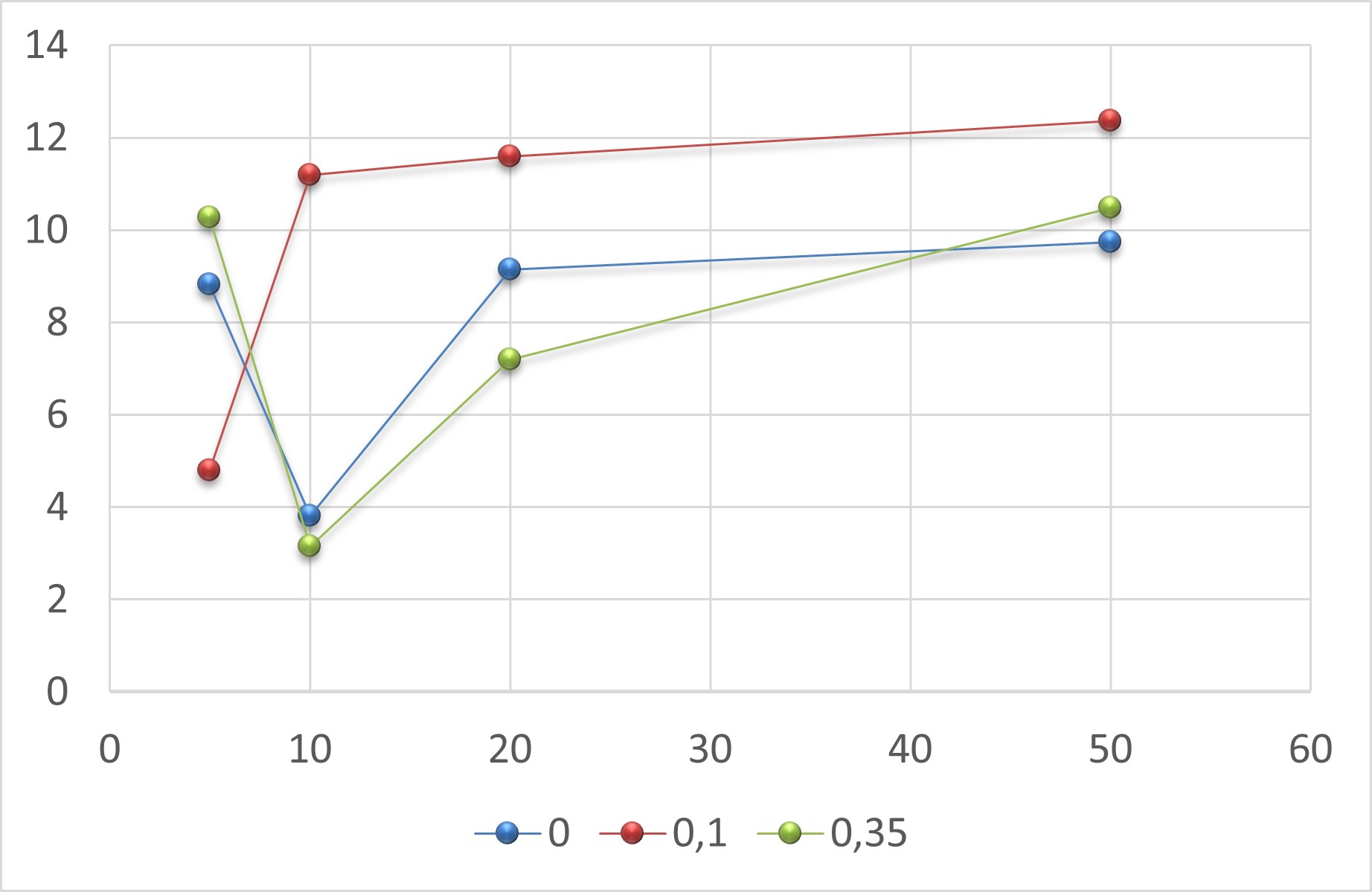} 
        \caption{$\gamma=5;-1\%$}
        \label{fig:image810min}
    \end{subfigure}

    \caption{Multi-step (10-month)  Sharpe ratios (Y-axis) for different number of predictors $M$ (X-axis), diversity parameter $\varepsilon=0,0.1,0.35$ (colours), multiplicative factor $\gamma$, and $-1\%\leq T\leq -0.5\%$. s-RBFN with radial basis functions.}
    \label{fig:large_figure10min}
\end{figure}

Figures \ref{fig:large_figure10m} and \ref{fig:large_figure10min} present the out-of-sample Sharpe ratios for different numbers of stocks \(M\); various values of the asset selection diversity parameter \(\gamma\); the learning stage diversity parameter \(\varepsilon\); 
and threshold values \(T\) used in the asset selection process based on forecast rankings, for the multi-step sequential (10-month strategy) case. Figures \ref{fig:large_figure1m} and \ref{fig:large_figure1mmin} in the Appendix show the same results for the one-step decision case (1-month, 100 simulations).

Within the proposed PSEM framework, the Sharpe ratios, along with the average Sharpe ratio for all levels of diversity determined by the structured ensemble diversity parameter \((\varepsilon)\), can also be analyzed to examine the impact of diversity in asset or hypothesis selection on out-of-sample portfolio diversification. These results are presented in Table \ref{tab:rmse_summary_combined1m} for the one-step-ahead case (1-month) and Table \ref{tab:rmse_summary_combined_multistep} for the multi-step experiments (10-month strategy), both using the s-RBFN model with radial basis functions.

\begin{table}[H]
\centering
\captionsetup{justification=centering}
\caption{One-step-ahead experiments (1-month; 100 simulations): Sharpe ratios for different diversity parameters in the learning stage \(\varepsilon = 0, 0.1, 0.35\), number of stocks \(M\), and multiplicative factor or diversity parameter in the asset selection stage \(\gamma = 1, 2, 3, 5\). \(T\) represents the threshold in the ranking of return predictions. Avg. Sh. refers to the average Sharpe ratio of the elements in the corresponding column.}
\label{tab:rmse_summary_combined1m}
\renewcommand{\arraystretch}{1.1}
\setlength{\tabcolsep}{2.5pt}
\normalsize
\begin{tabular}{l|ccccc|ccccc|ccccc|ccccc}
\textbf{Div. $(\varepsilon)$} & \multicolumn{5}{c|}{\textbf{($T=0.5\%$)}} & \multicolumn{5}{c|}{\textbf{($T=0.0\%$)}} & \multicolumn{5}{c|}{\textbf{($T=-0.5\%$)}} & \multicolumn{5}{c}{\textbf{($T=-1\%$)}} \\
\textbf{} & \textbf{2} & \textbf{5} & \textbf{10} & \textbf{20} & \textbf{50} & \textbf{2} & \textbf{5} & \textbf{10} & \textbf{20} & \textbf{50} & \textbf{2} & \textbf{5} & \textbf{10} & \textbf{20} & \textbf{50} & \textbf{2} & \textbf{5} & \textbf{10} & \textbf{20} & \textbf{50} \\\hline
\multicolumn{21}{c}{\textbf{Div. $(\gamma=1)$};\  \ \(m = M \times \gamma = \{2, 5, 10, 20, 50\}\)} \\\hline
0    & \textcolor{red}{3} & \textcolor{red}{5} & 6 & \textcolor{ForestGreen}{8} & \textcolor{ForestGreen}{10} & \textcolor{red}{3} & \textcolor{red}{4} & \textcolor{ForestGreen}{6} & \textcolor{ForestGreen}{8} & \textcolor{ForestGreen}{10} & \textcolor{red}{4} & \textcolor{red}{5} & \textcolor{ForestGreen}{7} & \textcolor{ForestGreen}{8} & \textcolor{ForestGreen}{9} & \textcolor{red}{3} & \textcolor{red}{5} & \textcolor{ForestGreen}{6} & \textcolor{ForestGreen}{8} & \textcolor{ForestGreen}{10} \\
0.1  & \textcolor{red}{4} & 6 & 6 & \textcolor{ForestGreen}{9} & \textcolor{ForestGreen}{10} & \textcolor{red}{4} & \textcolor{ForestGreen}{7} & \textcolor{ForestGreen}{7} & \textcolor{ForestGreen}{8} & \textcolor{ForestGreen}{10} & \textcolor{red}{4} & \textcolor{red}{5} & \textcolor{ForestGreen}{8} & \textcolor{ForestGreen}{8} & \textcolor{ForestGreen}{11} & \textcolor{red}{1} & \textcolor{red}{6} & \textcolor{ForestGreen}{7} & \textcolor{ForestGreen}{8} & \textcolor{ForestGreen}{10} \\
0.35 & \textcolor{ForestGreen}{5} & 6 & \textcolor{ForestGreen}{7} & \textcolor{ForestGreen}{9} & \textcolor{ForestGreen}{10} & \textcolor{ForestGreen}{5} & \textcolor{red}{5} & \textcolor{ForestGreen}{6} & \textcolor{ForestGreen}{8} & \textcolor{ForestGreen}{10} & \textcolor{red}{3} & \textcolor{red}{6} & \textcolor{ForestGreen}{7} & \textcolor{ForestGreen}{8} & \textcolor{ForestGreen}{10} & \textcolor{red}{0} & \textcolor{red}{6} & \textcolor{ForestGreen}{8} & \textcolor{ForestGreen}{9} & \textcolor{ForestGreen}{10} \\
\textbf{Avg. Sh.} & \textcolor{red}{4} & 6 & 6 & \textcolor{ForestGreen}{8} & \textcolor{ForestGreen}{10} & \textcolor{red}{4} & \textcolor{red}{5} & \textcolor{ForestGreen}{6} & \textcolor{ForestGreen}{8} & \textcolor{ForestGreen}{10} & \textcolor{red}{4} & \textcolor{red}{5} & \textcolor{ForestGreen}{7} & \textcolor{ForestGreen}{8} & \textcolor{ForestGreen}{10} & \textcolor{red}{1} & \textcolor{red}{6} & \textcolor{ForestGreen}{7} & \textcolor{ForestGreen}{8} & \textcolor{ForestGreen}{10} \\\hline
\multicolumn{21}{c}{\textbf{Div. $(\gamma=2)$};\  \ \(m = M \times \gamma = \{4, 10, 20, 40, 100\}\)} \\\hline
0    & \textcolor{red}{4} & 5 & 6 & \textcolor{ForestGreen}{8} & \textcolor{ForestGreen}{9} & \textcolor{red}{4} & \textcolor{ForestGreen}{6} & \textcolor{ForestGreen}{6} & \textcolor{ForestGreen}{7} & \textcolor{ForestGreen}{10} & \textcolor{red}{4} & \textcolor{red}{5} & \textcolor{ForestGreen}{6} & \textcolor{ForestGreen}{8} & \textcolor{ForestGreen}{10} & \textcolor{red}{4} & \textcolor{red}{5} & \textcolor{red}{5} & \textcolor{ForestGreen}{7} & \textcolor{ForestGreen}{9} \\
0.1  & \textcolor{red}{4} & 5 & \textcolor{ForestGreen}{8} & \textcolor{ForestGreen}{9} & \textcolor{ForestGreen}{11} & \textcolor{red}{4} & \textcolor{red}{5} & \textcolor{ForestGreen}{7} & \textcolor{ForestGreen}{9} & \textcolor{ForestGreen}{10} & \textcolor{red}{4} & \textcolor{red}{5} & \textcolor{ForestGreen}{7} & \textcolor{ForestGreen}{8} & \textcolor{ForestGreen}{10} & \textcolor{red}{3} & \textcolor{red}{5} & \textcolor{ForestGreen}{8} & \textcolor{ForestGreen}{9} & \textcolor{ForestGreen}{11} \\
0.35 & 3 & 5 & \textcolor{ForestGreen}{8} & \textcolor{ForestGreen}{9} & \textcolor{ForestGreen}{11} & \textcolor{ForestGreen}{6} & \textcolor{ForestGreen}{8} & \textcolor{ForestGreen}{9} & \textcolor{ForestGreen}{10} & \textcolor{ForestGreen}{10} & \textcolor{red}{4} & \textcolor{red}{6} & \textcolor{ForestGreen}{7} & \textcolor{ForestGreen}{8} & \textcolor{ForestGreen}{10} & \textcolor{red}{3} & \textcolor{red}{6} & \textcolor{ForestGreen}{7} & \textcolor{ForestGreen}{9} & \textcolor{ForestGreen}{11} \\
\textbf{Avg. Sh.} & \textcolor{red}{4} & 5 & \textcolor{ForestGreen}{7} & \textcolor{ForestGreen}{9} & \textcolor{ForestGreen}{11} & \textcolor{red}{5} & \textcolor{ForestGreen}{6} & \textcolor{ForestGreen}{7} & \textcolor{ForestGreen}{8} & \textcolor{ForestGreen}{10} & \textcolor{red}{4} & \textcolor{red}{5} & \textcolor{ForestGreen}{7} & \textcolor{ForestGreen}{8} & \textcolor{ForestGreen}{10} & \textcolor{red}{3} & \textcolor{red}{5} & \textcolor{ForestGreen}{7} & \textcolor{ForestGreen}{8} & \textcolor{ForestGreen}{10} \\\hline
\multicolumn{21}{c}{\textbf{Div. $(\gamma=3)$};\  \ \(m = M \times \gamma = \{6, 15, 30, 60, 150\}\)} \\\hline
0    & \textcolor{red}{4} & 5 & 6 & 7 & \textcolor{ForestGreen}{9} & \textcolor{red}{4} & \textcolor{ForestGreen}{6} & \textcolor{ForestGreen}{7} & \textcolor{ForestGreen}{8} & \textcolor{ForestGreen}{9} & \textcolor{red}{4} & \textcolor{red}{5} & \textcolor{ForestGreen}{6} & \textcolor{ForestGreen}{8} & \textcolor{ForestGreen}{10} & \textcolor{red}{5} & \textcolor{red}{5} & \textcolor{ForestGreen}{6} & \textcolor{ForestGreen}{8} & \textcolor{ForestGreen}{10} \\
0.1  & 5 & 5 & 7 & 9 & \textcolor{ForestGreen}{11} & \textcolor{ForestGreen}{6} & \textcolor{ForestGreen}{7} & \textcolor{ForestGreen}{8} & \textcolor{ForestGreen}{11} & \textcolor{red}{3} & \textcolor{red}{4} & \textcolor{red}{6} & \textcolor{ForestGreen}{7} & \textcolor{ForestGreen}{9} & \textcolor{red}{1} & \textcolor{red}{6} & \textcolor{ForestGreen}{7} & \textcolor{ForestGreen}{8} & \textcolor{ForestGreen}{11} \\
0.35 & 4 & 5 & 7 & 9 & \textcolor{ForestGreen}{11} & \textcolor{ForestGreen}{6} & \textcolor{ForestGreen}{6} & \textcolor{ForestGreen}{7} & \textcolor{ForestGreen}{9} & \textcolor{ForestGreen}{10} & \textcolor{red}{3} & \textcolor{red}{5} & \textcolor{ForestGreen}{7} & \textcolor{ForestGreen}{9} & \textcolor{ForestGreen}{11} & \textcolor{red}{3} & \textcolor{red}{5} & \textcolor{ForestGreen}{7} & \textcolor{ForestGreen}{9} & \textcolor{ForestGreen}{11} \\
\textbf{Avg. Sh.} & 4 & 5 & 7 & 9 & \textcolor{ForestGreen}{10} & \textcolor{red}{5} & \textcolor{ForestGreen}{6} & \textcolor{ForestGreen}{8} & \textcolor{ForestGreen}{9} & \textcolor{red}{7} & \textcolor{red}{3} & \textcolor{red}{5} & \textcolor{ForestGreen}{7} & \textcolor{ForestGreen}{9} & \textcolor{ForestGreen}{10} & \textcolor{red}{3} & \textcolor{red}{5} & \textcolor{ForestGreen}{7} & \textcolor{ForestGreen}{9} & \textcolor{ForestGreen}{10} \\\hline
\multicolumn{21}{c}{\textbf{Div. $(\gamma=5)$};\  \ \(m = M \times \gamma = \{25, 50, 100, 250\}\)} \\\hline
0    &     & \textcolor{ForestGreen}{6} & \textcolor{ForestGreen}{8} & \textcolor{ForestGreen}{7} & \textcolor{ForestGreen}{11} &     & \textcolor{red}{5} & \textcolor{ForestGreen}{6} & \textcolor{ForestGreen}{8} & \textcolor{ForestGreen}{10} &     & \textcolor{red}{5} & \textcolor{ForestGreen}{7} & \textcolor{ForestGreen}{8} & \textcolor{ForestGreen}{10} &     & \textcolor{red}{5} & \textcolor{ForestGreen}{6} & \textcolor{ForestGreen}{8} & \textcolor{ForestGreen}{8} \\
0.1  &     & \textcolor{ForestGreen}{6} & 7 & \textcolor{red}{3} & 6  &     & \textcolor{red}{5} & \textcolor{ForestGreen}{7} & \textcolor{ForestGreen}{9} & \textcolor{ForestGreen}{10} &     & \textcolor{red}{6} & \textcolor{ForestGreen}{8} & \textcolor{ForestGreen}{9} & \textcolor{ForestGreen}{11} &     & \textcolor{red}{5} & \textcolor{ForestGreen}{8} & \textcolor{ForestGreen}{9} & \textcolor{ForestGreen}{12} \\
0.35 &     & 5 & \textcolor{ForestGreen}{11} & \textcolor{ForestGreen}{9} & \textcolor{ForestGreen}{11} &     & \textcolor{ForestGreen}{6} & \textcolor{ForestGreen}{8} & \textcolor{ForestGreen}{10} & \textcolor{ForestGreen}{10} &     & \textcolor{red}{6} & \textcolor{ForestGreen}{8} & \textcolor{ForestGreen}{8} & \textcolor{ForestGreen}{11} &     & \textcolor{red}{6} & \textcolor{ForestGreen}{7} & \textcolor{ForestGreen}{9} & \textcolor{ForestGreen}{11} \\
\textbf{Avg. Sh} &     & 5 & \textcolor{ForestGreen}{9} & 6 & \textcolor{ForestGreen}{9}  &     & \textcolor{red}{5} & \textcolor{ForestGreen}{7} & \textcolor{ForestGreen}{9} & \textcolor{ForestGreen}{10} &     & \textcolor{red}{6} & \textcolor{ForestGreen}{8} & \textcolor{ForestGreen}{8} & \textcolor{ForestGreen}{11} &     & \textcolor{red}{5} & \textcolor{ForestGreen}{7} & \textcolor{ForestGreen}{9} & \textcolor{ForestGreen}{10} \\\hline
\end{tabular}
\end{table}

\begin{table}[H]
\centering
\captionsetup{justification=centering}
\caption{Multi-step-ahead experiments (10-month sequential strategy): Sharpe ratios for different diversity parameters in the learning stage \(\varepsilon = 0, 0.1, 0.35\), number of stocks \(M\), and multiplicative factor or diversity parameter in the asset selection stage \(\gamma = 1, 2, 3, 5\). \(T\) represents the threshold in the ranking of return predictions. Avg. Sh. refers to the average Sharpe ratio of the elements in the corresponding column.}
\label{tab:rmse_summary_combined_multistep}
\renewcommand{\arraystretch}{1.2}
\setlength{\tabcolsep}{3.5pt}
\footnotesize
\begin{tabular}{l|ccccc|ccccc|ccccc|ccccc}
\textbf{Div. $(\varepsilon)$} & \multicolumn{5}{c|}{\textbf{($T=0.5\%$)}} & \multicolumn{5}{c|}{\textbf{($T=0.0\%$)}} & \multicolumn{5}{c|}{\textbf{($T=-0.5\%$)}} & \multicolumn{5}{c}{\textbf{($T=-1\%$)}} \\\hline
\textbf{} & \textbf{2} & \textbf{5} & \textbf{10} & \textbf{20} & \textbf{50} & \textbf{2} & \textbf{5} & \textbf{10} & \textbf{20} & \textbf{50} & \textbf{2} & \textbf{5} & \textbf{10} & \textbf{20} & \textbf{50} & \textbf{2} & \textbf{5} & \textbf{10} & \textbf{20} & \textbf{50} \\\hline
\multicolumn{21}{c}{\textbf{Div. $(\gamma=1)$};\  \ \(m = M \times \gamma = \{2, 5, 10, 20, 50\}\)} \\\hline
0    & \textcolor{red}{-1} & \textcolor{red}{-1} & 3 & \textcolor{ForestGreen}{8} & \textcolor{ForestGreen}{6} & \textcolor{ForestGreen}{2} & \textcolor{ForestGreen}{5} & \textcolor{ForestGreen}{6} & \textcolor{ForestGreen}{8} & \textcolor{ForestGreen}{10} & \textcolor{ForestGreen}{5} & \textcolor{ForestGreen}{4} & \textcolor{ForestGreen}{9} & \textcolor{ForestGreen}{6} & \textcolor{ForestGreen}{10} & \textcolor{ForestGreen}{5} & \textcolor{ForestGreen}{6} & \textcolor{red}{-2} & \textcolor{ForestGreen}{10} & \textcolor{ForestGreen}{11} \\
0.1  & 0 & \textcolor{ForestGreen}{6} & 3 & \textcolor{ForestGreen}{15} & \textcolor{ForestGreen}{11} & \textcolor{ForestGreen}{10} & \textcolor{ForestGreen}{14} & \textcolor{ForestGreen}{10} & \textcolor{ForestGreen}{11} & \textcolor{ForestGreen}{11} & \textcolor{ForestGreen}{2} & \textcolor{ForestGreen}{8} & \textcolor{red}{1} & \textcolor{ForestGreen}{16} & \textcolor{ForestGreen}{10} & \textcolor{red}{1} & \textcolor{red}{1} & \textcolor{red}{1} & \textcolor{ForestGreen}{14} & \textcolor{ForestGreen}{9} \\
0.35 & \textcolor{red}{-3} & \textcolor{ForestGreen}{9} & \textcolor{ForestGreen}{13} & \textcolor{ForestGreen}{10} & \textcolor{ForestGreen}{12} & \textcolor{red}{-7} & \textcolor{ForestGreen}{5} & \textcolor{ForestGreen}{8} & \textcolor{ForestGreen}{9} & \textcolor{ForestGreen}{10} & \textcolor{ForestGreen}{8} & \textcolor{ForestGreen}{8} & \textcolor{ForestGreen}{12} & \textcolor{ForestGreen}{6} & \textcolor{ForestGreen}{13} & \textcolor{red}{0} & \textcolor{red}{-1} & \textcolor{red}{0} & \textcolor{ForestGreen}{5} & \textcolor{ForestGreen}{10} \\
\textbf{Avg. Sh.} & \textcolor{red}{-1} & 5 & \textcolor{ForestGreen}{7} & \textcolor{ForestGreen}{11} & \textcolor{ForestGreen}{10} & \textcolor{ForestGreen}{2} & \textcolor{ForestGreen}{8} & \textcolor{ForestGreen}{8} & \textcolor{ForestGreen}{10} & \textcolor{ForestGreen}{10} & \textcolor{ForestGreen}{5} & \textcolor{ForestGreen}{7} & \textcolor{ForestGreen}{7} & \textcolor{ForestGreen}{9} & \textcolor{ForestGreen}{11} & \textcolor{ForestGreen}{2} & \textcolor{ForestGreen}{2} & \textcolor{red}{0} & \textcolor{ForestGreen}{10} & \textcolor{ForestGreen}{10} \\\hline
\multicolumn{21}{c}{\textbf{Div. $(\gamma=2)$};\  \ \(m = M \times \gamma = \{4, 10, 20, 40, 100\}\)} \\\hline
0    & 3 & 7 & \textcolor{ForestGreen}{12} & \textcolor{ForestGreen}{14} & 7 & \textcolor{ForestGreen}{10} & \textcolor{ForestGreen}{11} & \textcolor{ForestGreen}{6} & \textcolor{red}{4} & \textcolor{ForestGreen}{11} & \textcolor{red}{0} & \textcolor{ForestGreen}{5} & \textcolor{ForestGreen}{6} & \textcolor{red}{1} & \textcolor{ForestGreen}{11} & \textcolor{red}{-4} & \textcolor{ForestGreen}{8} & \textcolor{ForestGreen}{8} & \textcolor{ForestGreen}{7} & \textcolor{red}{2} \\
0.1  & \textcolor{red}{-1} & \textcolor{ForestGreen}{11} & 8 & 9 & \textcolor{ForestGreen}{13} & \textcolor{ForestGreen}{5} & \textcolor{red}{3} & \textcolor{ForestGreen}{7} & \textcolor{red}{4} & \textcolor{ForestGreen}{11} & \textcolor{ForestGreen}{5} & \textcolor{ForestGreen}{4} & \textcolor{ForestGreen}{5} & \textcolor{ForestGreen}{7} & \textcolor{ForestGreen}{15} & \textcolor{red}{-6} & \textcolor{ForestGreen}{10} & \textcolor{ForestGreen}{14} & \textcolor{ForestGreen}{7} & \textcolor{ForestGreen}{11} \\
0.35 & 5 & 6 & 9 & \textcolor{ForestGreen}{10} & \textcolor{ForestGreen}{13} & \textcolor{ForestGreen}{8} & \textcolor{ForestGreen}{8} & \textcolor{ForestGreen}{16} & \textcolor{ForestGreen}{11} & \textcolor{ForestGreen}{10} & \textcolor{red}{-2} & \textcolor{ForestGreen}{5} & \textcolor{red}{3} & \textcolor{ForestGreen}{9} & \textcolor{ForestGreen}{7} & \textcolor{ForestGreen}{2} & \textcolor{ForestGreen}{13} & \textcolor{ForestGreen}{13} & \textcolor{ForestGreen}{11} & \textcolor{ForestGreen}{12} \\
\textbf{Avg. Sh.} & 3 & 8 & \textcolor{ForestGreen}{10} & \textcolor{ForestGreen}{11} & \textcolor{ForestGreen}{11} & \textcolor{ForestGreen}{8} & \textcolor{ForestGreen}{7} & \textcolor{ForestGreen}{10} & \textcolor{ForestGreen}{7} & \textcolor{ForestGreen}{10} & \textcolor{red}{1} & \textcolor{ForestGreen}{4} & \textcolor{ForestGreen}{5} & \textcolor{ForestGreen}{6} & \textcolor{ForestGreen}{11} & \textcolor{red}{-3} & \textcolor{ForestGreen}{10} & \textcolor{ForestGreen}{11} & \textcolor{ForestGreen}{8} & \textcolor{ForestGreen}{8} \\\hline
\multicolumn{21}{c}{\textbf{Div. $(\gamma=3)$};\  \ \(m = M \times \gamma = \{6, 15, 30, 60, 150\}\)} \\\hline
0    & \textcolor{ForestGreen}{6} & 5 & \textcolor{ForestGreen}{10} & 5 & \textcolor{ForestGreen}{8} & \textcolor{red}{-1} & \textcolor{ForestGreen}{2} & \textcolor{ForestGreen}{7} & \textcolor{red}{4} & \textcolor{ForestGreen}{10} & \textcolor{ForestGreen}{6} & \textcolor{ForestGreen}{8} & \textcolor{red}{0} & \textcolor{red}{4} & \textcolor{ForestGreen}{14} & \textcolor{ForestGreen}{4} & \textcolor{red}{-2} & \textcolor{ForestGreen}{11} & \textcolor{ForestGreen}{6} & \textcolor{ForestGreen}{10} \\
0.1  & \textcolor{ForestGreen}{11} & 8 & 6 & \textcolor{ForestGreen}{12} & \textcolor{ForestGreen}{10} & \textcolor{ForestGreen}{8} & \textcolor{ForestGreen}{11} & \textcolor{ForestGreen}{12} & \textcolor{ForestGreen}{9} & \textcolor{ForestGreen}{11} & \textcolor{red}{1} & \textcolor{ForestGreen}{12} & \textcolor{ForestGreen}{5} & \textcolor{ForestGreen}{6} & \textcolor{ForestGreen}{10} & \textcolor{red}{1} & \textcolor{ForestGreen}{5} & \textcolor{ForestGreen}{8} & \textcolor{ForestGreen}{8} & \textcolor{ForestGreen}{11} \\
0.35 & 4 & \textcolor{red}{1} & \textcolor{red}{0} & 7 & \textcolor{ForestGreen}{15} & \textcolor{ForestGreen}{8} & \textcolor{ForestGreen}{9} & \textcolor{ForestGreen}{13} & \textcolor{ForestGreen}{9} & \textcolor{ForestGreen}{11} & \textcolor{red}{-2} & \textcolor{ForestGreen}{12} & \textcolor{ForestGreen}{8} & \textcolor{ForestGreen}{12} & \textcolor{ForestGreen}{4} & \textcolor{red}{-2} & \textcolor{ForestGreen}{11} & \textcolor{ForestGreen}{6} & \textcolor{ForestGreen}{10} \\
\textbf{Avg. Sh.} & \textcolor{ForestGreen}{7} & 5 & 5 & \textcolor{ForestGreen}{8} & \textcolor{ForestGreen}{11} & \textcolor{ForestGreen}{7} & \textcolor{ForestGreen}{7} & \textcolor{ForestGreen}{11} & \textcolor{ForestGreen}{8} & \textcolor{ForestGreen}{11} & \textcolor{ForestGreen}{4} & \textcolor{ForestGreen}{8} & \textcolor{ForestGreen}{8} & \textcolor{ForestGreen}{9} & \textcolor{ForestGreen}{11} & \textcolor{red}{-1} & \textcolor{red}{3} & \textcolor{ForestGreen}{8} & \textcolor{ForestGreen}{8} & \textcolor{ForestGreen}{10} \\\hline
\multicolumn{21}{c}{\textbf{Div. $(\gamma=5)$};\  \ \(m = M \times \gamma = \{25, 50, 100, 250\}\)} \\\hline
0    &     & \textcolor{ForestGreen}{7} & \textcolor{ForestGreen}{9} & \textcolor{red}{0} & \textcolor{ForestGreen}{7} &     & \textcolor{ForestGreen}{6} & \textcolor{ForestGreen}{6} & \textcolor{ForestGreen}{11} & \textcolor{ForestGreen}{11} &     & \textcolor{ForestGreen}{7} & \textcolor{ForestGreen}{7} & \textcolor{ForestGreen}{8} & \textcolor{ForestGreen}{9} &     & \textcolor{ForestGreen}{9} & \textcolor{ForestGreen}{4} & \textcolor{ForestGreen}{9} & \textcolor{ForestGreen}{10} \\
0.1  &     & \textcolor{red}{1} & \textcolor{ForestGreen}{19} & 5 & 9 &     & \textcolor{ForestGreen}{7} & \textcolor{ForestGreen}{6} & \textcolor{ForestGreen}{11} & \textcolor{ForestGreen}{13} &     & \textcolor{red}{2} & \textcolor{ForestGreen}{10} & \textcolor{ForestGreen}{11} & \textcolor{ForestGreen}{12} &     & \textcolor{ForestGreen}{5} & \textcolor{ForestGreen}{11} & \textcolor{ForestGreen}{12} & \textcolor{ForestGreen}{12} \\
0.35 &     & 4 & \textcolor{ForestGreen}{13} & 2 & 8 &     & \textcolor{ForestGreen}{8} & \textcolor{ForestGreen}{8} & \textcolor{ForestGreen}{13} & \textcolor{ForestGreen}{11} &     & \textcolor{red}{0} & \textcolor{ForestGreen}{8} & \textcolor{ForestGreen}{8} & \textcolor{ForestGreen}{15} &     & \textcolor{ForestGreen}{10} & \textcolor{red}{3} & \textcolor{ForestGreen}{7} & \textcolor{ForestGreen}{10} \\
\textbf{Avg. Sh.} &     & 4 & \textcolor{ForestGreen}{14} & 3 & 8 &     & \textcolor{ForestGreen}{7} & \textcolor{ForestGreen}{7} & \textcolor{ForestGreen}{11} & \textcolor{ForestGreen}{12} &     & \textcolor{ForestGreen}{6} & \textcolor{ForestGreen}{6} & \textcolor{ForestGreen}{9} & \textcolor{ForestGreen}{12} &     & \textcolor{ForestGreen}{8} & \textcolor{ForestGreen}{6} & \textcolor{ForestGreen}{9} & \textcolor{ForestGreen}{11} \\\hline
\end{tabular}
\end{table}

In the one-step ahead case, a generalized increase in out-of-sample performance, represented by the Sharpe ratio, can be observed when selecting more diverse assets or hypotheses during the stock selection process prior to weight optimization, even if they have worse average predictions. This can be seen in Table \ref{tab:rmse_summary_combined1m}, where the average Sharpe ratio improves as the values of \(m\) grow, driven by the increasing value of \(\gamma\), allowing for greater diversity of predictors with different out-of-sample performances as part of the stock selection. 

This indicates that a higher number of candidates with greater diversity in predictions, and worse average 1-month out-of-sample return predictions (with relatively similar RMSE in their predictions), leads to portfolios with better out-of-sample Sharpe ratios and greater diversification, not due to the learning stage, but due to better asset selection. This pattern holds for both thresholds \(T = 0.5\%\) and \(T = 0.0\%\), meaning that the variety of 1-month cumulative return predictions from the sample of \(m\) candidates includes positive predictions above 0.5\% and 0.0\%, respectively. 

In the case of negative thresholds \(T = -0.5\%\) and \(T = -1\%\), it can be observed that the pattern is more pronounced, which surprisingly indicates that, by adding negative return predictions to the \(M\) portfolio constituents improve portfolio generalization performance and out-of-sample diversification. This occurs because it adds diversity to the hypothesis selection, even if it downgrades the average performance of the group, compared to other selections with more positive predictions but less diversity. This represents the diversity-quality trade-off of return predictions in asset selection discussed in Section \ref{assetselection}. It should be highlighted that \(\gamma = 2, 3, 5\) and \(M = 20, 50\) portfolio cases have constituents selected randomly from samples of \(m = 40\) to \(m = 250\). As the index consists of 500 stocks, this indicates that the spectrum of negative predictions is not small at all.

Table \ref{tab:rmse_summary_combined1m} also shows that, as \(T\) decreases, increasing diversity improves the average Sharpe ratio up to a limit, with \(T = -0.5\%\) yielding better values than \(T = -1.0\%\) and the other thresholds for the one-step-ahead case (1-month). This is consistent with the limits of diversification, as shown in Figure \ref{DiverRayd} \citep{dalio2018principles}.

Table \ref{tab:rmse_summary_combined_multistep} shows similar results, confirming the same hypothesis for sequential multi-step 10-month cases. It is important to note that this scenario is much more restrictive, as it involves making ten consecutive decisions over 10-months instead of just one. One might expect to prioritize the positiveness of predictions (quality) over the diversity of the prediction set during asset selection in such problems, as this would seem the most logical approach. However, the experiments validate the initial hypothesis and confirm the existence of this trade-off, with a significant impact on generalization performance and out-of-sample risk diversification. In Table \ref{tab:rmse_summary_combined_multistep} it can be observed that with a threshold \(T=-0.5\%\) for ranking predictions versus \(T=0.5\%\), the average out-of-sample Sharpe ratios are better for experiments with negative \(T\) in most cases. Furthermore, in the case of a positive \(T\), the results are negative for portfolios with fewer assets, indicating the need for a minimum number of stocks in the portfolio (Table \ref{tab:rmse_summary_combined_multistep}, case \(T = 0.5\%\)). This issue can be addressed easily without increasing \(M\) by adding diversity in asset or hypothesis selection through \(\gamma\). This is very useful for practitioners, as with constraints on the number of constituents, they can increase the level of diversification of their portfolio with the asset selection diversity component.

Finally, it can be shown that both in the one-step ahead case (1-month), which is comparable to the MPT setting and other plug-in methods, and in the multi-step case (10-month), the diversity parameter in the ensemble \((\varepsilon=0,0.1,0.35)\), which relates to weight optimization but not asset selection, shows improvement in all tables for \((\gamma=1)\). Without applying diversity in asset selection and only in optimization, the ratios improve with the parameter and the number of assets, consistent with the results shown in more detail in the previous section.

However, in cases where diversity is included in asset selection prior to optimization with \((\gamma=2,3,5)\), it can be seen that the parameter $\varepsilon$ still has a strong impact, although slightly reduced, on the out-of-sample Sharpe ratio and portfolio diversification. This can only be explained as diversification resulting from a combination of diversity in hypothesis selection, given by \(\gamma\), and the parametrically introduced diversity in the optimization process through \(\varepsilon\) during the learning of individual predictors.

\ref{Sensisection} presents additional experiments across different market regimes and a fixed income data set comprising 1,336 globally diversified bonds from 2014 to 2018, as shown in Figure~\ref{fig:bond_pie_subplots}. Specifically, out-of-sample portfolio performance metrics—including the Sharpe, Sortino, and Omega ratios, as well as maximum drawdown—are reported for the full range of the diversity parameter \( \varepsilon \).

Results for U.S. equity portfolios under bull and bear market regimes are shown in Tables~\ref{tab:compact_metric_summary} and~\ref{tab:compact_metric_summary_withM}, respectively. Table~\ref{tab:compact_metric_summary_fixed_income} presents the corresponding results for the fixed income data set. Across all cases, the diversity parameter \( \varepsilon \) improves performance metrics up to an optimal value. In the bull market regime, this optimal value is around 0.5, while in the bear market regime (2008 Credit Crunch), two optimal ranges are observed, typically around 0.5 and 0.95. This suggests that, while bull markets exhibit a well-defined optimal level of diversification, bear markets may benefit from multiple diversity regimes, indicating a more complex relationship between ensemble diversity and out-of-sample portfolio diversification under crisis conditions.

For the fixed income dataset, a pattern similar to the U.S. equity bull market case is observed. In both scenarios, the diversity parameter \( \varepsilon \) exhibits a clearly defined optimal range, and deviations from this range are primarily influenced by the number of portfolio constituents. As reported in Table~\ref{tab:compact_metric_summary_fixed_income}, the optimal value of \( \varepsilon \) is approximately 0.35 for small and large portfolios (5 and 100 assets or more), and around 0.1 for medium-sized portfolios (20-60 assets).

\subsection{Comparative Performance Analysis}
This section presents a comparative analysis between the proposed method, implemented with the s-RBFN as the PSEM model, and several well-established, robust and data-driven "plug-in" portfolio optimization methods. These consist of Inverse Volatility (IV) \citep{https://doi.org/10.1111/j.1540-6261.2006.00836.x}, CVaR Risk Parity (CVaR RP) \citep{kapsos2018robust}, Maximum Diversification (MD) \citep{Choueifaty2008}, Hierarchical Risk Parity (HRP) \citep{Lopez2016} and Hierarchical Equal Risk Contribution (HERC) \citep{Raffinot2018}. The comparison is conducted over a two-year investment horizon with monthly portfolio reallocation. The performance metric used is the monthly standard Sharpe Ratio, calculated using a time series of 1-month cumulative returns. This is computed as the average return over the 24-month period divided by its standard deviation and then annualized. It is important to note that the Sharpe ratios reported in this section are generally lower because they are calculated using average returns, whereas in previous sections, a modified version of the Sharpe ratio was used, based on cumulative returns over 1-month and 10-month periods.

Table \ref{tab:model_perf_by_gamma_threshold} reports the average Sharpe Ratios across various threshold levels $T$ and values of the asset selection diversity parameter $\gamma$. It can be shown how the presented method to select assets based on prediction diversity (diversify-quality trade-off of predictions as in Section \ref{gammasection}) makes all the methods improve generalization performance, due to increased out-of-sample diversification measured by the Sharpe ratios. Also, the s-RBFN is the top-performing method, as can be seen in the average value across all the experiments at the bottom of the same table. For selections of assets with less than -0.5\% cumulative monthly return, as in Table \ref{suba}, diversity is not of clear value, but for selections above 0.0\% and 0.3\% it is a clear winning selection choice for all methods.

\begin{table}[H]
\centering
\caption{Average Sharpe ratios across all configurations of the optimization diversity parameter $\varepsilon \in \{0, 0.1, 0.35\}$ and number of portfolio constituents $M \in \{5, 10, 20, 35\}$, grouped by threshold $T$ and the asset selection diversity parameter (multiplicative factor) $\gamma \in \{1, 2, 3\}$.}
\label{tab:model_perf_by_gamma_threshold}
\definecolor{ForestGreen}{RGB}{34,139,34}
\definecolor{red}{RGB}{220,0,0}
\renewcommand{\arraystretch}{1.1}

\begin{subtable}[t]{0.95\textwidth}
\centering
\caption{Threshold $T = -0.5$}
\label{suba}
\begin{tabular}{c|cccccc}
\toprule
$\gamma$ & s-RBFN & IV & CVaR RP & MD & HRP & HERC \\
\midrule
1 & \textcolor{ForestGreen}{0.79} & \textcolor{ForestGreen}{0.59} & \textcolor{ForestGreen}{0.74} & \textcolor{ForestGreen}{0.41} & \textcolor{ForestGreen}{0.54} & \textcolor{ForestGreen}{0.55} \\
2 & \textcolor{red}{0.64} & \textcolor{red}{0.40} & \textcolor{red}{0.50} & \textcolor{red}{0.29} & \textcolor{red}{0.36} & \textcolor{red}{0.28} \\
3 & 0.55 & 0.41 & 0.46 & 0.36 & 0.37 & 0.42 \\
\bottomrule
\end{tabular}
\end{subtable}

\vspace{0.8em}

\begin{subtable}[t]{0.95\textwidth}
\centering
\caption{Threshold $T = 0.0$}
\begin{tabular}{c|cccccc}
\toprule
$\gamma$ & s-RBFN & IV & CVaR RP & MD & HRP & HERC \\
\midrule
1 & 0.70 & 0.40 & 0.59 & 0.31 & 0.36 & 0.39 \\
2 & \textcolor{ForestGreen}{0.85} & \textcolor{ForestGreen}{0.56} & \textcolor{ForestGreen}{0.63} & \textcolor{ForestGreen}{0.55} & \textcolor{ForestGreen}{0.55} & 0.41 \\
3 & 0.69 & 0.43 & 0.49 & 0.29 & 0.43 & \textcolor{ForestGreen}{0.51} \\
\bottomrule
\end{tabular}
\end{subtable}

\vspace{0.8em}

\begin{subtable}[t]{0.95\textwidth}
\centering
\caption{Threshold $T = 0.3$}
\begin{tabular}{c|cccccc}
\toprule
$\gamma$ & s-RBFN & IV & CVaR RP & MD & HRP & HERC \\
\midrule
1 & 0.55 & 0.45 & 0.64 & 0.35 & \textcolor{ForestGreen}{0.45} & 0.42 \\
2 & 0.58 & 0.40 & 0.57 & 0.32 & 0.37 & \textcolor{ForestGreen}{0.52} \\
3 & \textcolor{ForestGreen}{0.83} & \textcolor{ForestGreen}{0.49} & \textcolor{ForestGreen}{0.65} & \textcolor{ForestGreen}{0.33} & \textcolor{ForestGreen}{0.48} & 0.47 \\
\midrule
\textbf{Avg.} & \textbf{0.69} & \textbf{0.46} & \textbf{0.58} & \textbf{0.35} & \textbf{0.44} & \textbf{0.44} \\
\bottomrule
\end{tabular}
\end{subtable}

\end{table}

\begin{figure}[t]
    \centering
    \includegraphics[width=0.65\linewidth]{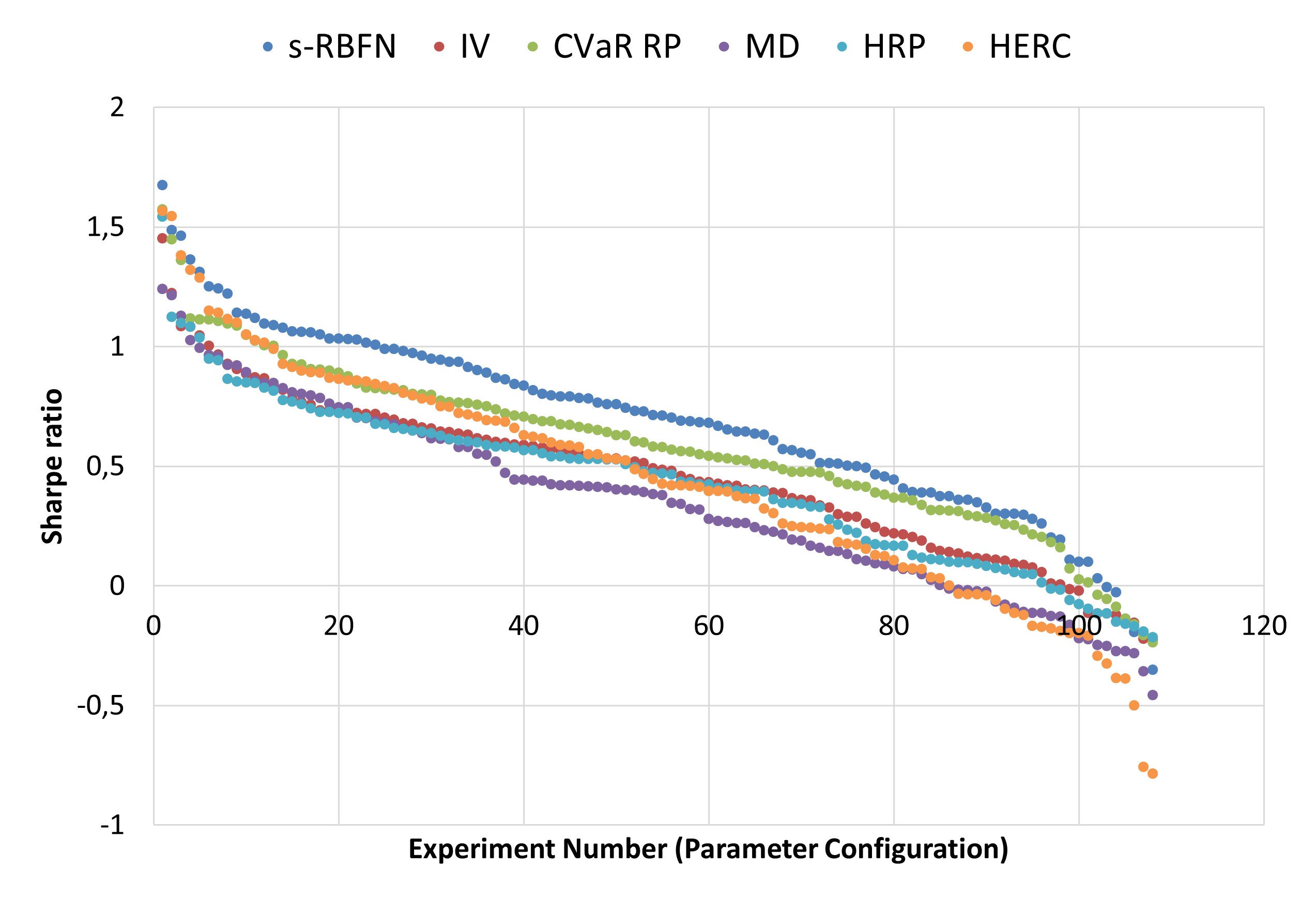}
    \caption{Sharpe ratios distributions across all experiments (108 configurations for $\varepsilon$,$\gamma$,$M$,$T$) for each portfolio optimization model.}
    \label{fig:enter-label}
\end{figure}

Figure \ref{fig:enter-label} displays the distribution of Sharpe ratios, ordered from highest to lowest, across all experiments conducted in this section (108 in total). These experiments include various configurations of the diversity parameters $\varepsilon$ and $\gamma$, threshold $T$, and the number of portfolio constituents $M$. The horizontal axis represents individual experiments, sorted by decreasing Sharpe ratio. Figure \ref{fig:enter-label} clearly demonstrates that the s-RBFN consistently outperforms all other models—often by a significant margin—highlighting the robustness and accuracy of the proposed method.

The results indicate that the s-RBFN method consistently outperforms the other portfolio optimization approaches across all threshold levels and values of the asset selection diversity parameter ($\gamma$), achieving the highest average Sharpe ratios in most cases. Furthermore, performance generally improves for all methods as $\gamma$ increases, suggesting that greater diversity in asset selection positively contributes to out-of-sample performance.

In \ref{compdifreg}, the proposed method is compared against alternative portfolio optimization approaches across distinct market regimes—specifically, the bear market during the Global Financial Crisis (2007–2009) and the bull market of the subsequent U.S. equity rally (2009–2016). Additionally, a diversified global bond data set comprising 1,336 instruments from 2014 to 2018, characterized in Figure~\ref{fig:bond_pie_subplots}, is used to evaluate performance across varying portfolio sizes.

As shown in Tables~\ref{tab:performance_metrics_rally} and~\ref{tab:metrics_credit_crunch}, the proposed s-RBFN model consistently achieves the highest average performance across all portfolio sizes under both bull and bear market regimes. Table~\ref{tab:all_metrics_stackedbonds} presents the results for the fixed income data set, where the s-RBFN attains superior Sharpe, Sortino, and Omega ratios, along with the second-lowest maximum drawdown among all evaluated methods. Notably, the s-RBFN model with 10 portfolio constituents delivers the best performance across all metrics for both the U.S. equity portfolios during the Credit Crunch (Table~\ref{tab:metrics_credit_crunch}) and the fixed income data set (Table~\ref{tab:all_metrics_stackedbonds}).

Overall, the proposed solution demonstrates significantly stronger performance for small-sized portfolios compared to other methods, which can be attributed to the enhanced diversification induced by the diversity parameters.

\section{Conclusion and Future Work}
\label{concl}

This work explored the relationship between the diversity of individual predictors in structured ensembles and portfolio diversification, offering a novel perspective within a multiple-hypothesis predict-then-optimize framework. One key contribution was the use of the ensemble combiner rule as the portfolio target during optimization—for example, the equal-weighted portfolio with the MSE loss. To the best of the authors’ knowledge, this represents the first learning-based plug-in method that enables parametric control of out-of-sample diversification prior to decision-making. Experimental results confirmed that out-of-sample diversification and generalization performance depend not only on the optimization procedure but also on the asset selection stage. Notably, when RMSE ranges were similar, asset selections with greater diversity in return predictions consistently outperformed those with less diversity, highlighting the importance of the diversity-quality trade-off. This forms a third contribution, and both sources of diversity were shown to extend the boundaries of portfolio diversification beyond previously suggested limits \citep{dalio2018principles}.

For future work, it would be valuable to explore deeper architectures for individual predictors, as well as the integration of multimodal data sets, enabling hypotheses to incorporate information beyond stock time series. Furthermore, extending the methodology to settings where each hypothesis represents a portfolio or a group of assets could open new avenues for application. Another promising direction involves investigating the interplay between ensemble complexity, prediction diversity, and portfolio risk. Additionally, exploring alternative loss functions and ensemble combiners tailored to various portfolio risk objectives may further enhance the flexibility and effectiveness of the proposed framework. 

\section*{Acknowledgements}
The authors gratefully acknowledge Miraltabank Asset Management, and in particular CIO Ignacio Fuertes Aguirre, for providing the fixed income data used in the experiments.

\appto{\bibsetup}{\sloppy}
\bibliographystyle{elsarticle-harv} 
\bibliography{cas-refs}

\begin{thebibliography}{49}
\expandafter\ifx\csname natexlab\endcsname\relax\def\natexlab#1{#1}\fi
\providecommand{\url}[1]{\texttt{#1}}
\providecommand{\href}[2]{#2}
\providecommand{\path}[1]{#1}
\providecommand{\DOIprefix}{doi:}
\providecommand{\ArXivprefix}{arXiv:}
\providecommand{\URLprefix}{URL: }
\providecommand{\Pubmedprefix}{pmid:}
\providecommand{\doi}[1]{\href{http://dx.doi.org/#1}{\path{#1}}}
\providecommand{\Pubmed}[1]{\href{pmid:#1}{\path{#1}}}
\providecommand{\bibinfo}[2]{#2}
\ifx\xfnm\relax \def\xfnm[#1]{\unskip,\space#1}\fi
\bibitem[{Akopov(2014)}]{Akopov2014}
\bibinfo{author}{Akopov, A.}, \bibinfo{year}{2014}.
\newblock \bibinfo{title}{Parallel genetic algorithm with fading selection}.
\newblock \bibinfo{journal}{International Journal of Computer Applications in Technology} \bibinfo{volume}{49}, \bibinfo{pages}{325--331}.
\newblock \DOIprefix\doi{10.1504/IJCAT.2014.062368}.
\bibitem[{ANG et~al.(2006)ANG, HODRICK, XING and ZHANG}]{https://doi.org/10.1111/j.1540-6261.2006.00836.x}
\bibinfo{author}{ANG, A.}, \bibinfo{author}{HODRICK, R.J.}, \bibinfo{author}{XING, Y.}, \bibinfo{author}{ZHANG, X.}, \bibinfo{year}{2006}.
\newblock \bibinfo{title}{The cross-section of volatility and expected returns}.
\newblock \bibinfo{journal}{The Journal of Finance} \bibinfo{volume}{61}, \bibinfo{pages}{259--299}.
\newblock \DOIprefix\doi{https://doi.org/10.1111/j.1540-6261.2006.00836.x}.
\bibitem[{Avramov and Zhou(2010)}]{Avramov2010BayesianPA}
\bibinfo{author}{Avramov, D.}, \bibinfo{author}{Zhou, G.}, \bibinfo{year}{2010}.
\newblock \bibinfo{title}{Bayesian portfolio analysis}.
\newblock \bibinfo{journal}{Review of Financial Economics} \bibinfo{volume}{2}, \bibinfo{pages}{25--47}.
\newblock \URLprefix \url{https://api.semanticscholar.org/CorpusID:45787980}.
\bibitem[{Ben-Tal et~al.(2009)Ben-Tal, Ghaoui and Nemirovski}]{RobustOptimization2009}
\bibinfo{author}{Ben-Tal, A.}, \bibinfo{author}{Ghaoui, L.}, \bibinfo{author}{Nemirovski, A.}, \bibinfo{year}{2009}.
\newblock \bibinfo{title}{Robust Optimization}.
\newblock \DOIprefix\doi{10.1515/9781400831050}.
\bibitem[{Bertsimas et~al.(2011)Bertsimas, Brown and Caramanis}]{doi:10.1137/080734510}
\bibinfo{author}{Bertsimas, D.}, \bibinfo{author}{Brown, D.B.}, \bibinfo{author}{Caramanis, C.}, \bibinfo{year}{2011}.
\newblock \bibinfo{title}{Theory and applications of robust optimization}.
\newblock \bibinfo{journal}{SIAM Review} \bibinfo{volume}{53}, \bibinfo{pages}{464--501}.
\newblock \URLprefix \url{https://doi.org/10.1137/080734510}, \DOIprefix\doi{10.1137/080734510}, \href{http://arxiv.org/abs/https://doi.org/10.1137/080734510}{{\tt arXiv:https://doi.org/10.1137/080734510}}.
\bibitem[{Bertsimas et~al.(2013)Bertsimas, Gupta and Kallus}]{articleBertsimas}
\bibinfo{author}{Bertsimas, D.}, \bibinfo{author}{Gupta, V.}, \bibinfo{author}{Kallus, N.}, \bibinfo{year}{2013}.
\newblock \bibinfo{title}{Data-driven robust optimization}.
\newblock \bibinfo{journal}{Mathematical Programming} \bibinfo{volume}{167}.
\newblock \DOIprefix\doi{10.1007/s10107-017-1125-8}.
\bibitem[{Best and Grauer(1991)}]{b8473306-c76a-342a-a678-7bb18db4be6a}
\bibinfo{author}{Best, M.J.}, \bibinfo{author}{Grauer, R.R.}, \bibinfo{year}{1991}.
\newblock \bibinfo{title}{On the sensitivity of mean-variance-efficient portfolios to changes in asset means: Some analytical and computational results}.
\newblock \bibinfo{journal}{The Review of Financial Studies} \bibinfo{volume}{4}, \bibinfo{pages}{315--342}.
\newblock \URLprefix \url{http://www.jstor.org/stable/2962107}.
\bibitem[{Bodnar et~al.(2024)Bodnar, Parolya and Thorsen}]{JMLR:v25:22-1337}
\bibinfo{author}{Bodnar, T.}, \bibinfo{author}{Parolya, N.}, \bibinfo{author}{Thorsen, E.}, \bibinfo{year}{2024}.
\newblock \bibinfo{title}{Two is better than one: Regularized shrinkage of large minimum variance portfolios}.
\newblock \bibinfo{journal}{Journal of Machine Learning Research} \bibinfo{volume}{25}, \bibinfo{pages}{1--32}.
\newblock \URLprefix \url{http://jmlr.org/papers/v25/22-1337.html}.
\bibitem[{Broadie(1993)}]{10.1007/BF02282040}
\bibinfo{author}{Broadie, M.}, \bibinfo{year}{1993}.
\newblock \bibinfo{title}{Computing efficient frontiers using estimated parameters}.
\newblock \bibinfo{journal}{Ann. Oper. Res.} \bibinfo{volume}{45}, \bibinfo{pages}{21–58}.
\newblock \URLprefix \url{https://doi.org/10.1007/BF02282040}, \DOIprefix\doi{10.1007/BF02282040}.
\bibitem[{Cai et~al.(2024)Cai, Cui, Lassance and Simaan}]{Cai2024}
\bibinfo{author}{Cai, Z.}, \bibinfo{author}{Cui, Z.}, \bibinfo{author}{Lassance, N.}, \bibinfo{author}{Simaan, M.}, \bibinfo{year}{2024}.
\newblock \bibinfo{title}{The economic value of mean squared error: Evidence from portfolio selection} .
\bibitem[{Chen and sha Zhou(2018)}]{CHEN2018165}
\bibinfo{author}{Chen, C.}, \bibinfo{author}{sha Zhou, Y.}, \bibinfo{year}{2018}.
\newblock \bibinfo{title}{Robust multiobjective portfolio with higher moments}.
\newblock \bibinfo{journal}{Expert Systems with Applications} \bibinfo{volume}{100}, \bibinfo{pages}{165--181}.
\bibitem[{Choueifaty and Coignard(2008)}]{Choueifaty2008}
\bibinfo{author}{Choueifaty, Y.}, \bibinfo{author}{Coignard, Y.}, \bibinfo{year}{2008}.
\newblock \bibinfo{title}{Toward maximum diversification}.
\newblock \bibinfo{journal}{Journal of Portfolio Management - J PORTFOLIO MANAGE} \bibinfo{volume}{35}, \bibinfo{pages}{40--51}.
\newblock \DOIprefix\doi{10.3905/JPM.2008.35.1.40}.
\bibitem[{Dalio(2018)}]{dalio2018principles}
\bibinfo{author}{Dalio, R.}, \bibinfo{year}{2018}.
\newblock \bibinfo{title}{Principles}.
\newblock \bibinfo{publisher}{Simon \& Schuster}.
\newblock \URLprefix \url{https://books.google.es/books?id=qNNmDwAAQBAJ}.
\bibitem[{DeMiguel et~al.(2009)DeMiguel, Garlappi and Uppal}]{DeMiguel2009OptimalVN}
\bibinfo{author}{DeMiguel, V.}, \bibinfo{author}{Garlappi, L.}, \bibinfo{author}{Uppal, R.}, \bibinfo{year}{2009}.
\newblock \bibinfo{title}{Optimal versus naive diversification: How inefficient is the 1/n portfolio strategy?}
\newblock \bibinfo{journal}{Review of Financial Studies} \bibinfo{volume}{22}, \bibinfo{pages}{1915--1953}.
\newblock \URLprefix \url{https://api.semanticscholar.org/CorpusID:1073674}.
\bibitem[{DeMiguel and Nogales(2007)}]{DeMiguel2007PortfolioSW}
\bibinfo{author}{DeMiguel, V.}, \bibinfo{author}{Nogales, F.J.}, \bibinfo{year}{2007}.
\newblock \bibinfo{title}{Portfolio selection with robust estimation}.
\newblock \bibinfo{journal}{European Finance eJournal} \URLprefix \url{https://api.semanticscholar.org/CorpusID:263885541}.
\bibitem[{Dominguez et~al.(2025)Dominguez, Shahzad and Hong}]{10.1007/978-3-031-77915-2_7}
\bibinfo{author}{Dominguez, A.R.}, \bibinfo{author}{Shahzad, M.}, \bibinfo{author}{Hong, X.}, \bibinfo{year}{2025}.
\newblock \bibinfo{title}{Structured radial basis function network: Modelling diversity for multiple hypotheses prediction}, in: \bibinfo{editor}{Bramer, M.}, \bibinfo{editor}{Stahl, F.} (Eds.), \bibinfo{booktitle}{Artificial Intelligence XLI}, \bibinfo{publisher}{Springer Nature Switzerland}, \bibinfo{address}{Cham}. pp. \bibinfo{pages}{88--101}.
\bibitem[{Donti and Kolter(2017)}]{articleDonti}
\bibinfo{author}{Donti, P.}, \bibinfo{author}{Kolter, J.}, \bibinfo{year}{2017}.
\newblock \bibinfo{title}{Task-based end-to-end model learning} \DOIprefix\doi{10.48550/arXiv.1703.04529}.
\bibitem[{Elmachtoub and Grigas(2017)}]{Elmachtoub2017SmartT}
\bibinfo{author}{Elmachtoub, A.N.}, \bibinfo{author}{Grigas, P.}, \bibinfo{year}{2017}.
\newblock \bibinfo{title}{Smart "predict, then optimize"}.
\newblock \bibinfo{journal}{ArXiv} \bibinfo{volume}{abs/1710.08005}.
\newblock \URLprefix \url{https://api.semanticscholar.org/CorpusID:38399700}.
\bibitem[{Elmachtoub and Grigas(2022)}]{doi:10.1287/mnsc.2020.3922}
\bibinfo{author}{Elmachtoub, A.N.}, \bibinfo{author}{Grigas, P.}, \bibinfo{year}{2022}.
\newblock \bibinfo{title}{Smart “predict, then optimize”}.
\newblock \bibinfo{journal}{Management Science} \bibinfo{volume}{68}, \bibinfo{pages}{9--26}.
\newblock \DOIprefix\doi{10.1287/mnsc.2020.3922}.
\bibitem[{Esfahani and Kuhn(2015)}]{Esfahani2015DatadrivenDR}
\bibinfo{author}{Esfahani, P.M.}, \bibinfo{author}{Kuhn, D.}, \bibinfo{year}{2015}.
\newblock \bibinfo{title}{Data-driven distributionally robust optimization using the wasserstein metric: performance guarantees and tractable reformulations}.
\newblock \bibinfo{journal}{Mathematical Programming} \bibinfo{volume}{171}, \bibinfo{pages}{115 -- 166}.
\newblock \URLprefix \url{https://api.semanticscholar.org/CorpusID:14542431}.
\bibitem[{Giglio et~al.(2022)Giglio, Kelly and Xiu}]{annurev:/content/journals/10.1146/annurev-financial-101521-104735}
\bibinfo{author}{Giglio, S.}, \bibinfo{author}{Kelly, B.}, \bibinfo{author}{Xiu, D.}, \bibinfo{year}{2022}.
\newblock \bibinfo{title}{Factor models, machine learning, and asset pricing}.
\newblock \bibinfo{journal}{Annual Review of Financial Economics} \bibinfo{volume}{14}, \bibinfo{pages}{337--368}.
\newblock \DOIprefix\doi{https://doi.org/10.1146/annurev-financial-101521-104735}.
\bibitem[{Goh and Jaillet(2016)}]{Goh2016StructuredPB}
\bibinfo{author}{Goh, C.Y.}, \bibinfo{author}{Jaillet, P.}, \bibinfo{year}{2016}.
\newblock \bibinfo{title}{Structured prediction by least squares estimated conditional risk minimization}.
\newblock \bibinfo{journal}{ArXiv} \bibinfo{volume}{abs/1611.07096}.
\newblock \URLprefix \url{https://api.semanticscholar.org/CorpusID:2579082}.
\bibitem[{Kan and Zhou(2007)}]{Kan_Zhou_2007}
\bibinfo{author}{Kan, R.}, \bibinfo{author}{Zhou, G.}, \bibinfo{year}{2007}.
\newblock \bibinfo{title}{Optimal portfolio choice with parameter uncertainty}.
\newblock \bibinfo{journal}{Journal of Financial and Quantitative Analysis} \bibinfo{volume}{42}, \bibinfo{pages}{621–656}.
\newblock \DOIprefix\doi{10.1017/S0022109000004129}.
\bibitem[{Kapsos et~al.(2018)Kapsos, Christofides and Rustem}]{kapsos2018robust}
\bibinfo{author}{Kapsos, M.}, \bibinfo{author}{Christofides, N.}, \bibinfo{author}{Rustem, B.}, \bibinfo{year}{2018}.
\newblock \bibinfo{title}{Robust risk budgeting}.
\newblock \bibinfo{journal}{Annals of Operations Research} \bibinfo{volume}{266}, \bibinfo{pages}{199--221}.
\newblock \URLprefix \url{https://doi.org/10.1007/s10479-017-2469-4}, \DOIprefix\doi{10.1007/s10479-017-2469-4}.
\bibitem[{Kotary et~al.(2023)Kotary, Vito, Christopher, Hentenryck and Fioretto}]{kotary2023predictthenoptimizeproxylearningjoint}
\bibinfo{author}{Kotary, J.}, \bibinfo{author}{Vito, V.D.}, \bibinfo{author}{Christopher, J.}, \bibinfo{author}{Hentenryck, P.V.}, \bibinfo{author}{Fioretto, F.}, \bibinfo{year}{2023}.
\newblock \bibinfo{title}{Predict-then-optimize by proxy: Learning joint models of prediction and optimization}.
\newblock \URLprefix \url{https://arxiv.org/abs/2311.13087}, \href{http://arxiv.org/abs/2311.13087}{{\tt arXiv:2311.13087}}.
\bibitem[{Lassance et~al.(2023)Lassance, Martin-Utrera and Simaan}]{articleLassance}
\bibinfo{author}{Lassance, N.}, \bibinfo{author}{Martin-Utrera, A.}, \bibinfo{author}{Simaan, M.}, \bibinfo{year}{2023}.
\newblock \bibinfo{title}{The risk of expected utility under parameter uncertainty}.
\newblock \bibinfo{journal}{Management Science} \bibinfo{volume}{forthcoming}.
\newblock \DOIprefix\doi{10.1287/mnsc.2023.00178}.
\bibitem[{Lasse Heje~Pedersen and Levine(2021)}]{doi:10.1080/0015198X.2020.1854543}
\bibinfo{author}{Lasse Heje~Pedersen, A.B.}, \bibinfo{author}{Levine, A.}, \bibinfo{year}{2021}.
\newblock \bibinfo{title}{Enhanced portfolio optimization}.
\newblock \bibinfo{journal}{Financial Analysts Journal} \bibinfo{volume}{77}, \bibinfo{pages}{124--151}.
\newblock \DOIprefix\doi{10.1080/0015198X.2020.1854543}, \href{http://arxiv.org/abs/https://doi.org/10.1080/0015198X.2020.1854543}{{\tt arXiv:https://doi.org/10.1080/0015198X.2020.1854543}}.
\bibitem[{Ledoit and Wolf(2003)}]{Ledoit2003}
\bibinfo{author}{Ledoit, O.}, \bibinfo{author}{Wolf, M.}, \bibinfo{year}{2003}.
\newblock \bibinfo{title}{Honey, i shrunk the sample covariance matrix}.
\newblock \bibinfo{journal}{The Journal of Portfolio Management} \bibinfo{volume}{30}.
\newblock \DOIprefix\doi{10.2139/ssrn.433840}.
\bibitem[{Lwin et~al.(2014)Lwin, Qu and Kendall}]{LWIN2014757}
\bibinfo{author}{Lwin, K.}, \bibinfo{author}{Qu, R.}, \bibinfo{author}{Kendall, G.}, \bibinfo{year}{2014}.
\newblock \bibinfo{title}{A learning-guided multi-objective evolutionary algorithm for constrained portfolio optimization}.
\newblock \bibinfo{journal}{Applied Soft Computing} \bibinfo{volume}{24}, \bibinfo{pages}{757--772}.
\newblock \DOIprefix\doi{https://doi.org/10.1016/j.asoc.2014.08.026}.
\bibitem[{Ma et~al.(2021)Ma, Han and Wang}]{MA2021113973}
\bibinfo{author}{Ma, Y.}, \bibinfo{author}{Han, R.}, \bibinfo{author}{Wang, W.}, \bibinfo{year}{2021}.
\newblock \bibinfo{title}{Portfolio optimization with return prediction using deep learning and machine learning}.
\newblock \bibinfo{journal}{Expert Systems with Applications} \bibinfo{volume}{165}, \bibinfo{pages}{113973}.
\newblock \DOIprefix\doi{https://doi.org/10.1016/j.eswa.2020.113973}.
\bibitem[{Mahadi et~al.(2022)Mahadi, Ballal, Moinuddin and Al-Saggaf}]{file-23SKslwIN3yLKCPK1veP3z7i}
\bibinfo{author}{Mahadi, M.}, \bibinfo{author}{Ballal, T.}, \bibinfo{author}{Moinuddin, M.}, \bibinfo{author}{Al-Saggaf, U.M.}, \bibinfo{year}{2022}.
\newblock \bibinfo{title}{Portfolio optimization using a consistent vector-based mse estimation approach}.
\newblock \bibinfo{journal}{IEEE Access} \bibinfo{volume}{10}, \bibinfo{pages}{86635--86645}.
\newblock \DOIprefix\doi{10.1109/ACCESS.2022.3197896}.
\bibitem[{Markowitz(1952)}]{https://doi.org/10.1111/j.1540-6261.1952.tb01525.x}
\bibinfo{author}{Markowitz, H.}, \bibinfo{year}{1952}.
\newblock \bibinfo{title}{Portfolio selection}.
\newblock \bibinfo{journal}{The Journal of Finance} \bibinfo{volume}{7}, \bibinfo{pages}{77--91}.
\newblock \DOIprefix\doi{https://doi.org/10.1111/j.1540-6261.1952.tb01525.x}.
\bibitem[{Nguyen and Lo(2012)}]{NGUYEN2012407}
\bibinfo{author}{Nguyen, T.D.}, \bibinfo{author}{Lo, A.W.}, \bibinfo{year}{2012}.
\newblock \bibinfo{title}{Robust ranking and portfolio optimization}.
\newblock \bibinfo{journal}{European Journal of Operational Research} \bibinfo{volume}{221}, \bibinfo{pages}{407--416}.
\newblock \DOIprefix\doi{https://doi.org/10.1016/j.ejor.2012.03.023}.
\bibitem[{Ortega et~al.(2022)Ortega, Caba\~nas and Masegosa}]{pmlr-v151-ortega22a}
\bibinfo{author}{Ortega, L.A.}, \bibinfo{author}{Caba\~nas, R.}, \bibinfo{author}{Masegosa, A.}, \bibinfo{year}{2022}.
\newblock \bibinfo{title}{Diversity and generalization in neural network ensembles}, in: \bibinfo{editor}{Camps-Valls, G.}, \bibinfo{editor}{Ruiz, F.J.R.}, \bibinfo{editor}{Valera, I.} (Eds.), \bibinfo{booktitle}{Proceedings of The 25th International Conference on Artificial Intelligence and Statistics}, \bibinfo{publisher}{PMLR}. pp. \bibinfo{pages}{11720--11743}.
\newblock \URLprefix \url{https://proceedings.mlr.press/v151/ortega22a.html}.
\bibitem[{Ozelim et~al.(2023)Ozelim, Ribeiro, Schiavon, Domingues and Queiroz}]{Ozelim2023}
\bibinfo{author}{Ozelim, L.}, \bibinfo{author}{Ribeiro, D.}, \bibinfo{author}{Schiavon, J.}, \bibinfo{author}{Domingues, V.}, \bibinfo{author}{Queiroz, P.}, \bibinfo{year}{2023}.
\newblock \bibinfo{title}{Hposs: A hierarchical portfolio optimization stacking strategy to reduce the generalization error of ensembles of models}.
\newblock \bibinfo{journal}{PLOS ONE} \bibinfo{volume}{18}, \bibinfo{pages}{e0290331}.
\newblock \DOIprefix\doi{10.1371/journal.pone.0290331}.
\bibitem[{Lopez~de Prado(2016)}]{Lopez2016}
\bibinfo{author}{Lopez~de Prado, M.}, \bibinfo{year}{2016}.
\newblock \bibinfo{title}{Building diversified portfolios that outperform out of sample:}.
\newblock \bibinfo{journal}{The Journal of Portfolio Management} \bibinfo{volume}{42}, \bibinfo{pages}{59--69}.
\newblock \DOIprefix\doi{10.3905/jpm.2016.42.4.059}.
\bibitem[{Raffinot(2018)}]{Raffinot2018}
\bibinfo{author}{Raffinot, T.}, \bibinfo{year}{2018}.
\newblock \bibinfo{title}{The hierarchical equal risk contribution portfolio}.
\newblock \bibinfo{journal}{SSRN Electronic Journal} \DOIprefix\doi{10.2139/ssrn.3237540}.
\bibitem[{Rahimian and Mehrotra(2022)}]{OJMO_2022__3__A4_0}
\bibinfo{author}{Rahimian, H.}, \bibinfo{author}{Mehrotra, S.}, \bibinfo{year}{2022}.
\newblock \bibinfo{title}{Frameworks and {Results} in {Distributionally} {Robust} {Optimization}}.
\newblock \bibinfo{journal}{Open Journal of Mathematical Optimization} \bibinfo{volume}{3}, \bibinfo{pages}{1--85}.
\newblock \URLprefix \url{http://www.numdam.org/articles/10.5802/ojmo.15/}, \DOIprefix\doi{10.5802/ojmo.15}.
\bibitem[{Roncalli(2013)}]{Roncalli2013IntroductionTR}
\bibinfo{author}{Roncalli, T.}, \bibinfo{year}{2013}.
\newblock \bibinfo{title}{Introduction to risk parity and budgeting}.
\newblock \bibinfo{journal}{CGN: Sovereign Wealth Funds as Investors (Sub-Topic)} \URLprefix \url{https://api.semanticscholar.org/CorpusID:154472565}.
\bibitem[{Shapiro et~al.(2009)Shapiro, Dentcheva and Ruszczyński}]{bookShapiro2009}
\bibinfo{author}{Shapiro, A.}, \bibinfo{author}{Dentcheva, D.}, \bibinfo{author}{Ruszczyński, A.}, \bibinfo{year}{2009}.
\newblock \bibinfo{title}{Lectures on stochastic programming. Modeling and theory}.
\newblock \DOIprefix\doi{10.1137/1.9780898718751}.
\bibitem[{Sharpe(1994)}]{Sharpe1994TheSR}
\bibinfo{author}{Sharpe, W.F.}, \bibinfo{year}{1994}.
\newblock \bibinfo{title}{The sharpe ratio}.
\newblock \URLprefix \url{https://api.semanticscholar.org/CorpusID:55394403}.
\bibitem[{{S\&P 500}(2005)}]{wikiSP500}
\bibinfo{author}{{S\&P 500}}, \bibinfo{year}{2005}.
\newblock \bibinfo{title}{S\&p 500 --- {W}ikipedia{,} the free encyclopedia}.
\newblock \bibinfo{note}{[Online; accessed 02-September-2024]}.
\bibitem[{{S\&P 500}(2024)}]{Bloomberg}
\bibinfo{author}{{S\&P 500}}, \bibinfo{year}{2024}.
\newblock \bibinfo{title}{S\&p 500 --- bloomberg l.p.}
\newblock \bibinfo{note}{[Online; accessed 02-June-2024]}.
\bibitem[{Vanderschueren et~al.(2022)Vanderschueren, Verdonck, Baesens and Verbeke}]{Vanderschueren2022PredictthenoptimizeOP}
\bibinfo{author}{Vanderschueren, T.}, \bibinfo{author}{Verdonck, T.}, \bibinfo{author}{Baesens, B.}, \bibinfo{author}{Verbeke, W.}, \bibinfo{year}{2022}.
\newblock \bibinfo{title}{Predict-then-optimize or predict-and-optimize? an empirical evaluation of cost-sensitive learning strategies}.
\newblock \bibinfo{journal}{Inf. Sci.} \bibinfo{volume}{594}, \bibinfo{pages}{400--415}.
\newblock \URLprefix \url{https://api.semanticscholar.org/CorpusID:247061868}.
\bibitem[{Wilder et~al.(2019)Wilder, Dilkina and Tambe}]{articleWilder}
\bibinfo{author}{Wilder, B.}, \bibinfo{author}{Dilkina, B.}, \bibinfo{author}{Tambe, M.}, \bibinfo{year}{2019}.
\newblock \bibinfo{title}{Melding the data-decisions pipeline: Decision-focused learning for combinatorial optimization}.
\newblock \bibinfo{journal}{Proceedings of the AAAI Conference on Artificial Intelligence} \bibinfo{volume}{33}, \bibinfo{pages}{1658--1665}.
\newblock \DOIprefix\doi{10.1609/aaai.v33i01.33011658}.
\bibitem[{Wood et~al.(2024)Wood, Mu, Webb, Reeve, Luj\'{a}n and Brown}]{wood2023unified}
\bibinfo{author}{Wood, D.}, \bibinfo{author}{Mu, T.}, \bibinfo{author}{Webb, A.M.}, \bibinfo{author}{Reeve, H.W.J.}, \bibinfo{author}{Luj\'{a}n, M.}, \bibinfo{author}{Brown, G.}, \bibinfo{year}{2024}.
\newblock \bibinfo{title}{A unified theory of diversity in ensemble learning}.
\newblock \bibinfo{journal}{J. Mach. Learn. Res.} \bibinfo{volume}{24}.
\bibitem[{Wu and Liu(2012)}]{Wu2012}
\bibinfo{author}{Wu, X.}, \bibinfo{author}{Liu, Y.K.}, \bibinfo{year}{2012}.
\newblock \bibinfo{title}{Optimizing fuzzy portfolio selection problems by parametric quadratic programming}.
\newblock \bibinfo{journal}{Fuzzy Optimization and Decision Making} \bibinfo{volume}{11}.
\newblock \DOIprefix\doi{10.1007/s10700-012-9126-9}.
\bibitem[{Zhang et~al.(2018)Zhang, Li and Guo}]{article2018}
\bibinfo{author}{Zhang, Y.}, \bibinfo{author}{Li, X.}, \bibinfo{author}{Guo, S.}, \bibinfo{year}{2018}.
\newblock \bibinfo{title}{Portfolio selection problems with markowitz’s mean–variance framework: a review of literature}.
\newblock \bibinfo{journal}{Fuzzy Optimization and Decision Making} \bibinfo{volume}{17}.
\newblock \DOIprefix\doi{10.1007/s10700-017-9266-z}.
\bibitem[{Zhao et~al.(2022)Zhao, Liu, Wu, Zhang and Zhang}]{10.3389/fenrg.2022.852520}
\bibinfo{author}{Zhao, W.}, \bibinfo{author}{Liu, X.}, \bibinfo{author}{Wu, Y.}, \bibinfo{author}{Zhang, T.}, \bibinfo{author}{Zhang, L.}, \bibinfo{year}{2022}.
\newblock \bibinfo{title}{A learning-to-rank-based investment portfolio optimization framework for smart grid planning}.
\newblock \bibinfo{journal}{Frontiers in Energy Research} \bibinfo{volume}{10}.
\newblock \DOIprefix\doi{10.3389/fenrg.2022.852520}.

\end{thebibliography}






\appendix

\section{Diversity in Learning Stage: Radial Basis Functions}

\begin{figure}[H]
    \centering
    \begin{subfigure}{0.45\textwidth}
        \centering
        \includegraphics[width=\textwidth]{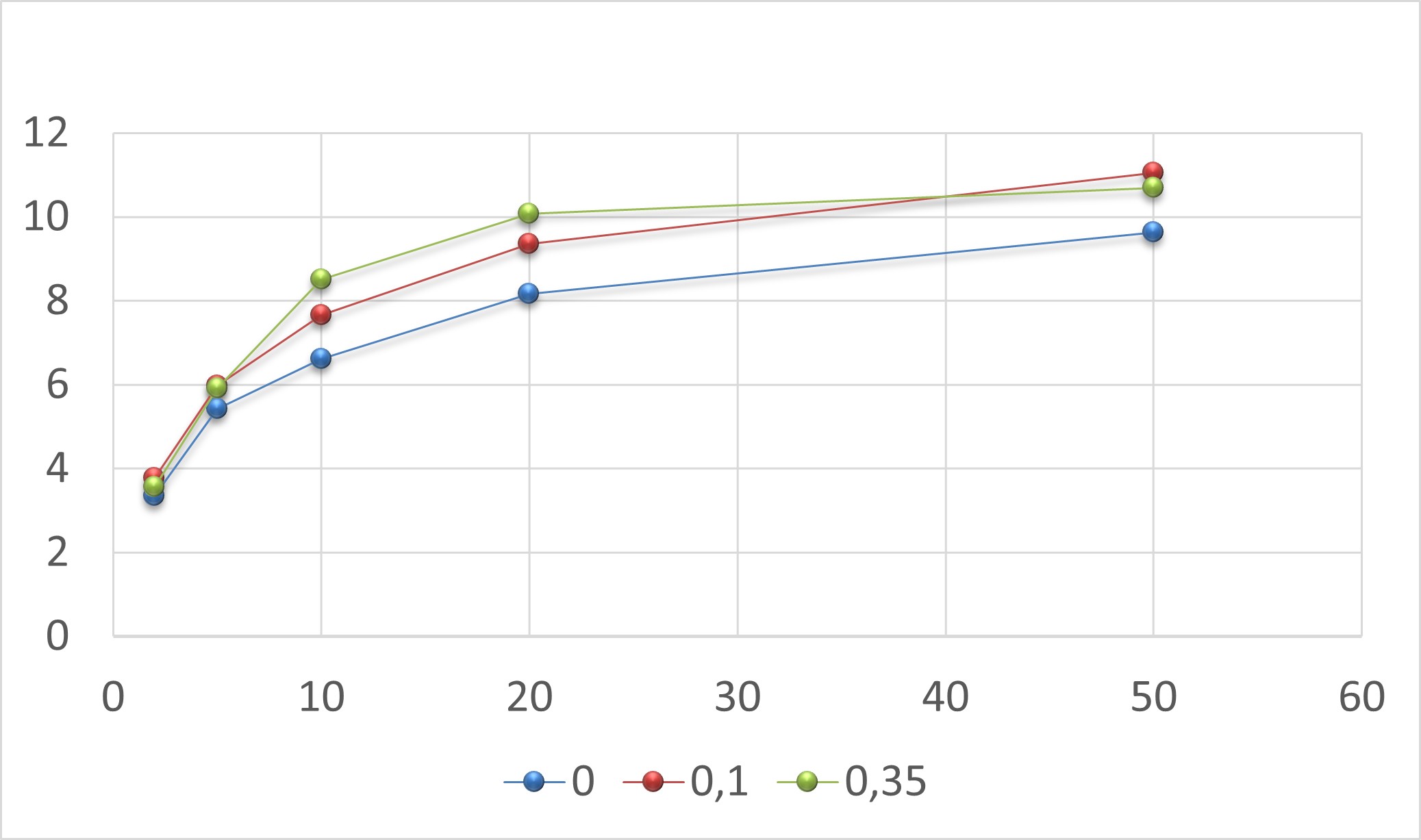}
        \caption{$\lambda_s=0$; $T=1\%$}
        \label{sub5R1}
    \end{subfigure}
    \hfill
    \begin{subfigure}{0.45\textwidth}
        \centering
        \includegraphics[width=\textwidth]{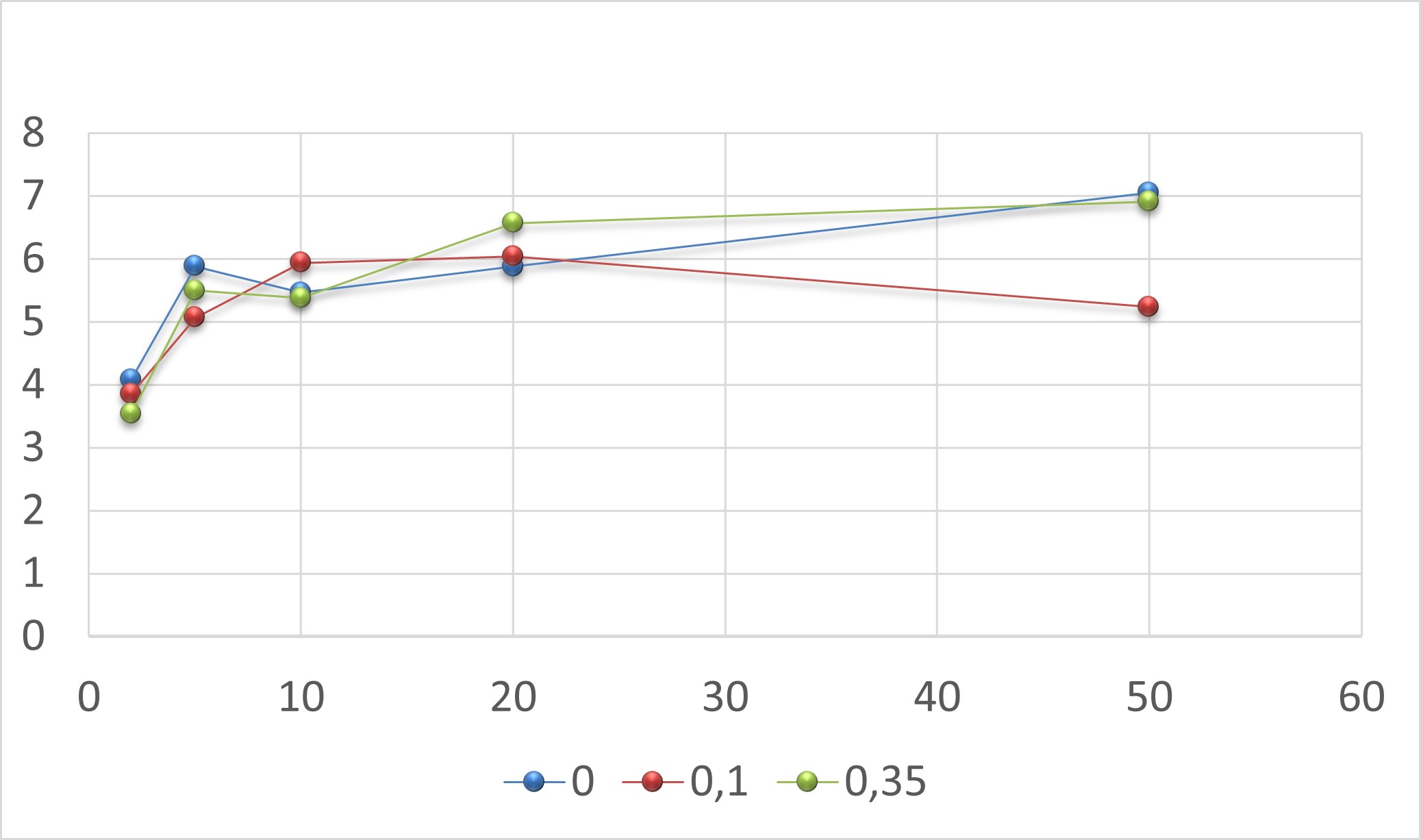}
        \caption{$\lambda_s=10$; $T=1\%$}
        \label{sub6R1}
    \end{subfigure}
    \vspace{0.5cm}    
    \begin{subfigure}{0.45\textwidth}
        \centering
        \includegraphics[width=\textwidth]{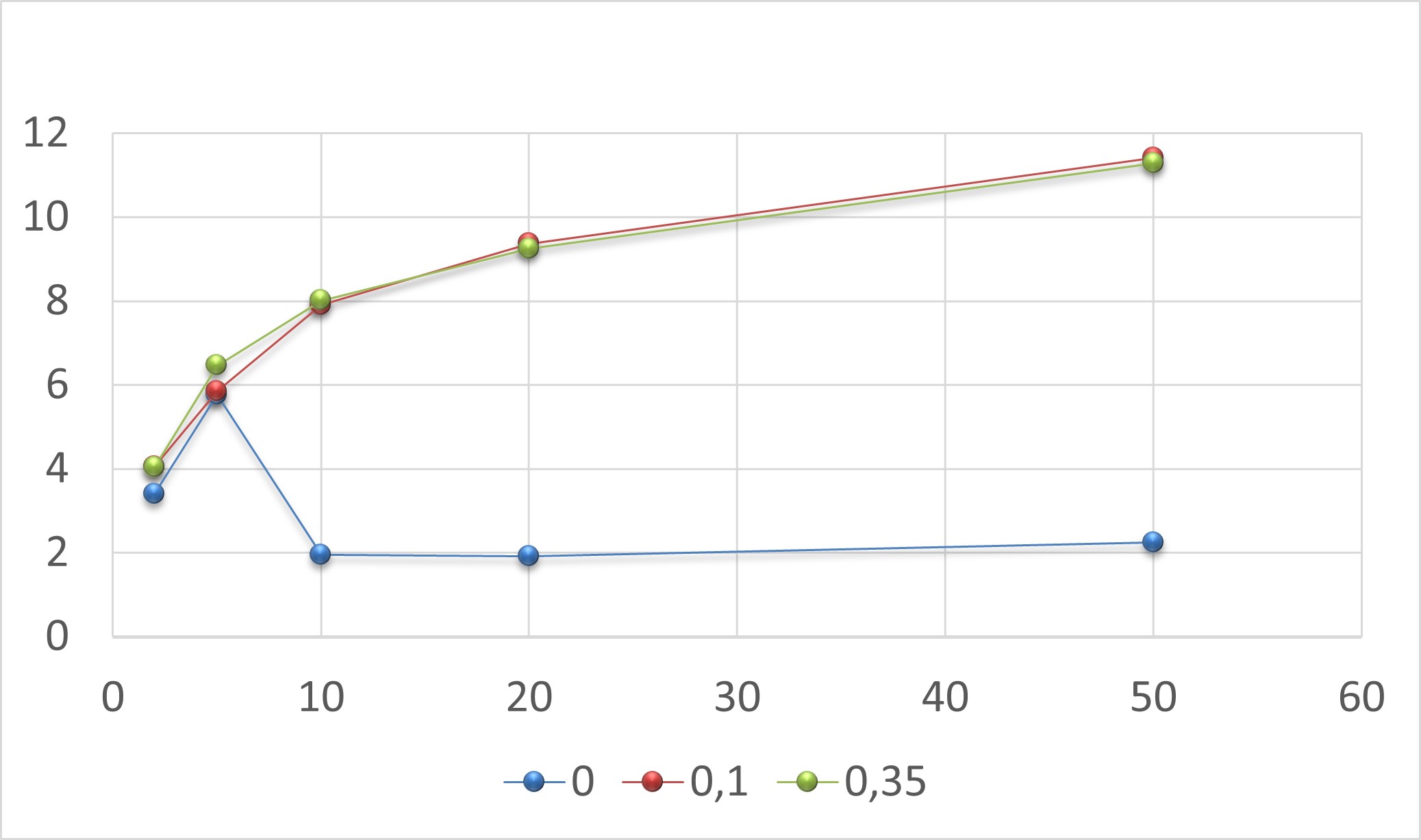}
        \caption{$\lambda_s=0$; $T=0.5\%$}
        \label{sub7R1}
    \end{subfigure}
    \hfill
    \begin{subfigure}{0.45\textwidth}
        \centering
        \includegraphics[width=\textwidth]{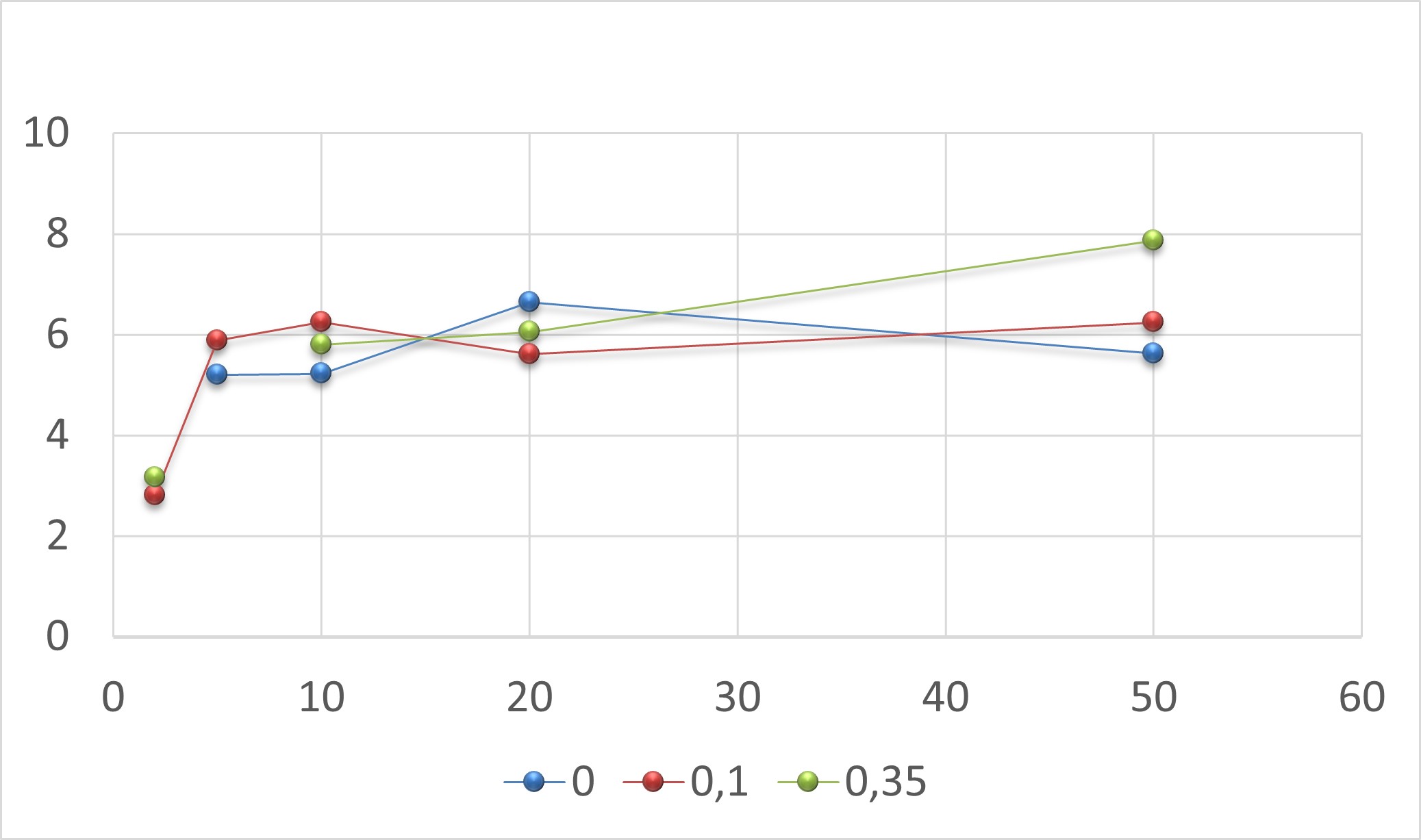}
        \caption{$\lambda_s=10$; $T=0.5\%$}
        \label{sub8R1}
    \end{subfigure}
    \vspace{0.5cm} 
    \begin{subfigure}{0.45\textwidth}
        \centering
        \includegraphics[width=\textwidth]{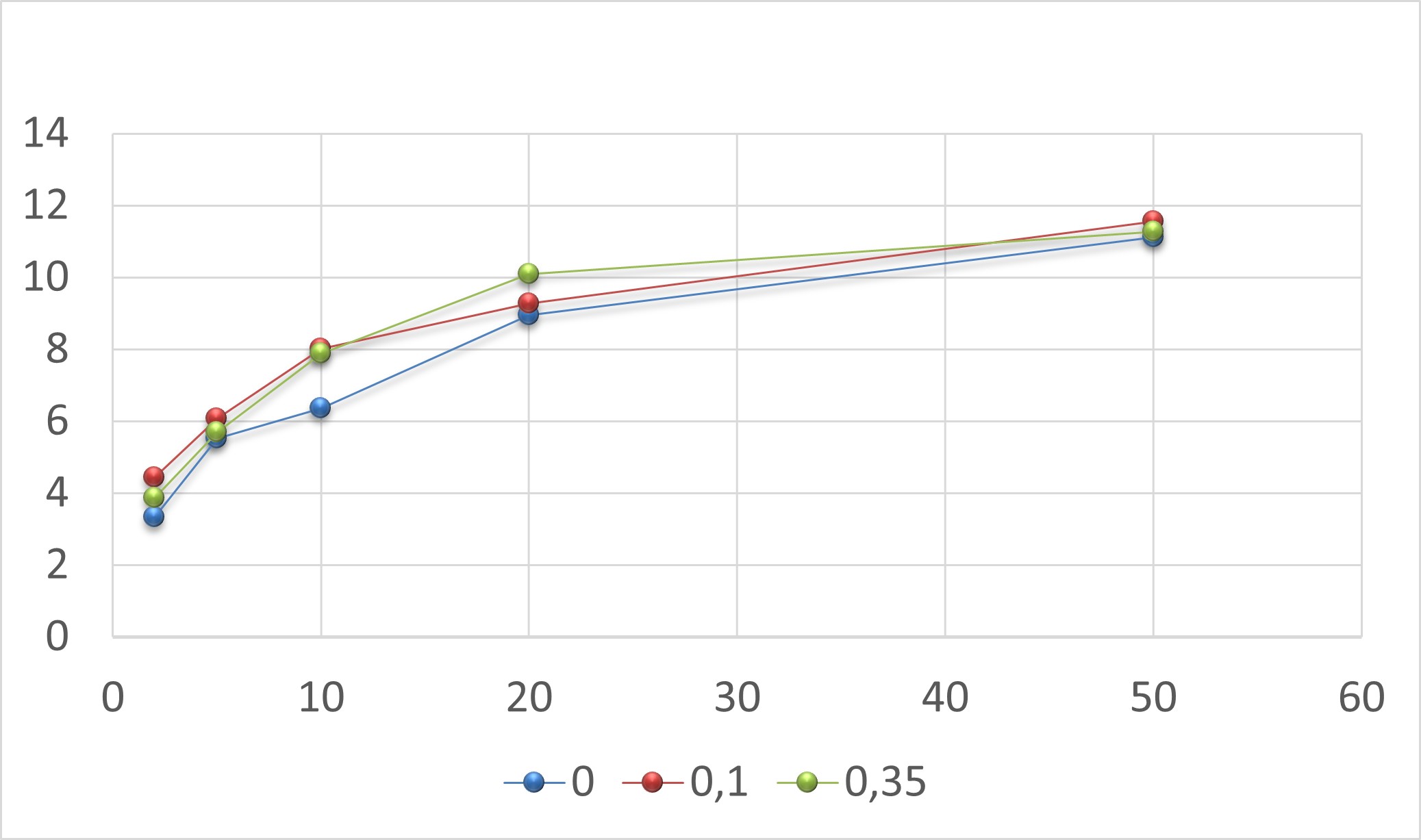}
        \caption{$\lambda_s=0$; $T=-0.5\%$}
        \label{sub9R1}
    \end{subfigure}
    \hfill
    \begin{subfigure}{0.45\textwidth}
        \centering
        \includegraphics[width=\textwidth]{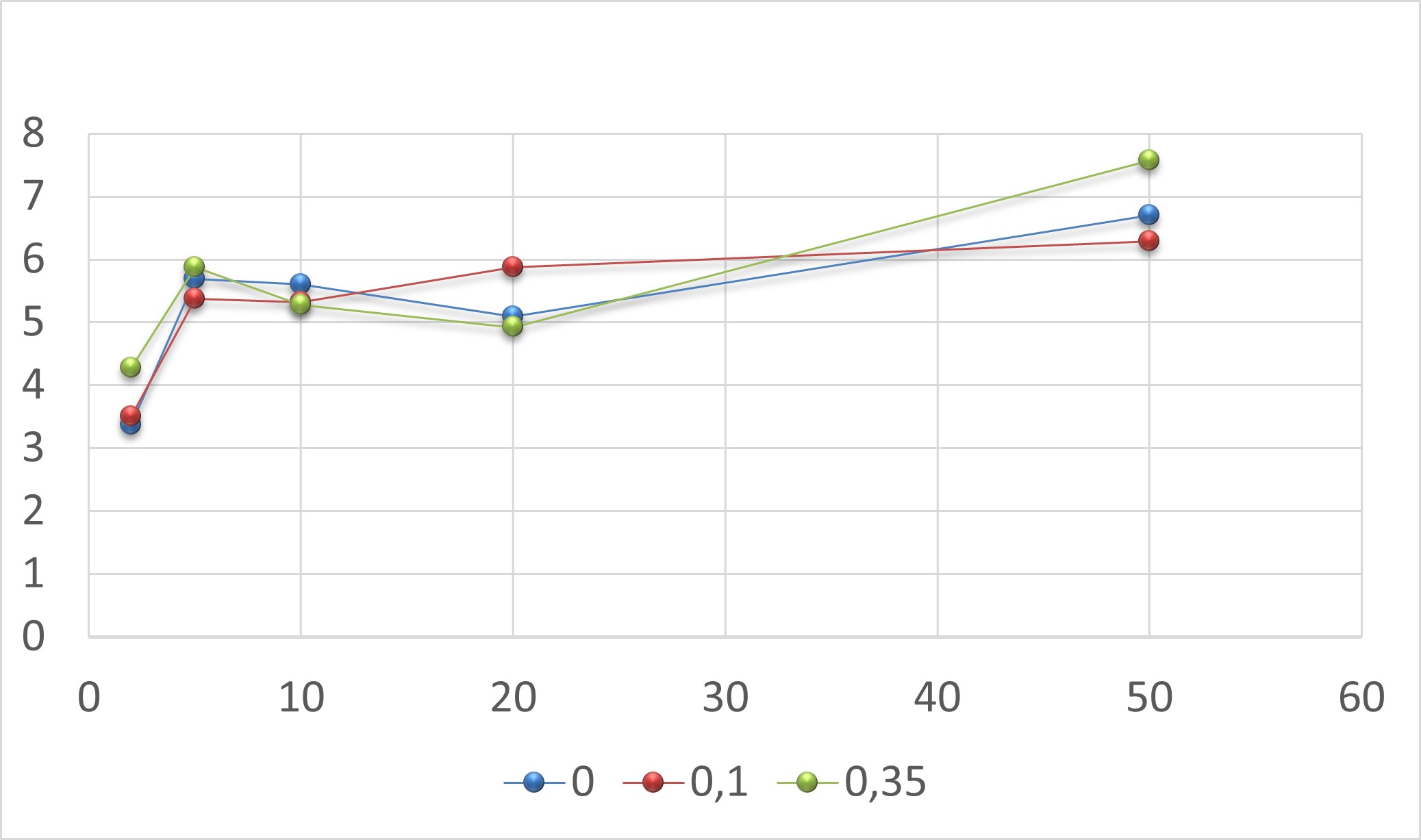}
        \caption{$\lambda_s=10$; $T=-0.5\%$}
        \label{sub10R1}
    \end{subfigure}
    \vspace{0.5cm}
    \begin{subfigure}{0.45\textwidth}
        \centering
        \includegraphics[width=\textwidth]{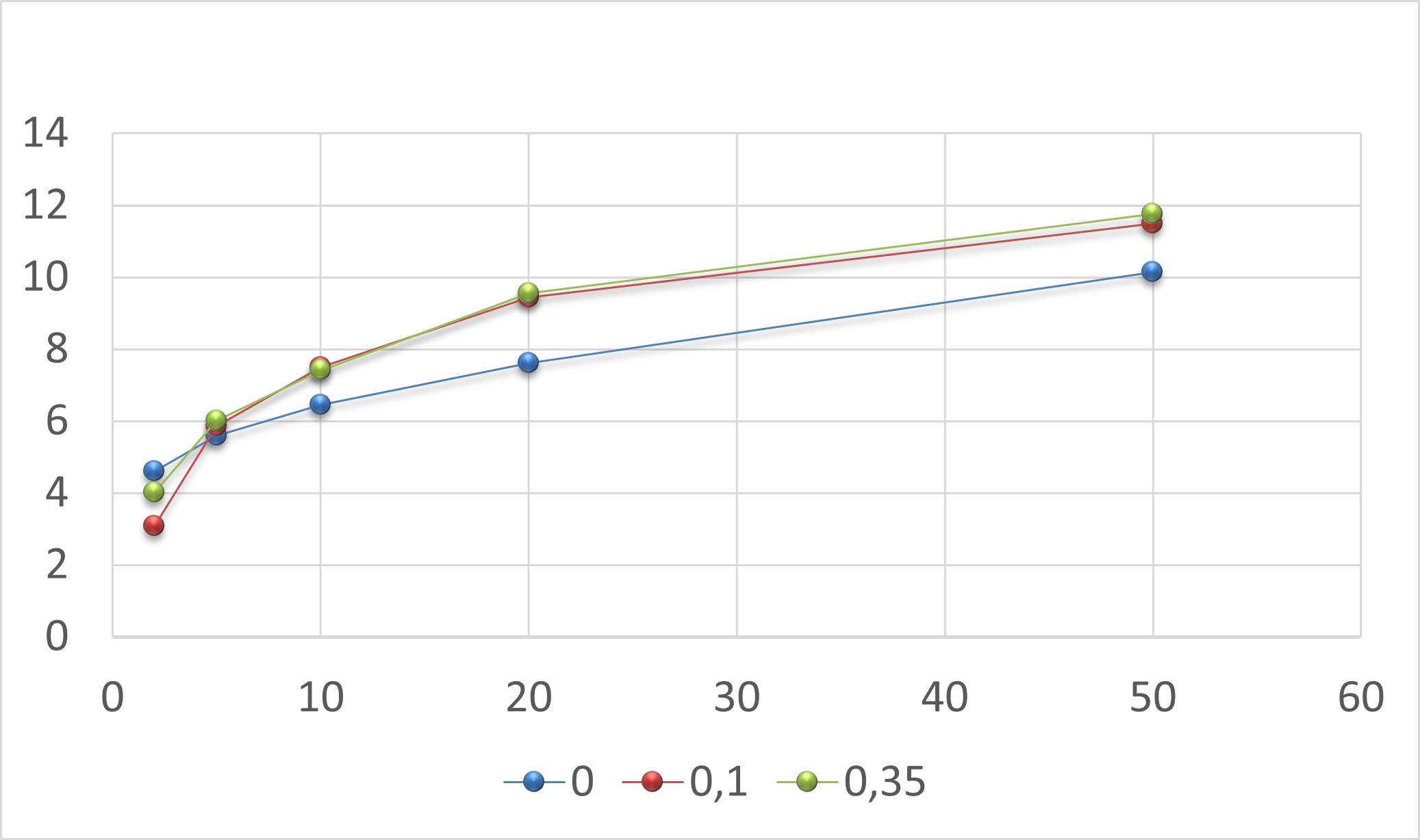}
        \caption{$\lambda_s=0$; $T=-1\%$}
        \label{sub11R1}
    \end{subfigure}
    \hfill
    \begin{subfigure}{0.45\textwidth}
        \centering
        \includegraphics[width=\textwidth]{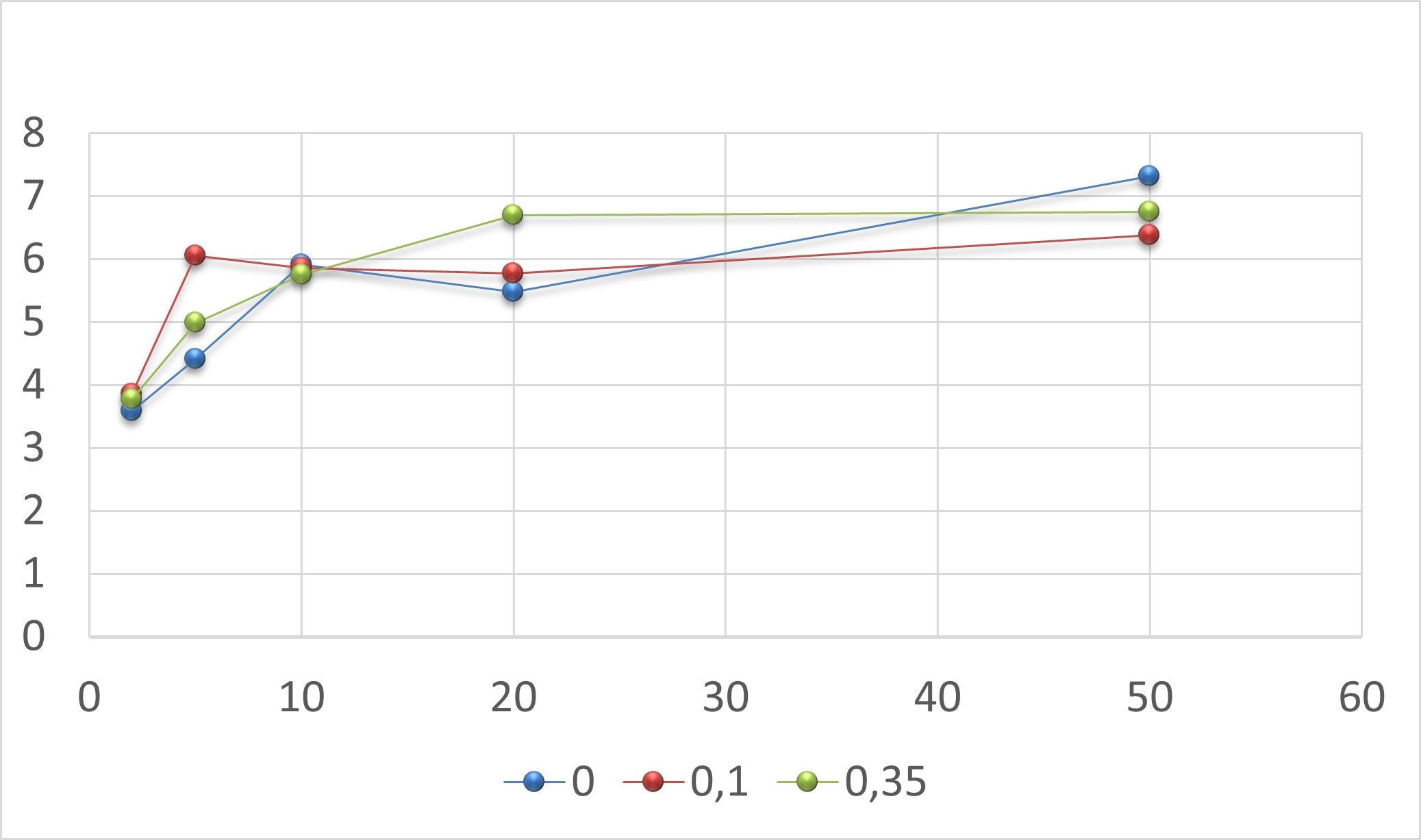}
        \caption{$\lambda_s=10$; $T=-1\%$}
        \label{sub12R1}
    \end{subfigure}
    \caption{One-step (1-month) 100 simulations Sharpe ratios (Y-axis) for different number of predictors $M$ (X-axis), diversity parameter $\varepsilon=0,0.1,0.35$ (colours), s-RBFN regularization parameter $\lambda_s$, and $-1\%\leq T\leq 1\%$. s-RBFN with radial basis functions.}
    \label{FigureSR_rad1-Simp}
\end{figure}

\begin{figure}[H]
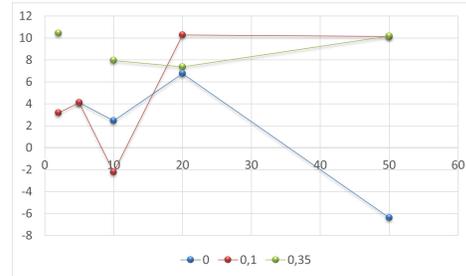
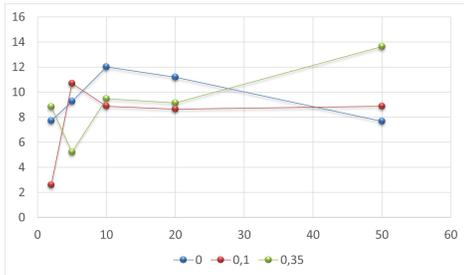
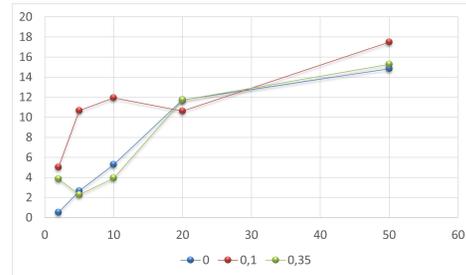
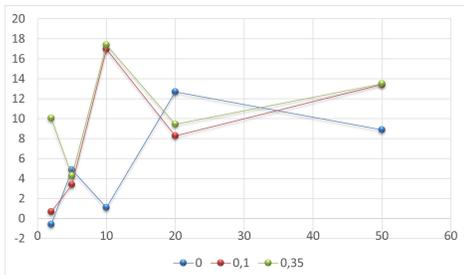
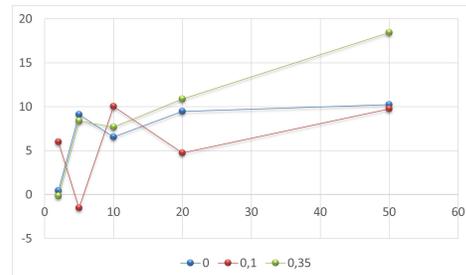

    \centering
    \begin{subfigure}{0.45\textwidth}
        \centering
        \includegraphics[width=\textwidth]{1m110G.jpg}
        \caption{$\lambda_s=0$; $T=1\%$}
        \label{sub5G10}
    \end{subfigure}
    \hfill
    \begin{subfigure}{0.45\textwidth}
        \centering
        \includegraphics[width=\textwidth]{1m210G.jpg}
        \caption{$\lambda_s=10$; $T=1\%$}
        \label{sub6R10}
    \end{subfigure}
    \vspace{0.5cm}    
    \begin{subfigure}{0.45\textwidth}
        \centering
        \includegraphics[width=\textwidth]{1m310G.jpg}
        \caption{$\lambda_s=0$; $T=0.5\%$}
        \label{sub7R10}
    \end{subfigure}
    \hfill
    \begin{subfigure}{0.45\textwidth}
        \centering
        \includegraphics[width=\textwidth]{1m410G.jpg}
        \caption{$\lambda_s=10$; $T=0.5\%$}
        \label{sub8R10}
    \end{subfigure}
    \vspace{0.5cm} 
    \begin{subfigure}{0.45\textwidth}
        \centering
        \includegraphics[width=\textwidth]{1m710G.jpg}
        \caption{$\lambda_s=0$; $T=-0.5\%$}
        \label{sub9R10}
    \end{subfigure}
    \hfill
    \begin{subfigure}{0.45\textwidth}
        \centering
        \includegraphics[width=\textwidth]{1m810G.jpg}
        \caption{$\lambda_s=10$; $T=-0.5\%$}
        \label{sub10R10}
    \end{subfigure}
    \vspace{0.5cm}
    \begin{subfigure}{0.45\textwidth}
        \centering
        \includegraphics[width=\textwidth]{1m910G.jpg}
        \caption{$\lambda_s=0$; $T=-1\%$}
        \label{sub11R10}
    \end{subfigure}
    \hfill
    \begin{subfigure}{0.45\textwidth}
        \centering
        \includegraphics[width=\textwidth]{1m1010G.jpg}
        \caption{$\lambda_s=10$; $T=-1\%$}
        \label{sub12R10}
    \end{subfigure}
    \caption{Multi-step (10-month)  Sharpe ratios (Y-axis) for different number of predictors $M$ (X-axis), diversity parameter $\varepsilon=0,0.1,0.35$ (colours), s-RBFN regularization parameter $\lambda_s$, and $-1\%\leq T\leq 1\%$. s-RBFN with radial basis functions.}
    \label{FigureSR_rad10-Simp}
\end{figure}

\section{Ensemble Regularization in the Gaussian Basis Functions Experiments (Multi-step ahead case)}

\begin{figure}[H]
    \centering
    \captionsetup{justification=centering}
    \begin{subfigure}{0.3\textwidth} 
        \centering
        \captionsetup{justification=centering}
        \includegraphics[width=\textwidth]{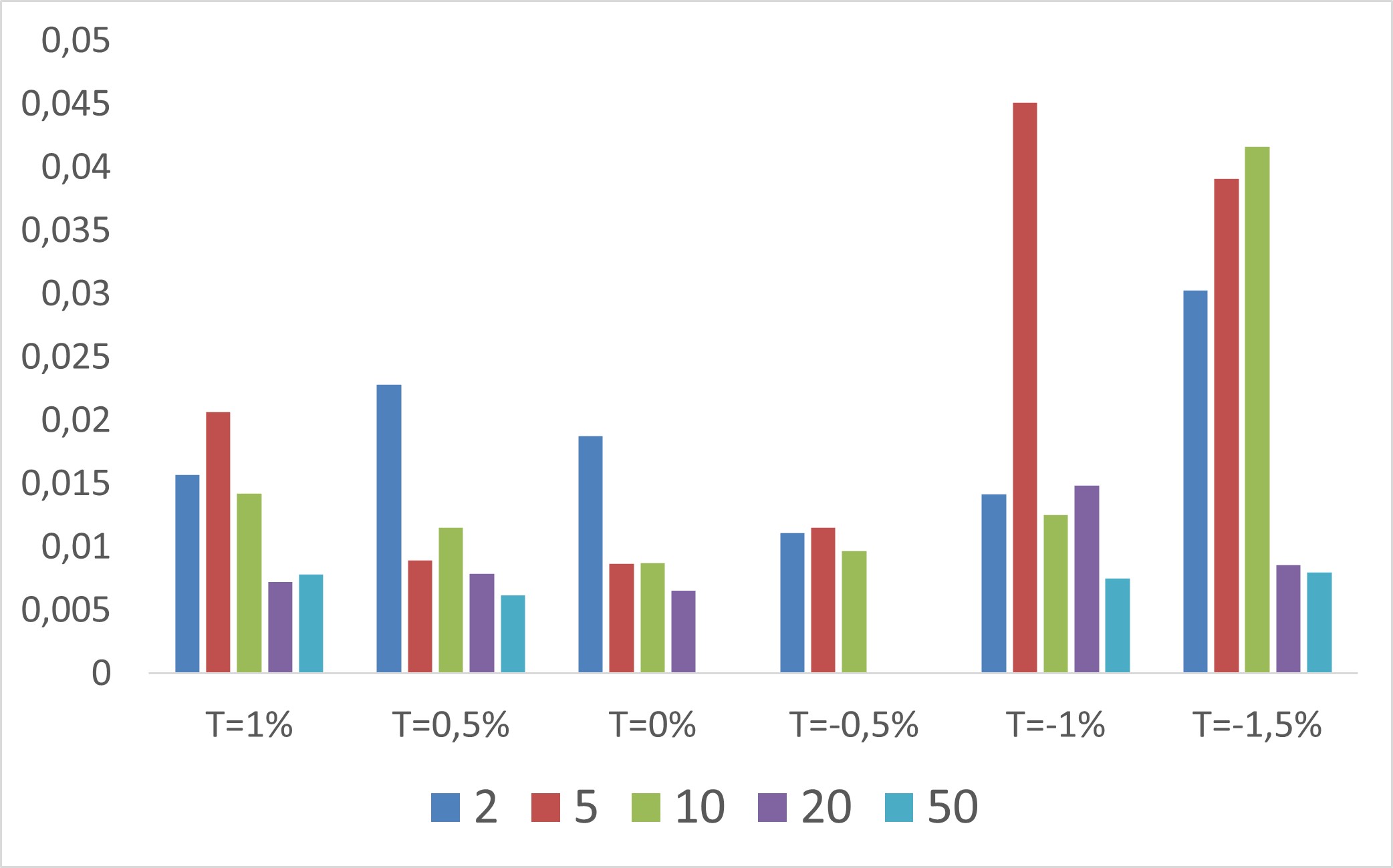}
        \subcaption{$\varepsilon = 0$ ; $\lambda_s= 0$}
        \label{regsub1}
    \end{subfigure}
    \hfill
    \begin{subfigure}{0.3\textwidth}
        \centering
        \includegraphics[width=\textwidth]{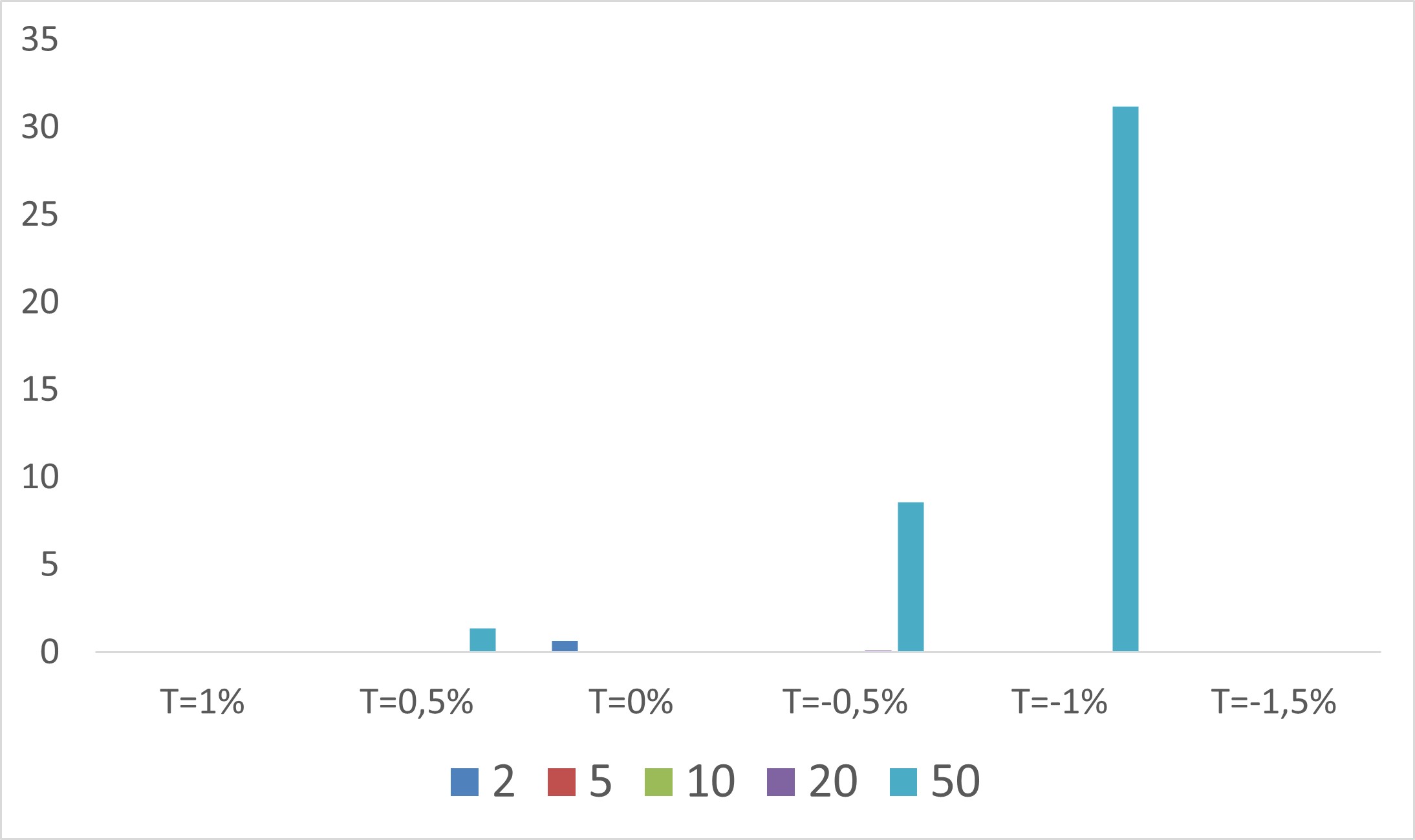}
        \subcaption{$\varepsilon = 0.1$ ; $\lambda_s = 0$}
        \label{regsub2}
    \end{subfigure}
    \hfill
    \begin{subfigure}{0.3\textwidth}
        \centering
        \includegraphics[width=\textwidth]{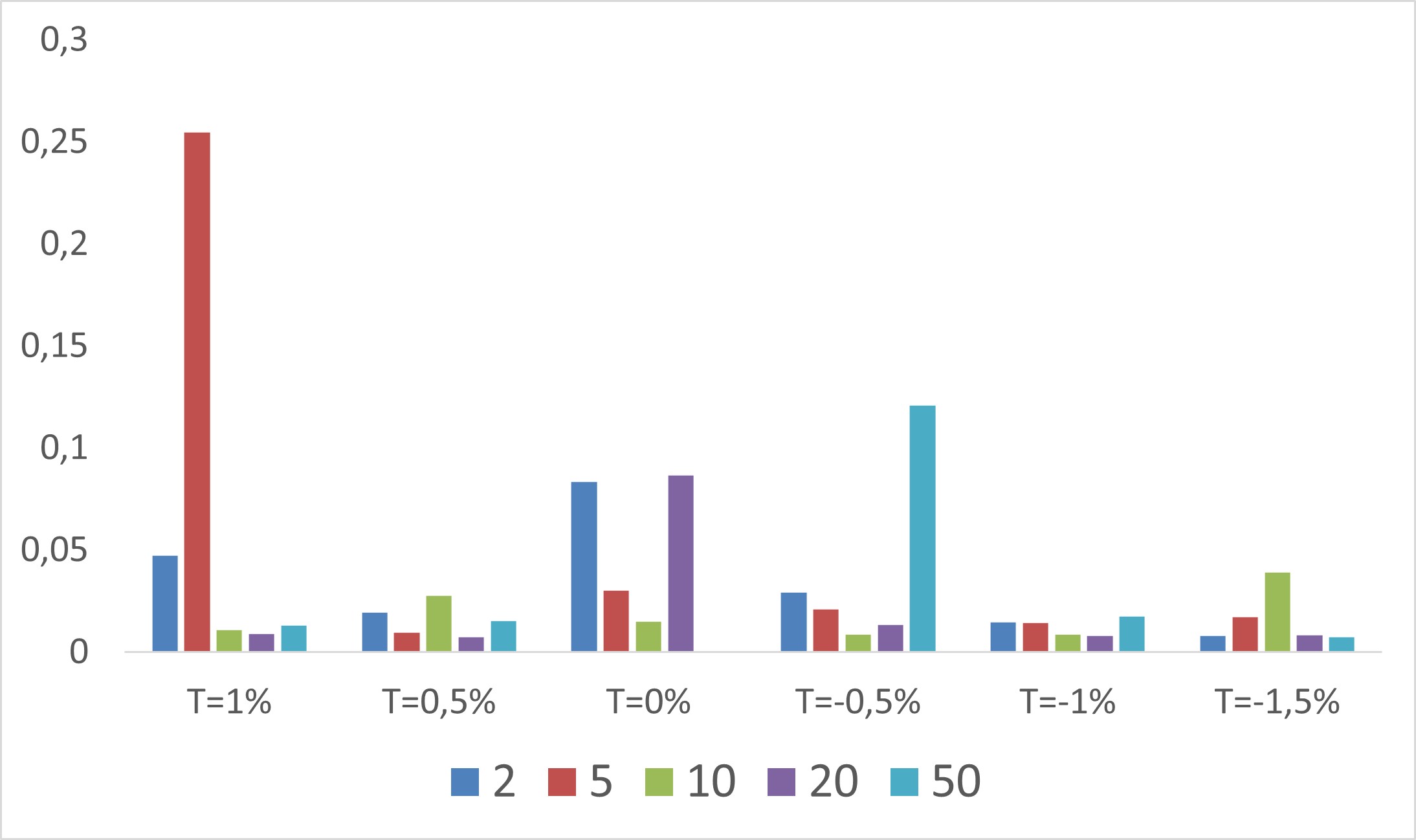}
        \subcaption{$\varepsilon = 0.35$ ; $\lambda_s= 0$}
        \label{regsub3}
    \end{subfigure}
    
    \vspace{0.3cm} 
    
    \begin{subfigure}{0.3\textwidth}
        \centering
        \includegraphics[width=\textwidth]{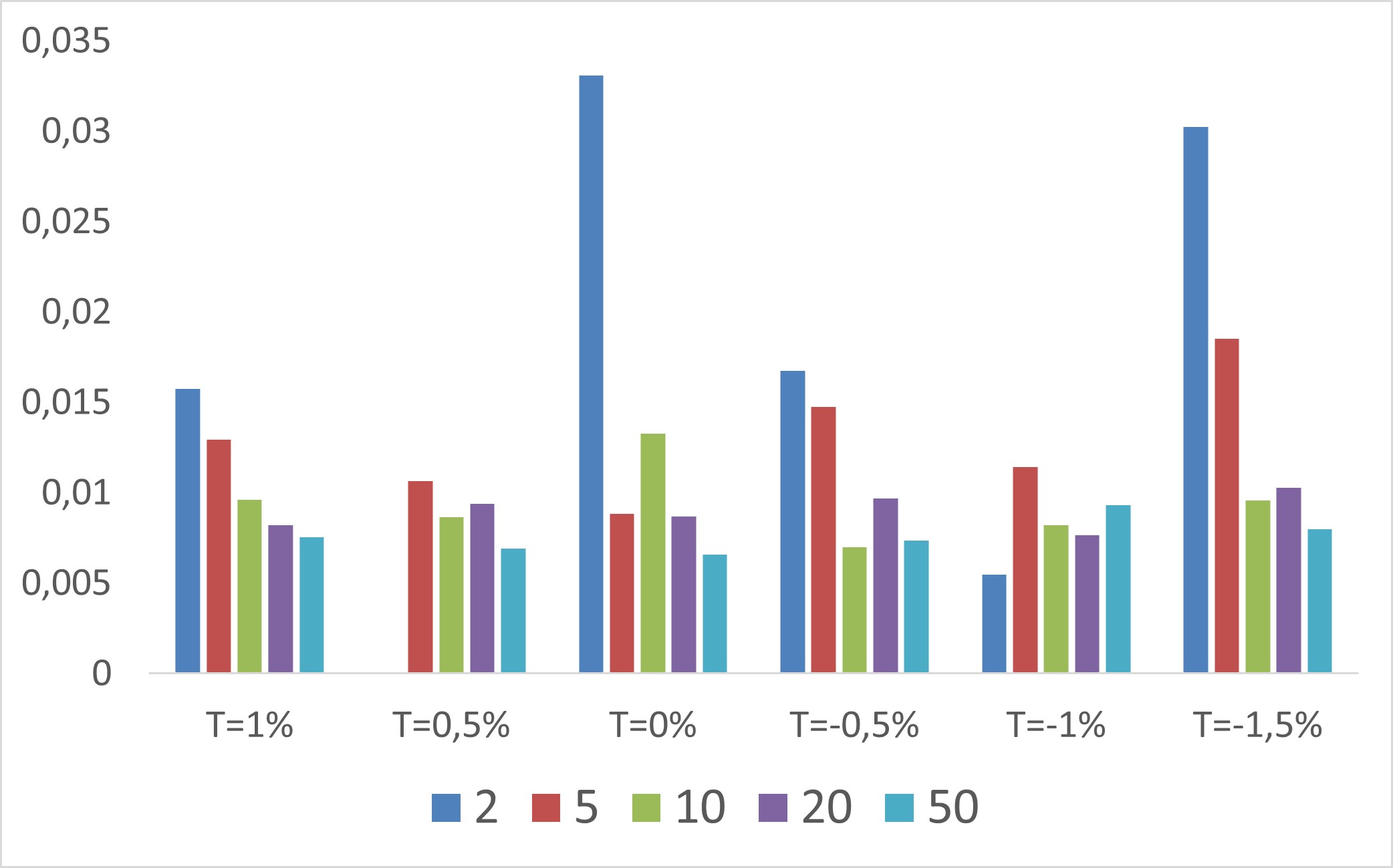}
        \subcaption{$\varepsilon = 0$ ; $\lambda_s= 10$}
        \label{regsub4}
    \end{subfigure}
    \hfill
    \begin{subfigure}{0.3\textwidth}
        \centering
        \includegraphics[width=\textwidth]{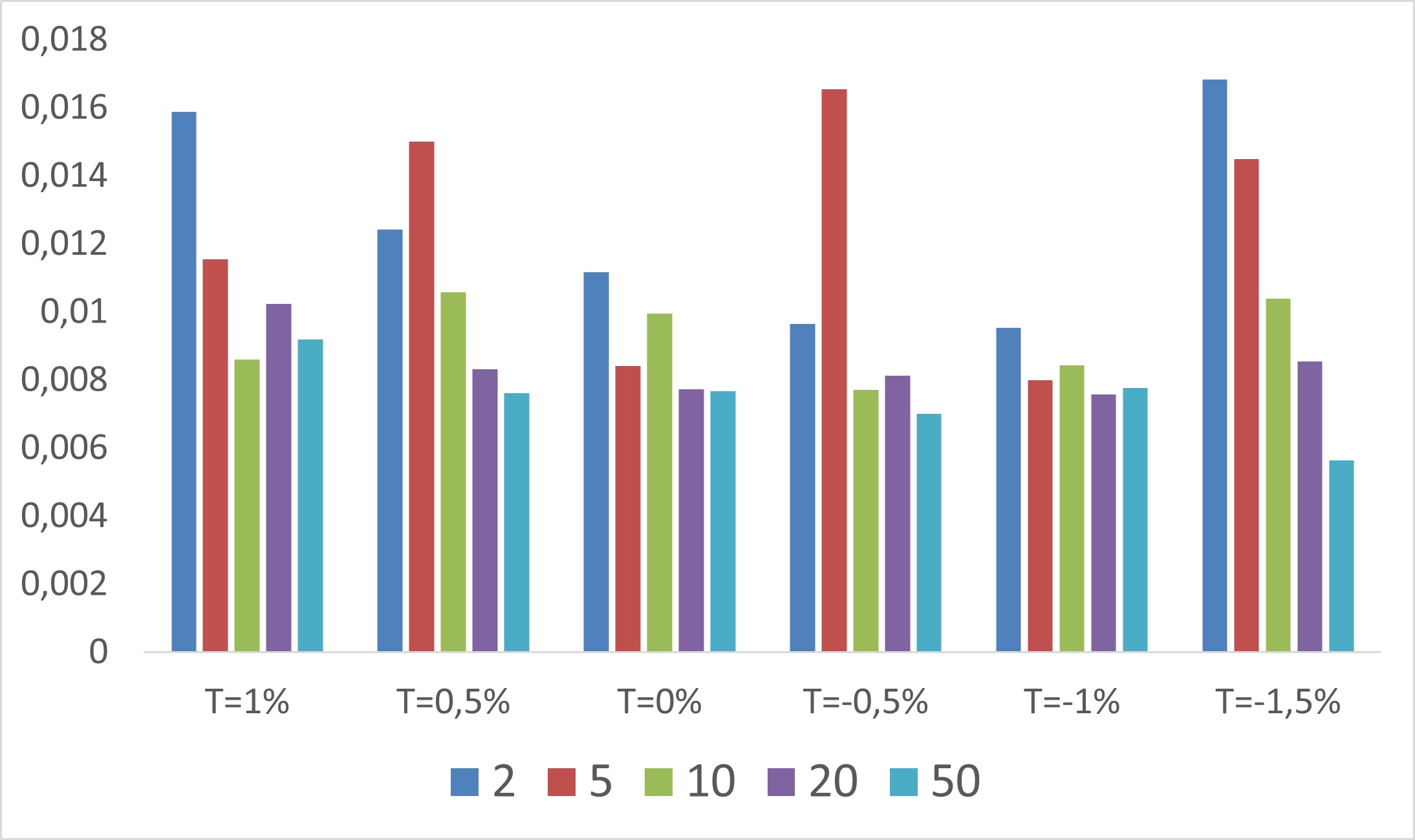}
        \subcaption{$\varepsilon = 0.1$ ; $\lambda_s= 10$}
        \label{regsub5}
    \end{subfigure}
    \hfill
    \begin{subfigure}{0.3\textwidth}
        \centering
        \includegraphics[width=\textwidth]{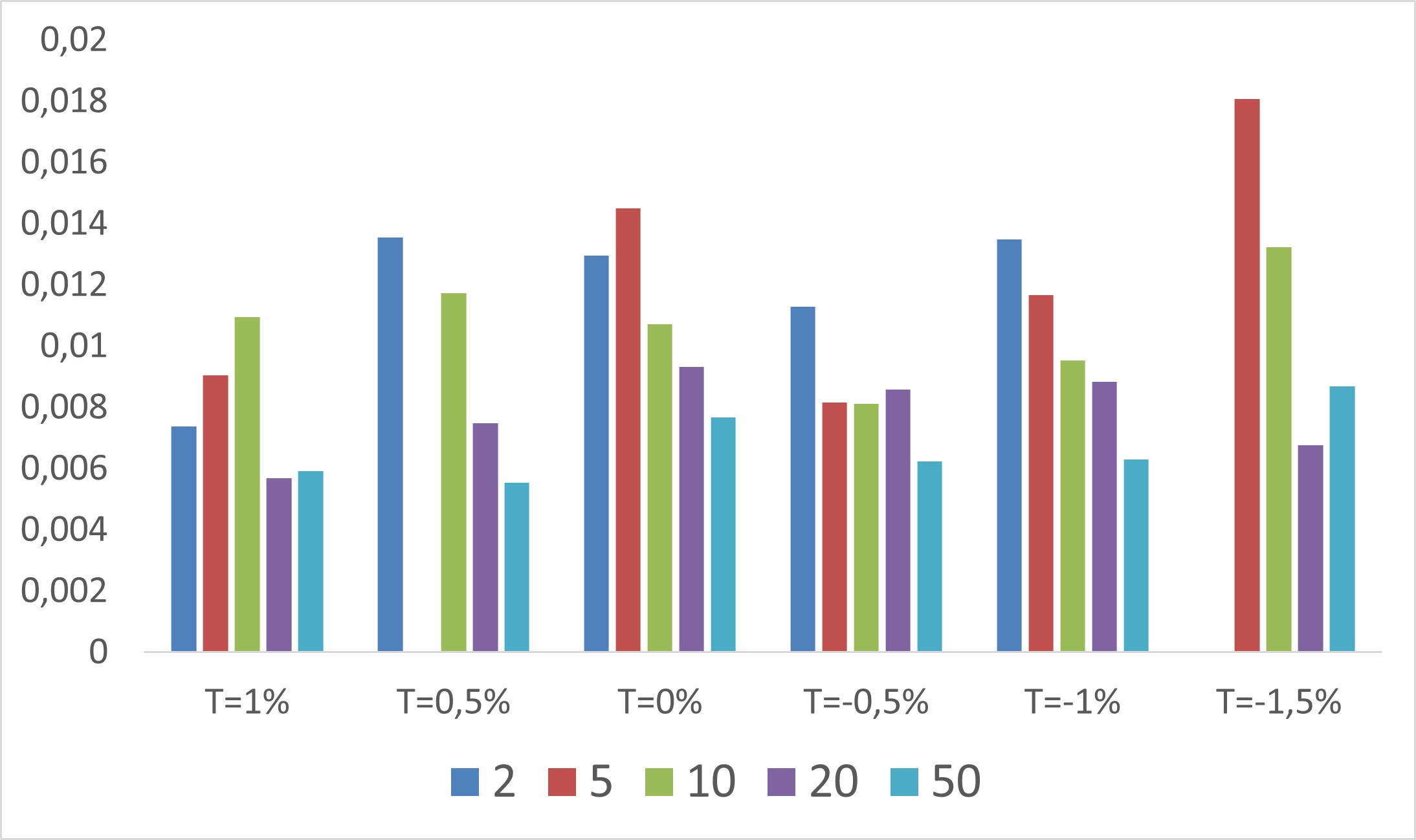}
        \subcaption{$\varepsilon = 0.35$ ; $\lambda_s= 10$}
        \label{regsub6}
    \end{subfigure}

    \caption{Portfolio ensemble 10-month (multi-step) average test RMSE with different diversity parameter $\varepsilon$, ensemble (s-RBFN) regularization parameter $\lambda_s$, threshold for asset selection $T$, number of portfolio constituents or predictors (2, 5, 10, 20, 50), and Gaussian basis functions.\\
    To perform a closer analysis of the regularization parameter, it can be seen how for $\lambda_s=10$ the generalization error is more stable for all model hyperparameters, and in the case of no regularization parameter, there are some cases in which the results are quite unstable (like in Figures \ref{regsub2} and \ref{regsub3}) where the test RMSE are displayed. It can be concluded that the regularization parameter reduces the uncertainty of the s-RBFN hyperparameters in the case of the Gaussian basis function. This instability in hyperparameter selection was not observed with the s-RBFN using radial basis functions.}
    \label{MSEGaussianFigures}
\end{figure}

\section{Including Diversity in the Asset or Hypothesis Selection Stage: One-step Decision-making}

\begin{figure}[H]
    \centering
    \begin{subfigure}{0.45\textwidth}
        \includegraphics[width=\linewidth]{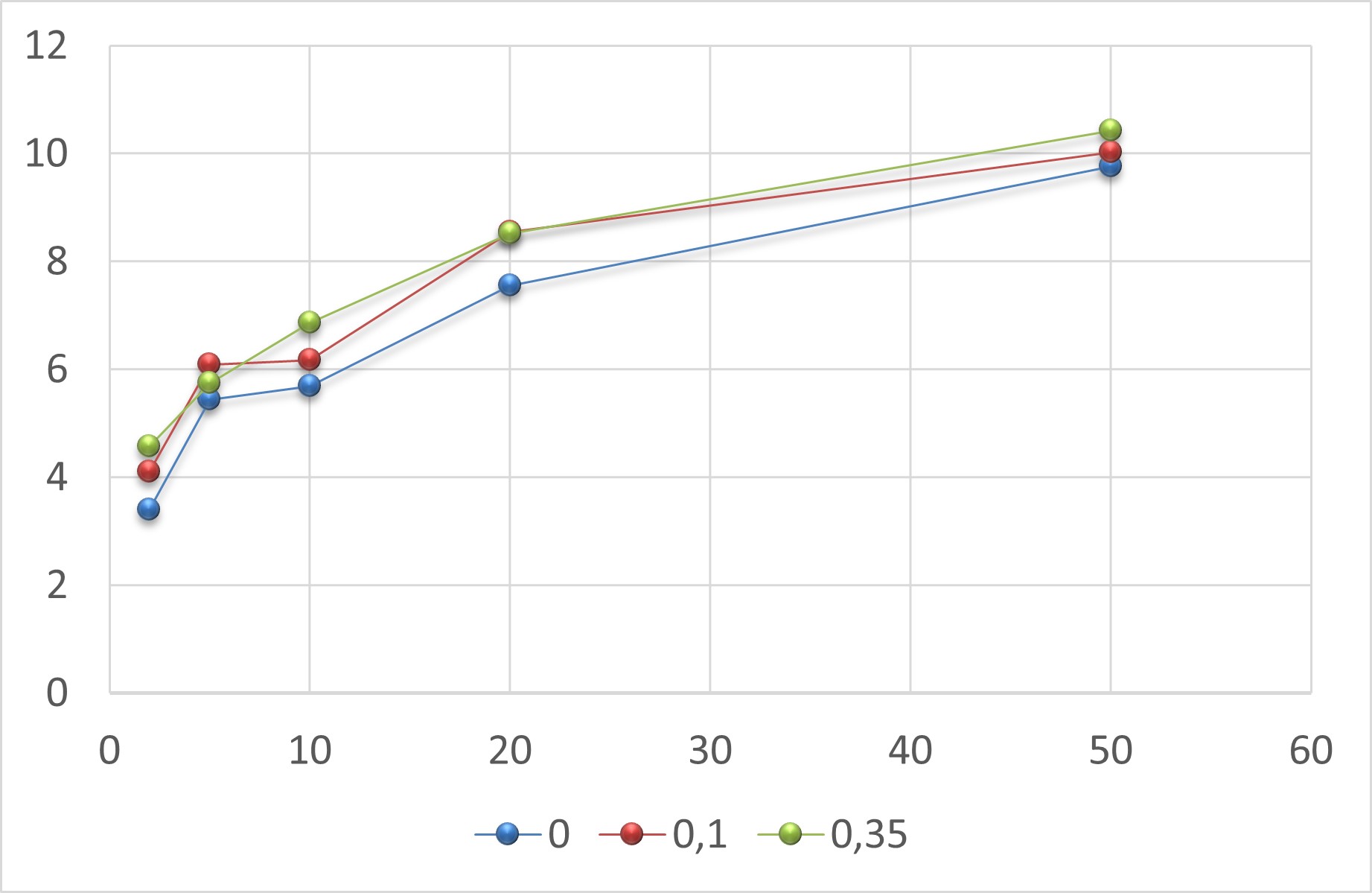} 
        \caption{$\gamma=1;0.5\%$}
        \label{fig:image11m}
    \end{subfigure}
    \begin{subfigure}{0.45\textwidth}
        \includegraphics[width=\linewidth]{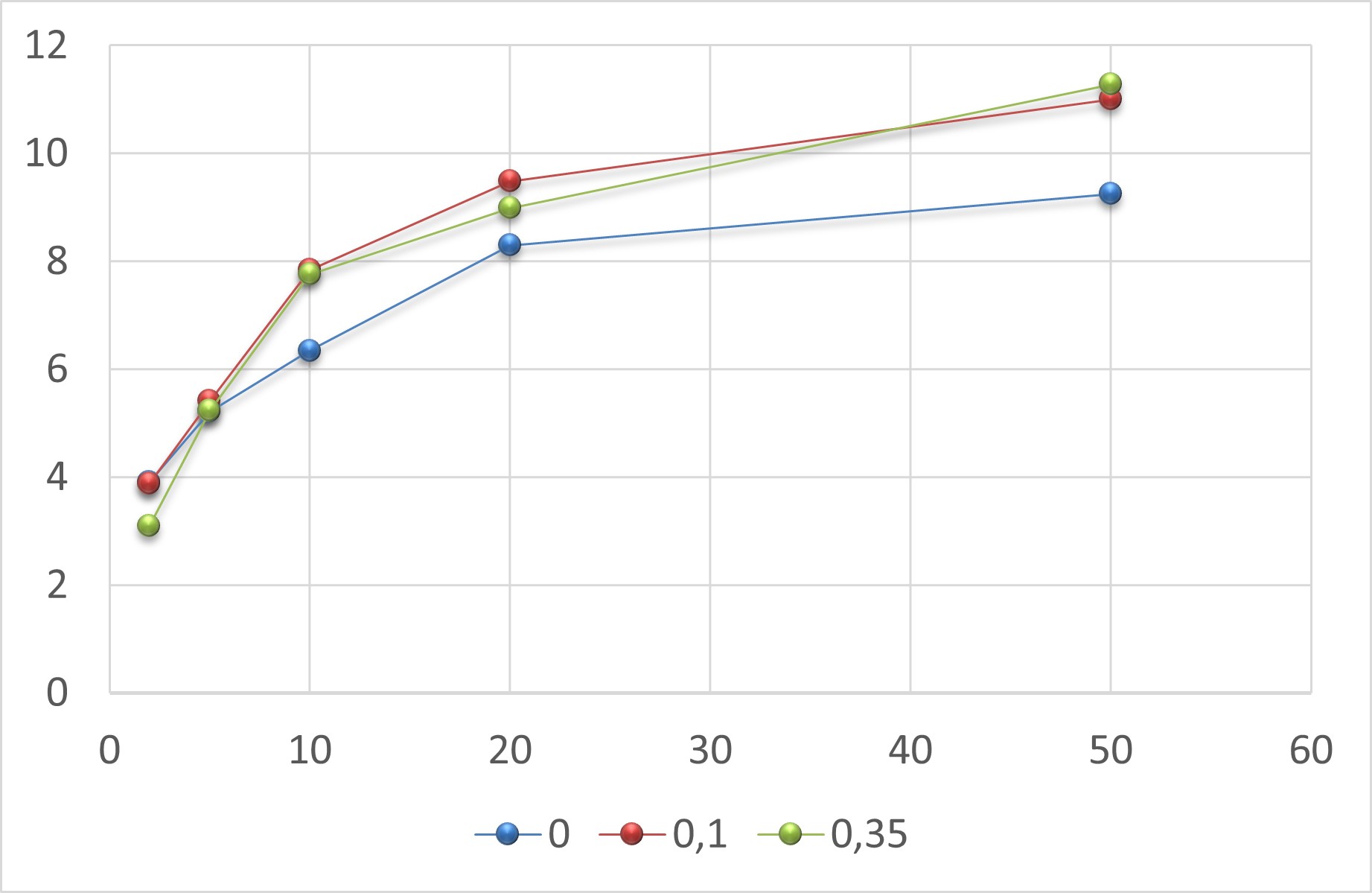} 
        \caption{$\gamma=2;0.5\%$}
        \label{fig:image21m}
    \end{subfigure}

    \begin{subfigure}{0.45\textwidth}
        \includegraphics[width=\linewidth]{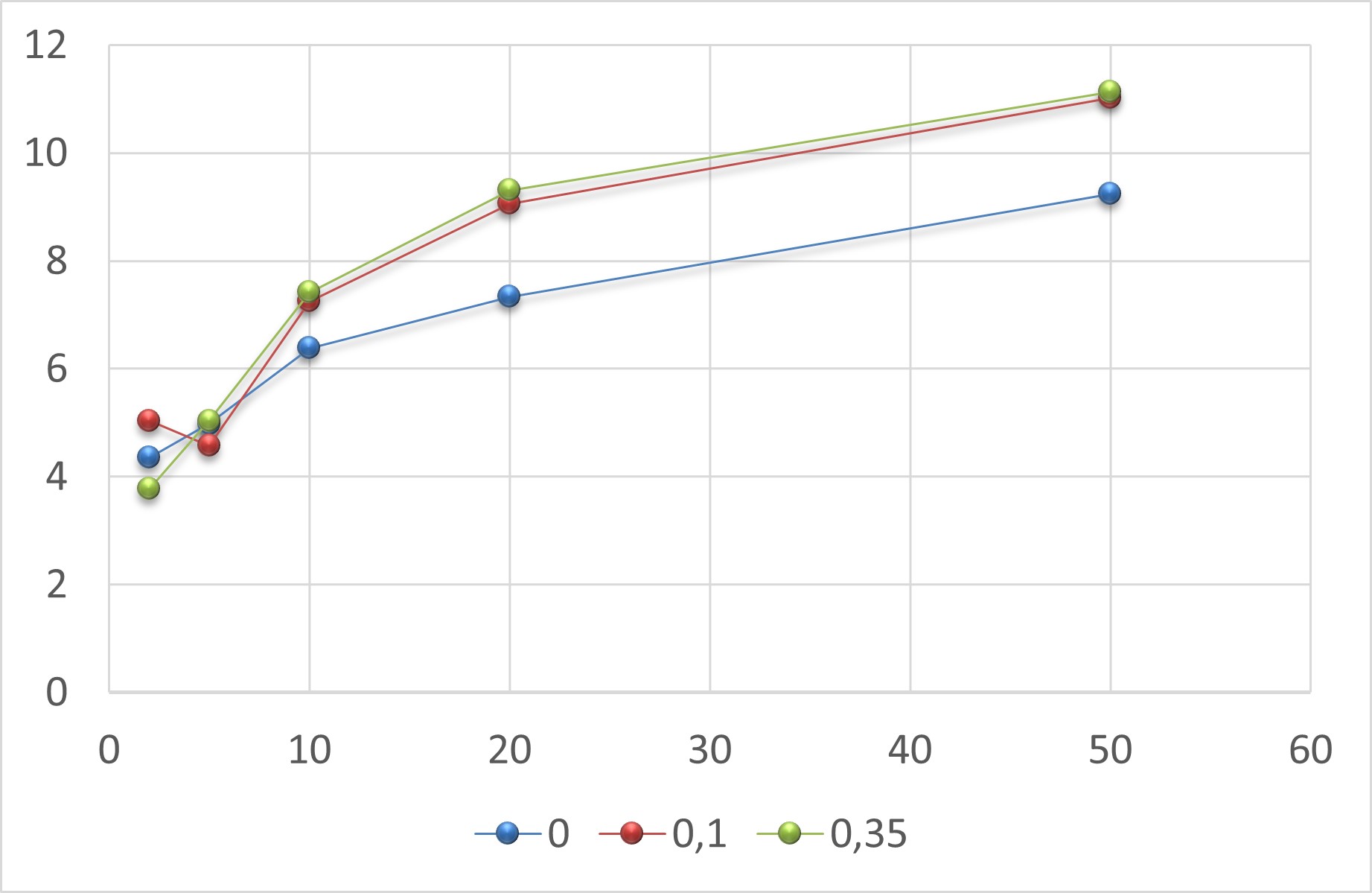} 
        \caption{$\gamma=3;0.5\%$}
        \label{fig:image31m}
    \end{subfigure}
    \begin{subfigure}{0.45\textwidth}
        \includegraphics[width=\linewidth]{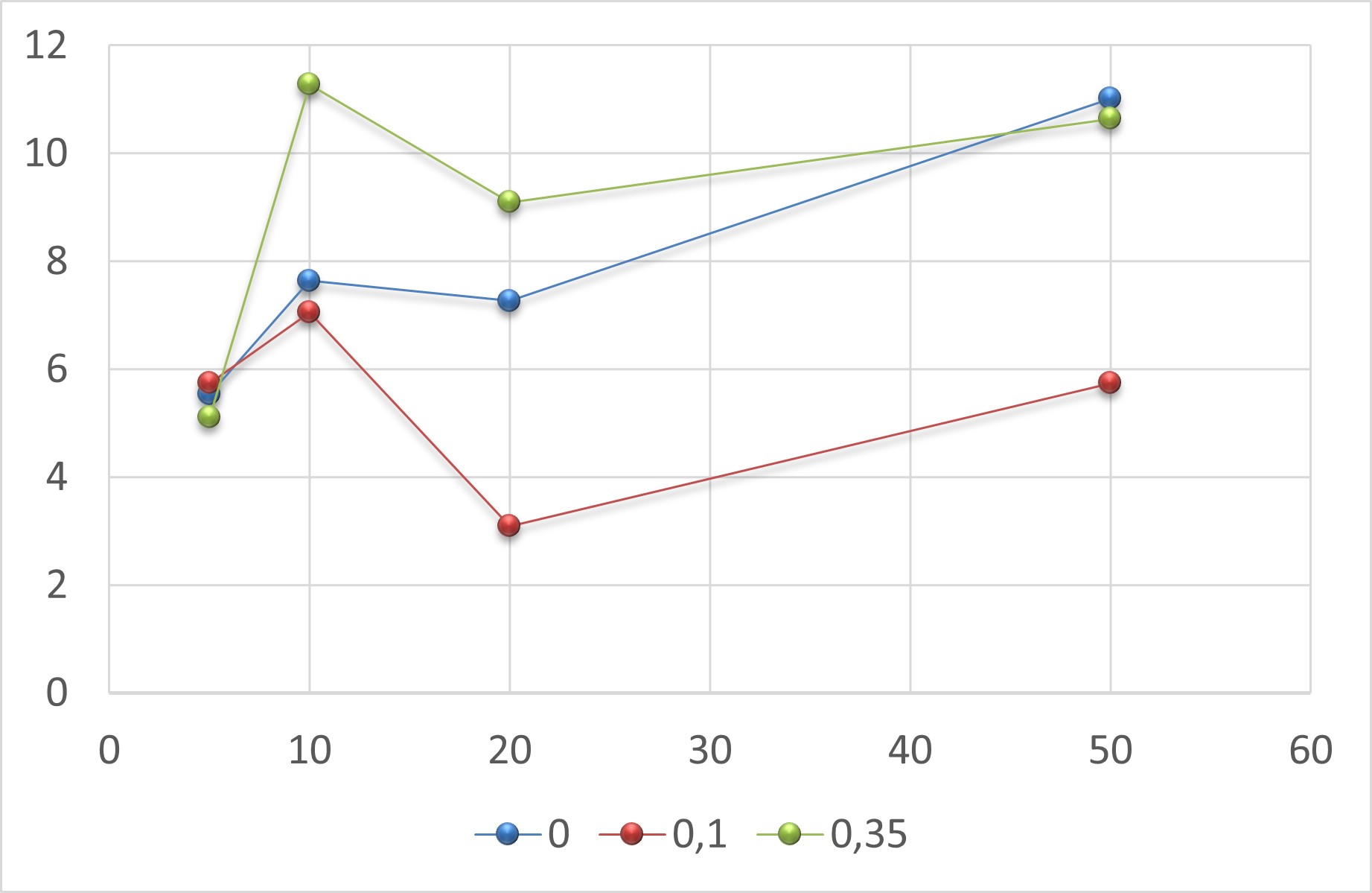} 
        \caption{$\gamma=5;0.5\%$}
        \label{fig:image41m}
    \end{subfigure}

    \begin{subfigure}{0.45\textwidth}
        \includegraphics[width=\linewidth]{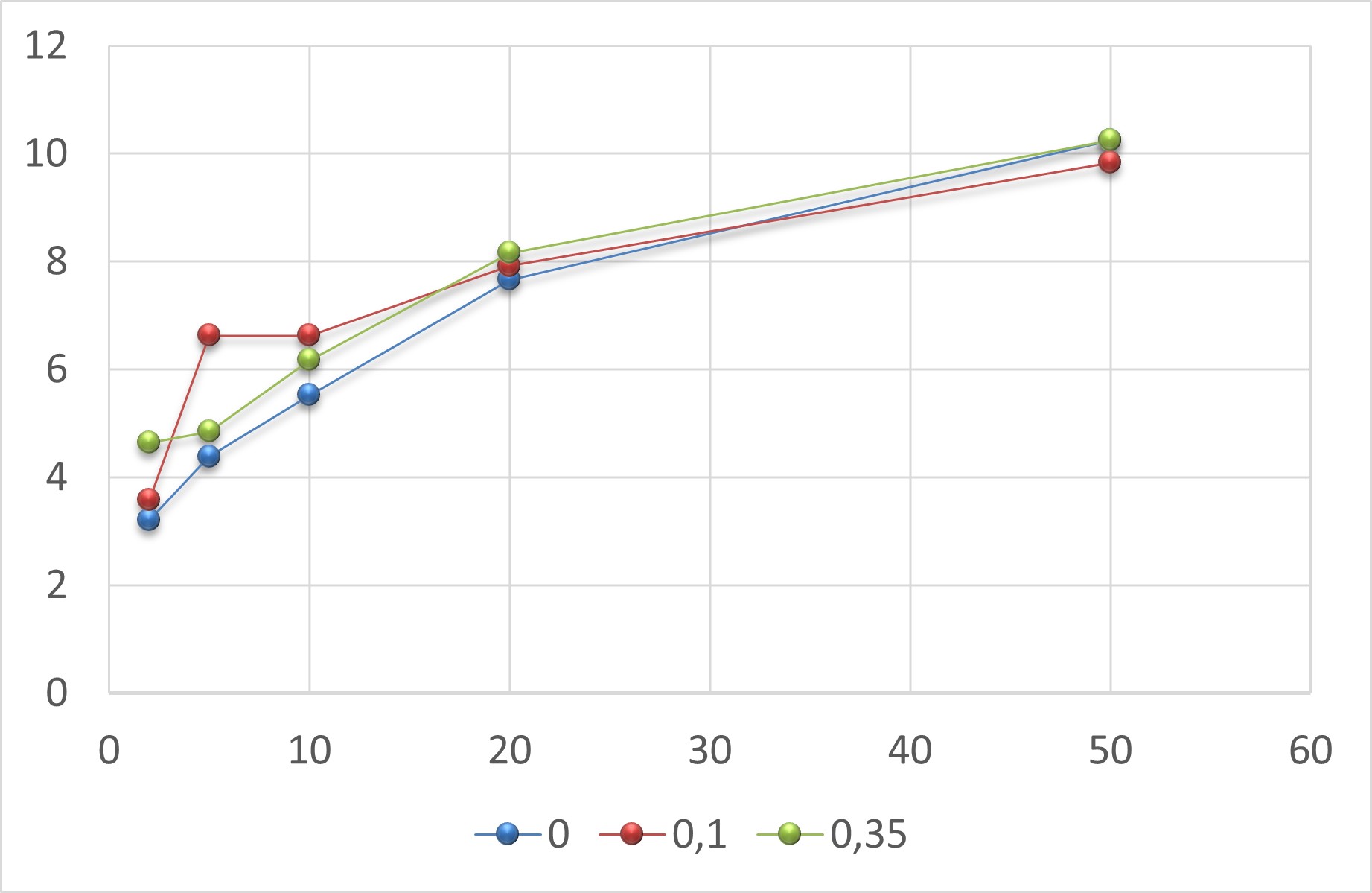} 
        \caption{$\gamma=1;0.0\%$}
        \label{fig:image51m}
    \end{subfigure}
    \begin{subfigure}{0.45\textwidth}
        \includegraphics[width=\linewidth]{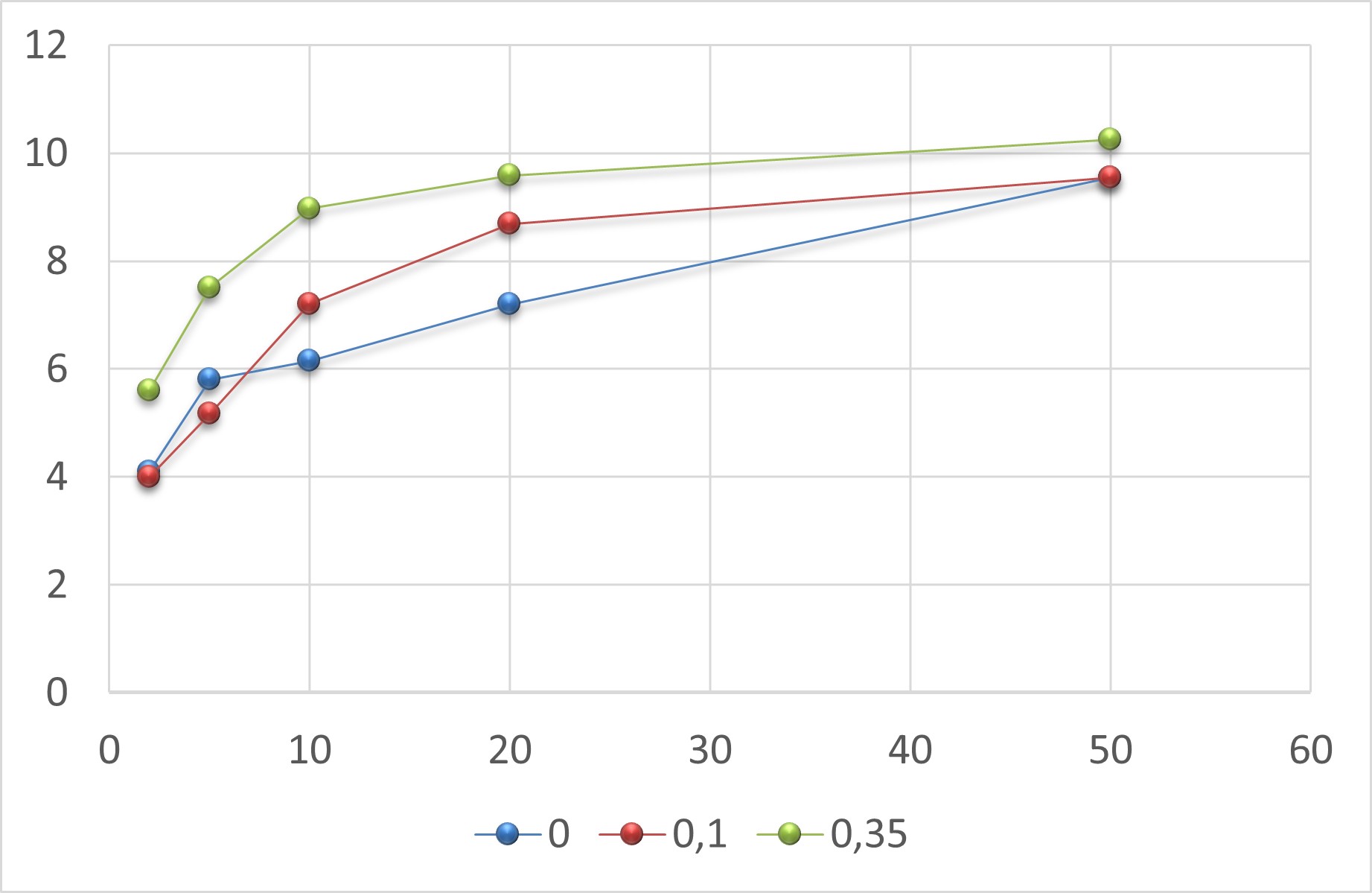} 
        \caption{$\gamma=2;0.0\%$}
        \label{fig:image61m}
    \end{subfigure}

    \begin{subfigure}{0.45\textwidth}
        \includegraphics[width=\linewidth]{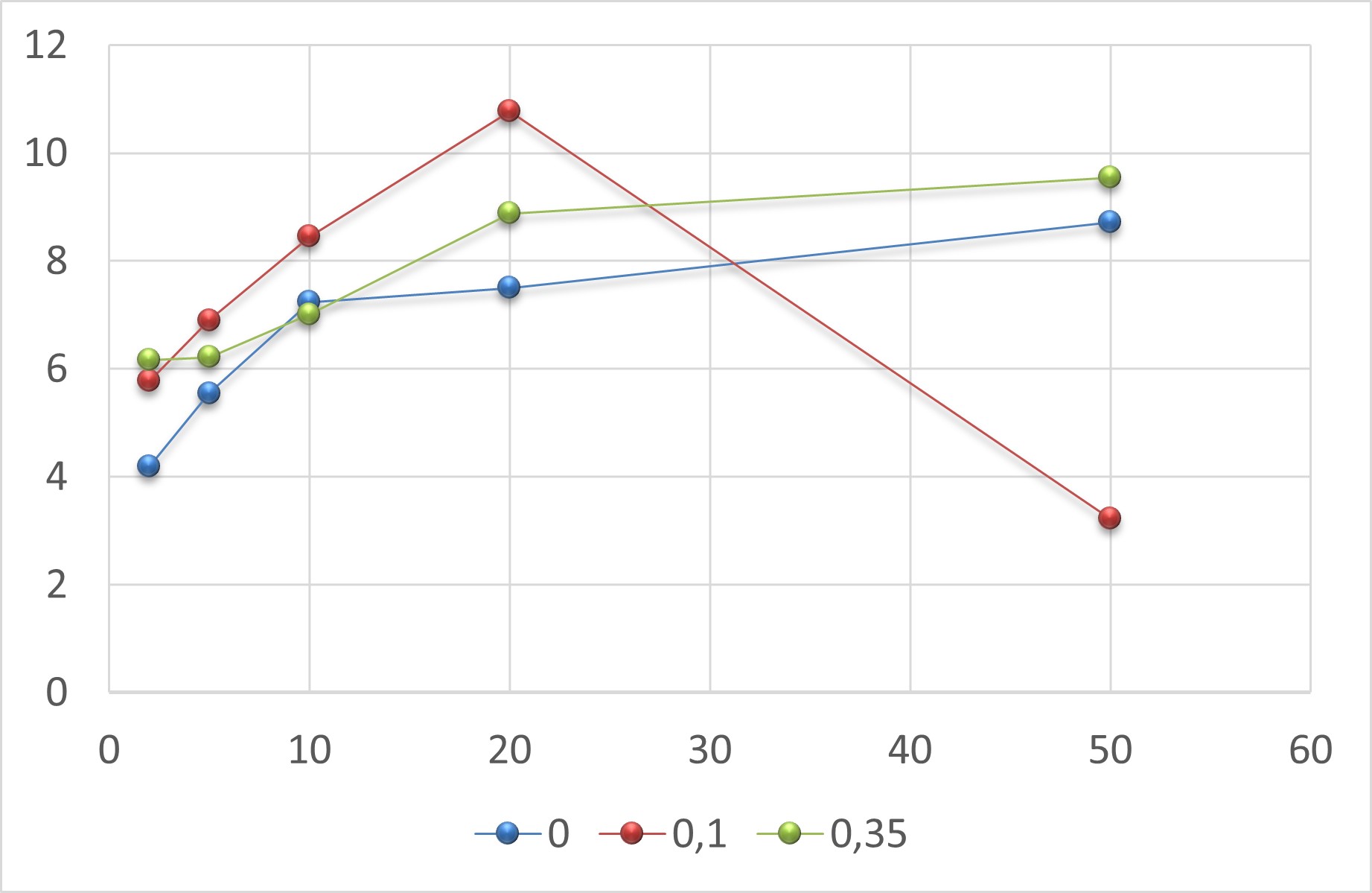} 
        \caption{$\gamma=3;0.0\%$}
        \label{fig:image71m}
    \end{subfigure}
    \begin{subfigure}{0.45\textwidth}
        \includegraphics[width=\linewidth]{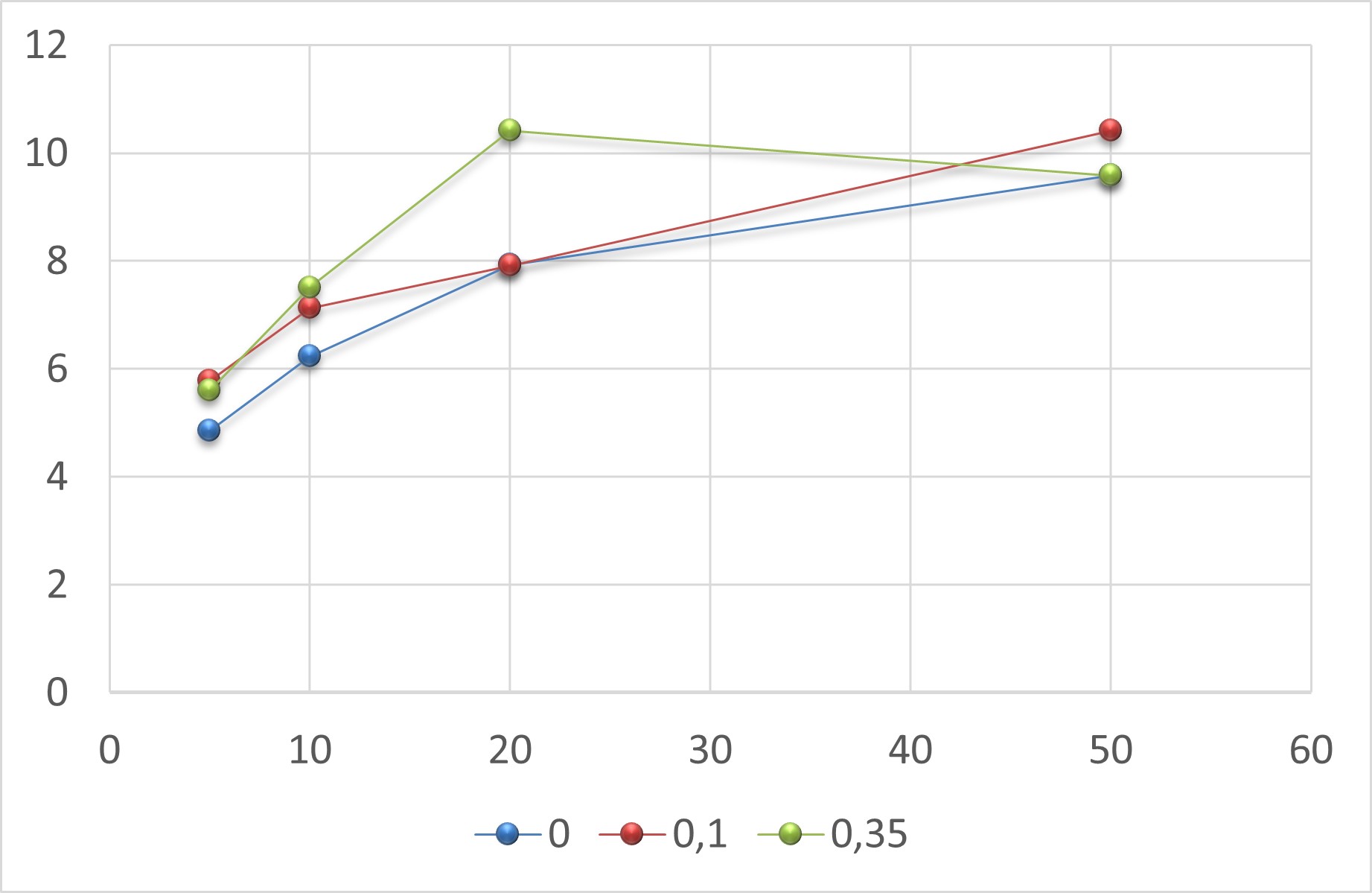} 
        \caption{$\gamma=5;0.0\%$}
        \label{fig:image81m}
    \end{subfigure}

    \caption{One-step (1-month)  Sharpe ratios (Y-axis) for different number of predictors $M$ (X-axis), diversity parameter $\varepsilon=0,0.1,0.35$ (colours),  multiplicative factor or diversity parameter in the asset selection stage $\gamma$, and $0\%\leq T\leq 0.5\%$. s-RBFN with radial basis functions.}
    \label{fig:large_figure1m}
\end{figure}

\begin{figure}[H]
    \centering
    \begin{subfigure}{0.45\textwidth}
        \includegraphics[width=\linewidth]{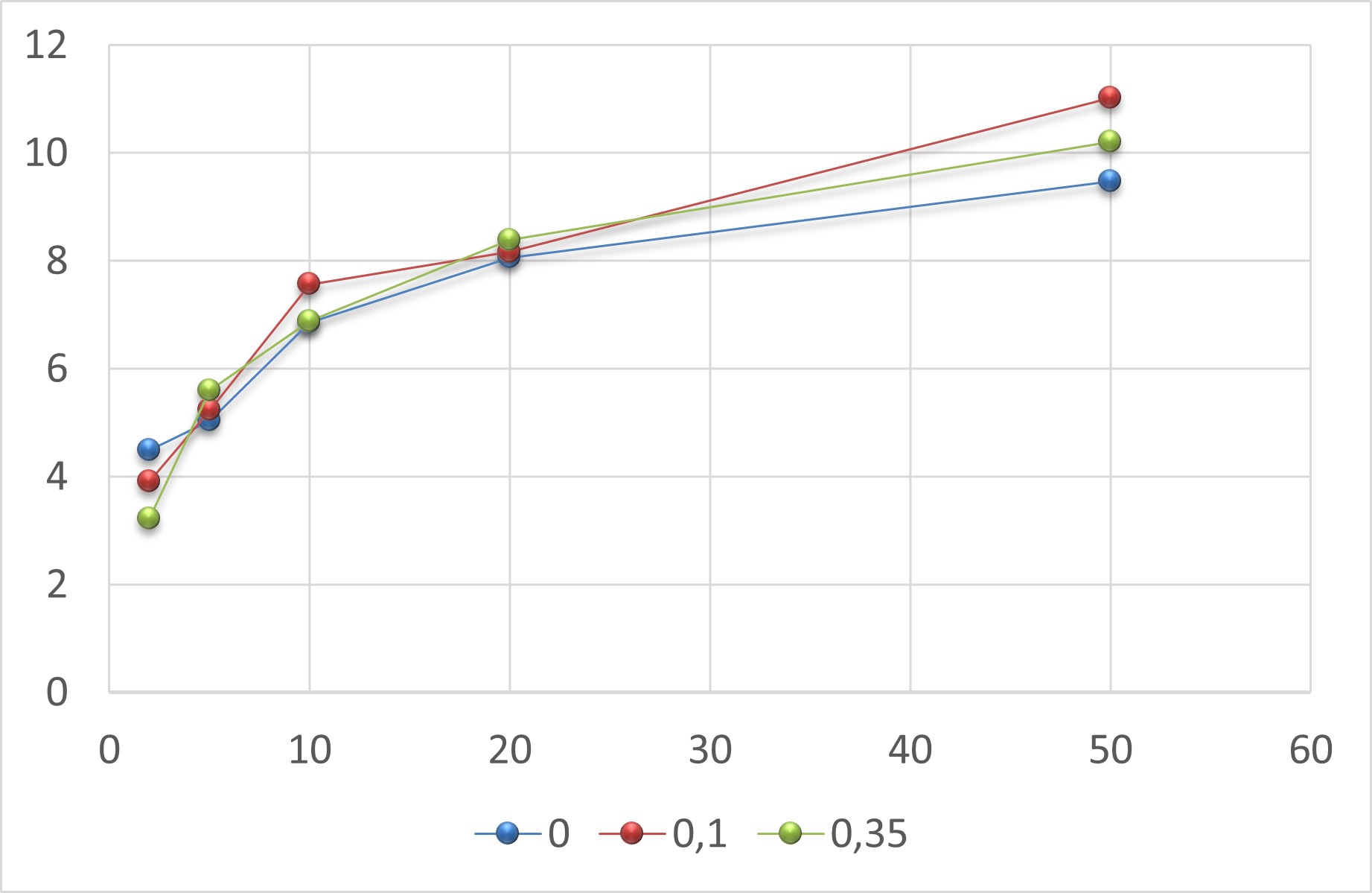} 
        \caption{$\gamma=1;T=-0.5\%$}
        \label{fig:image11m}
    \end{subfigure}
    \begin{subfigure}{0.45\textwidth}
        \includegraphics[width=\linewidth]{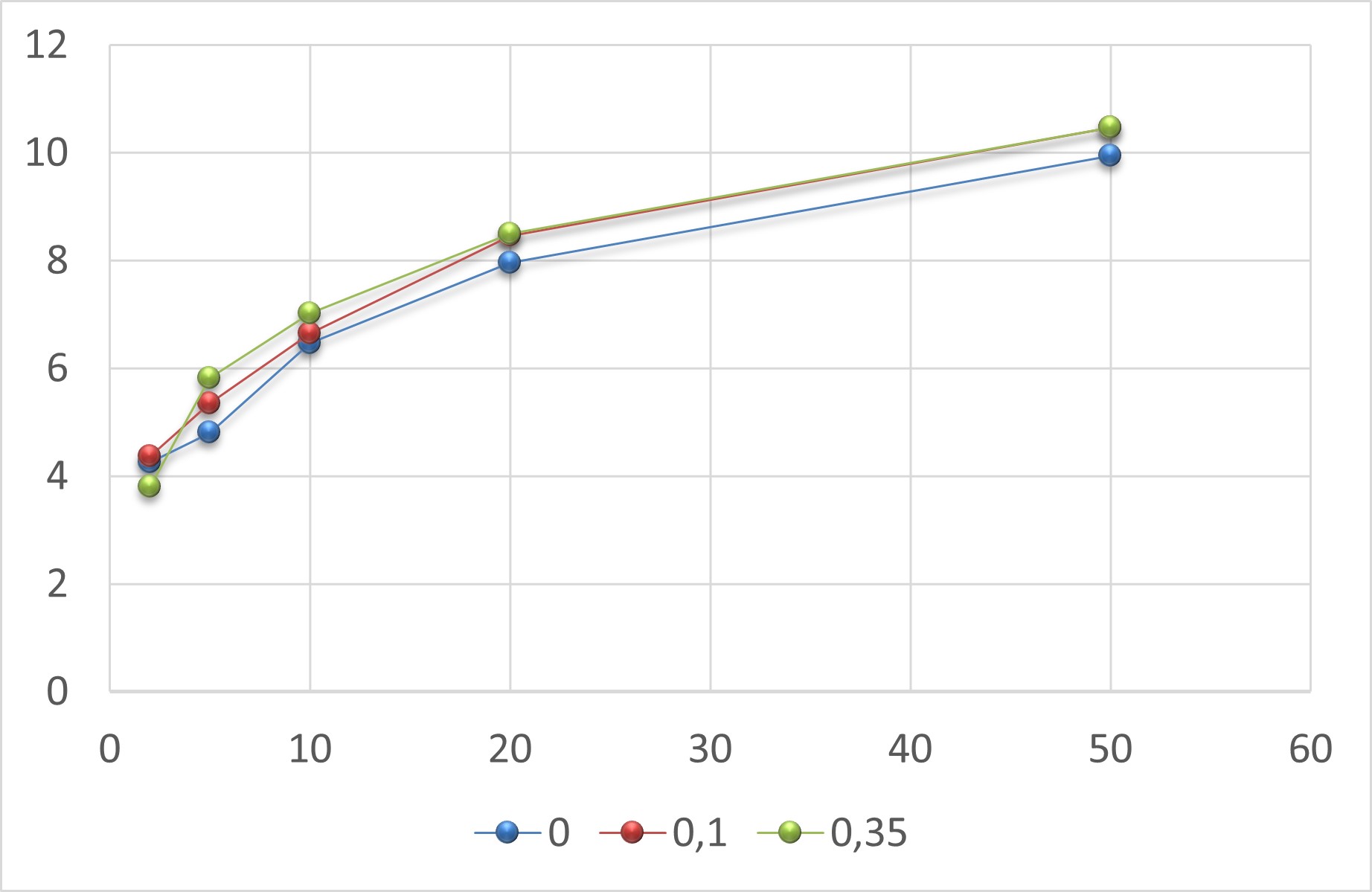} 
        \caption{$\gamma=2;T=-0.5\%$}
        \label{fig:image21m}
    \end{subfigure}

    \begin{subfigure}{0.45\textwidth}
        \includegraphics[width=\linewidth]{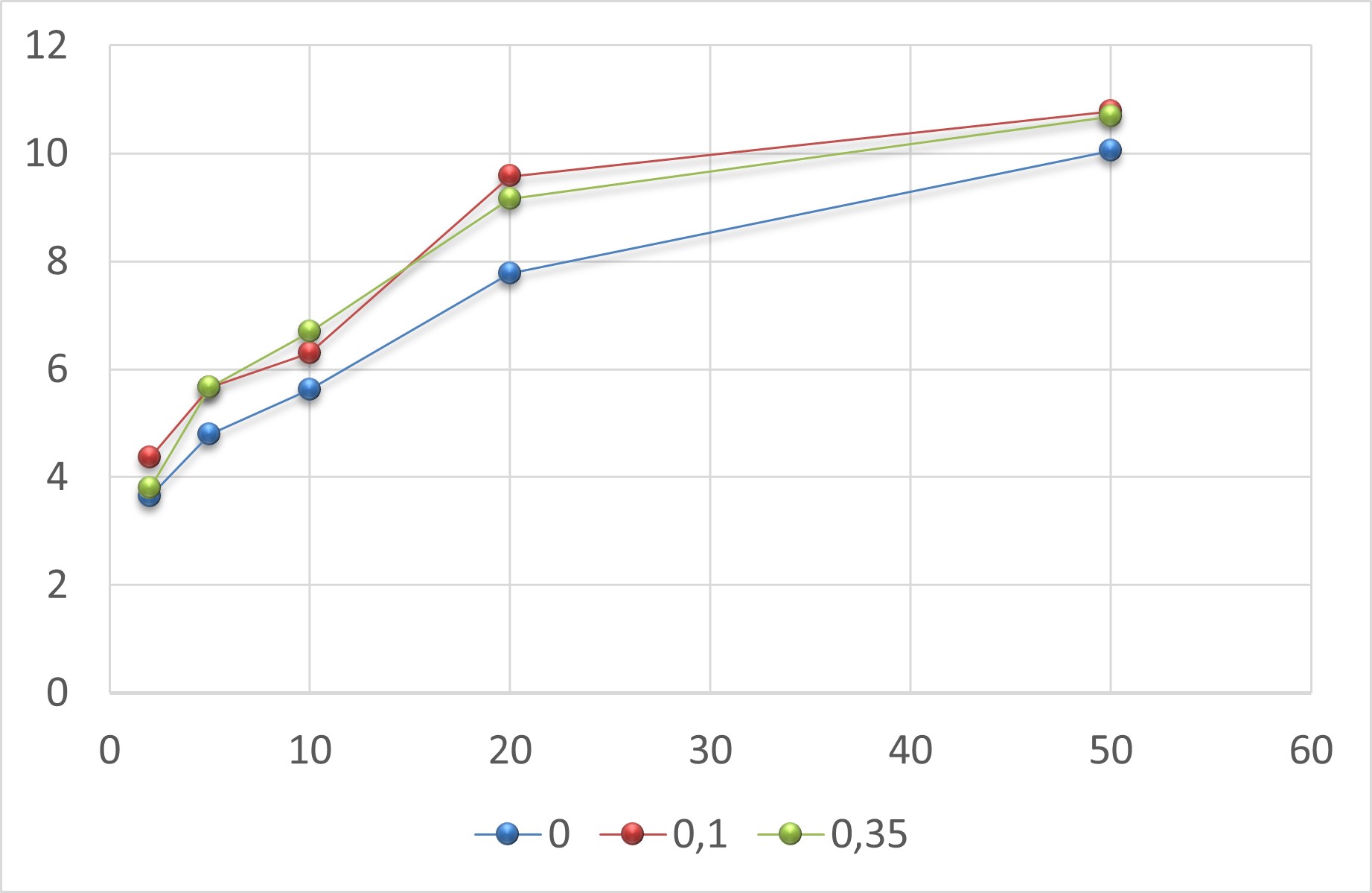} 
        \caption{$\gamma=3;T=-0.5\%$}
        \label{fig:image31m}
    \end{subfigure}
    \begin{subfigure}{0.45\textwidth}
        \includegraphics[width=\linewidth]{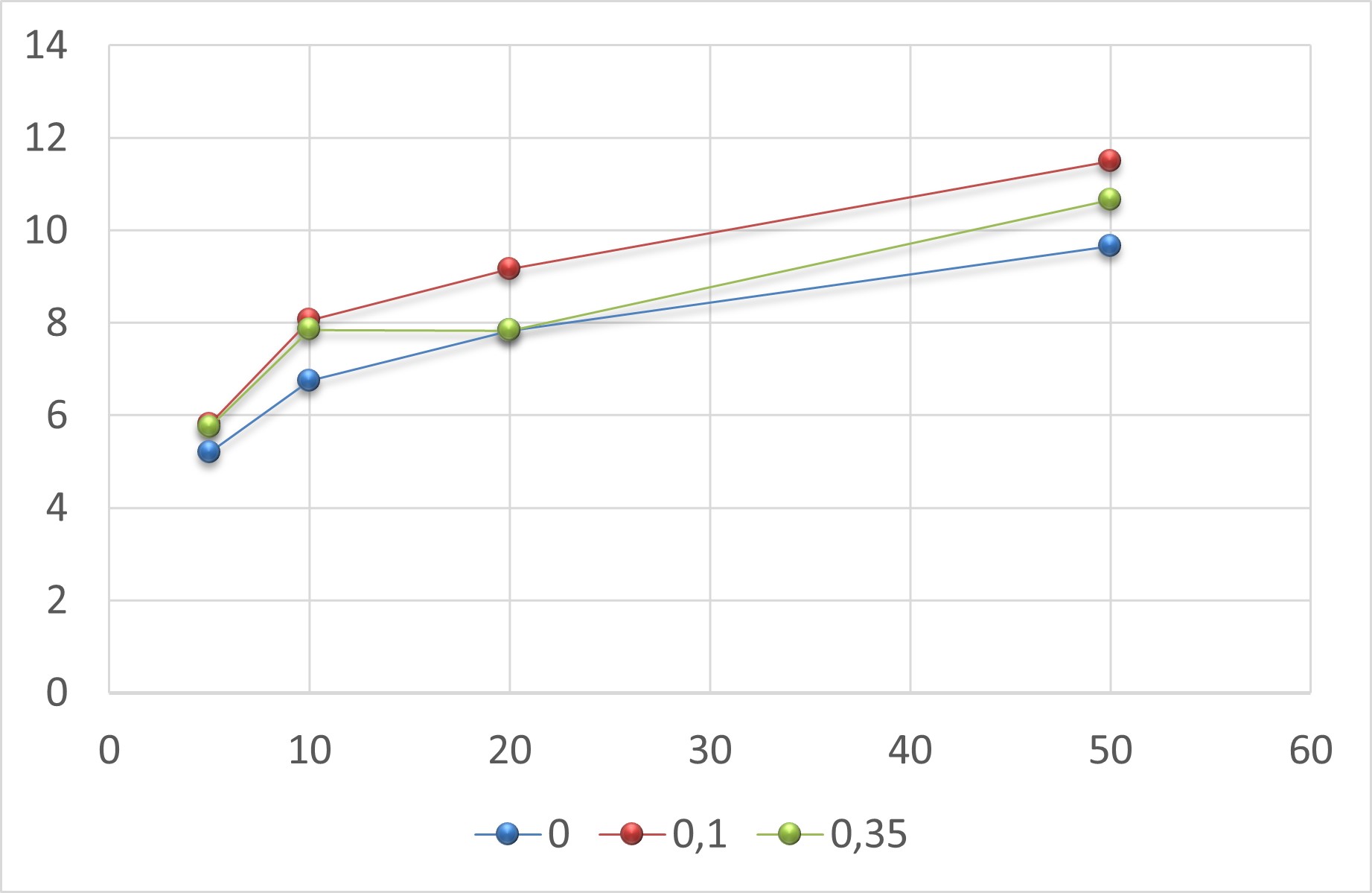} 
        \caption{$\gamma=5;T=-0.5\%$}
        \label{fig:image41m}
    \end{subfigure}

    \begin{subfigure}{0.45\textwidth}
        \includegraphics[width=\linewidth]{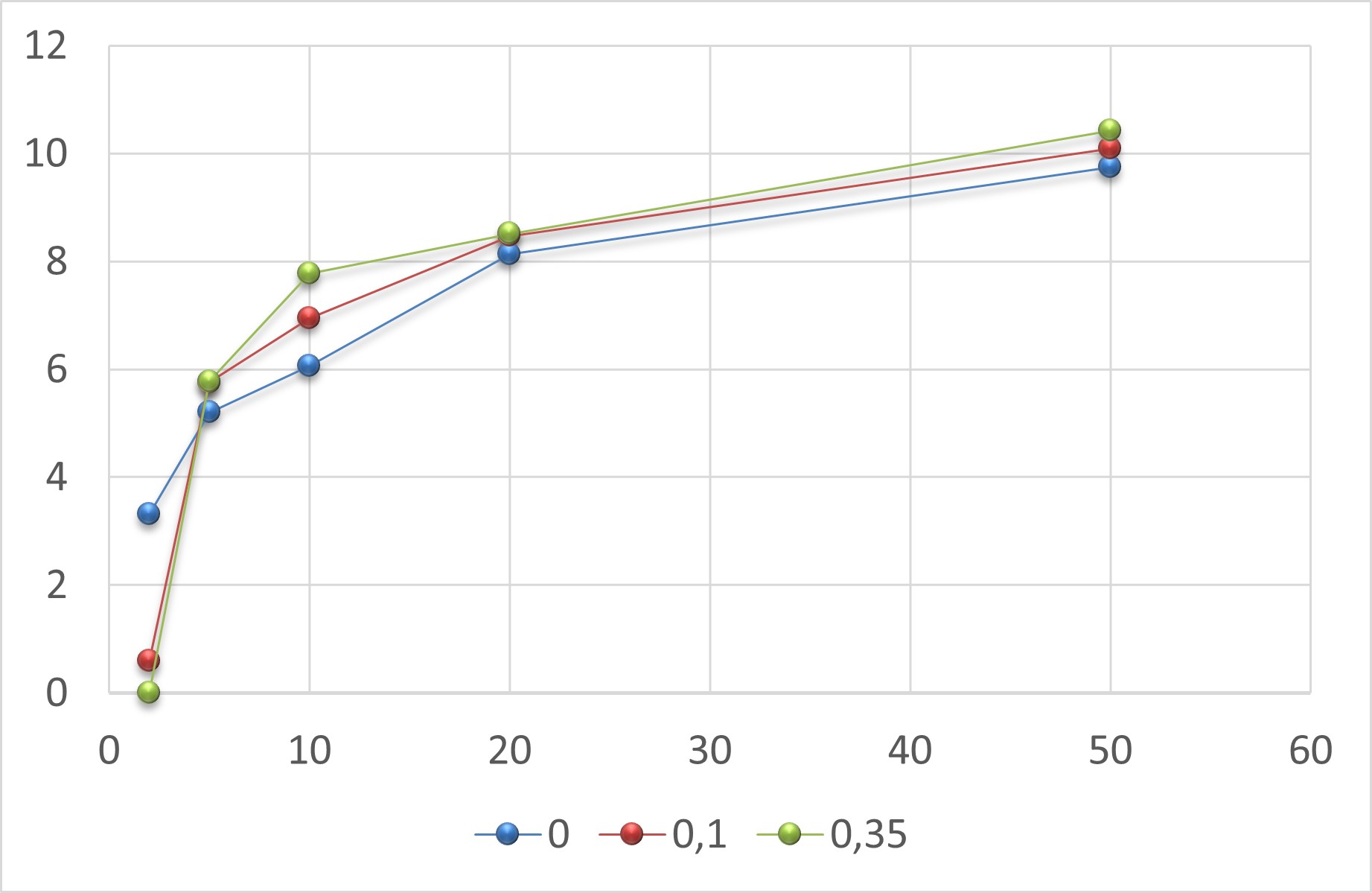} 
        \caption{$\gamma=1;T=-1\%$}
        \label{fig:image51m}
    \end{subfigure}
    \begin{subfigure}{0.45\textwidth}
        \includegraphics[width=\linewidth]{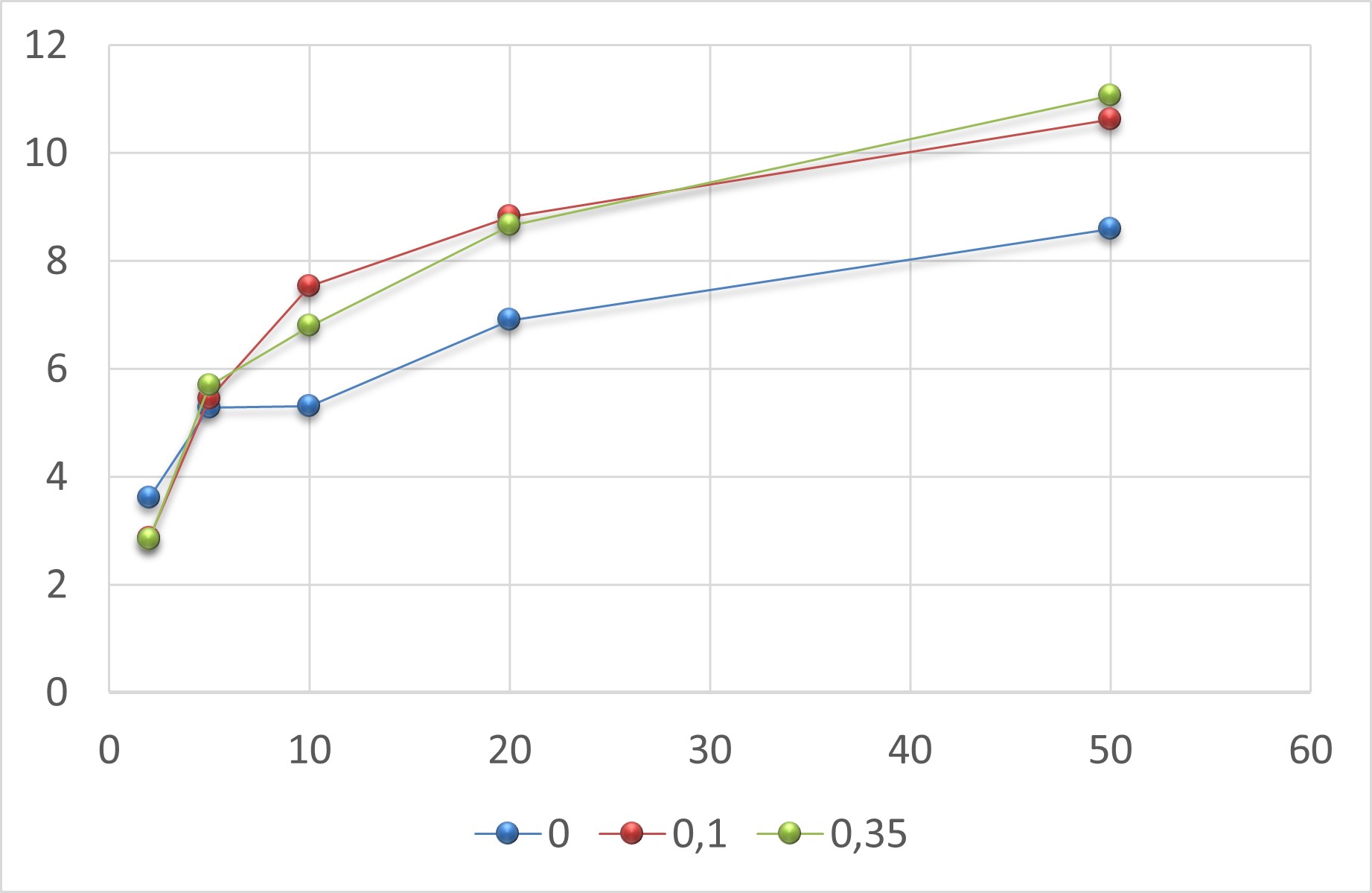} 
        \caption{$\gamma=2;T=-1\%$}
        \label{fig:image61m}
    \end{subfigure}

    \begin{subfigure}{0.45\textwidth}
        \includegraphics[width=\linewidth]{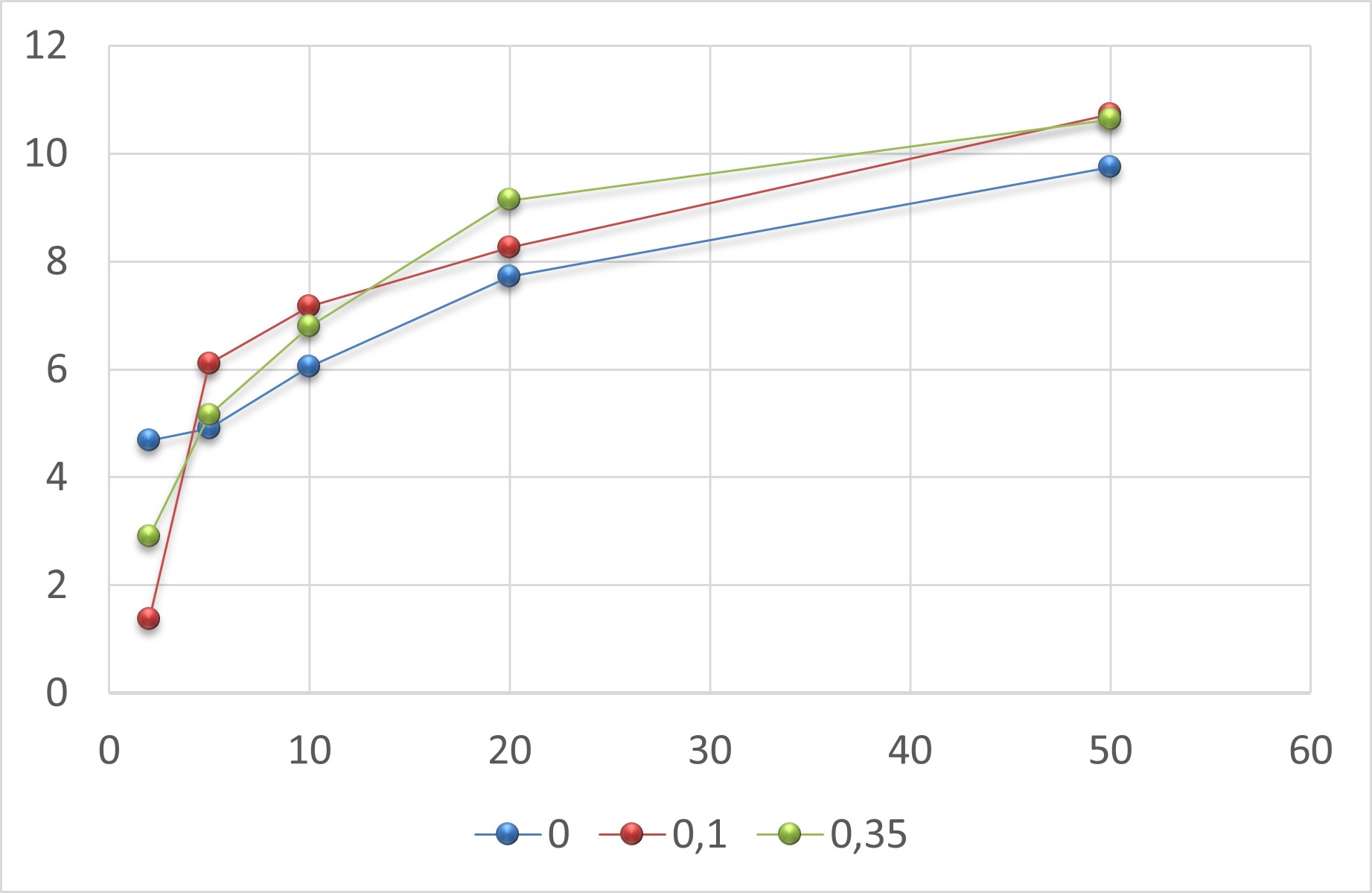} 
        \caption{$\gamma=3;T=-1\%$}
        \label{fig:image71m}
    \end{subfigure}
    \begin{subfigure}{0.45\textwidth}
        \includegraphics[width=\linewidth]{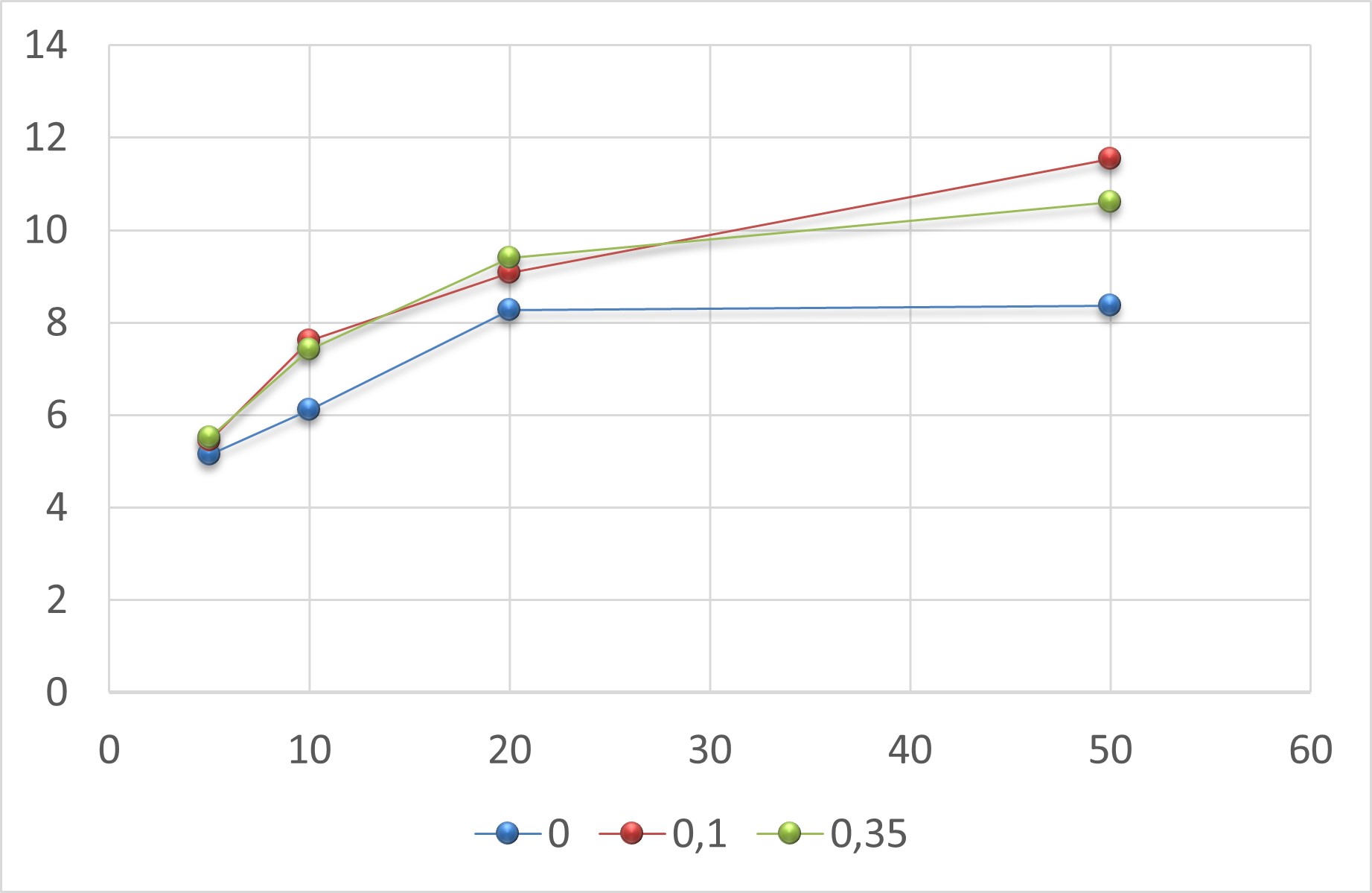} 
        \caption{$\gamma=5;T=-1\%$}
        \label{fig:image81m}
    \end{subfigure}

    \caption{One-step (1-month)  Sharpe ratios (Y-axis) for different number of predictors $M$ (X-axis), diversity parameter $\varepsilon=0,0.1,0.35$ (colours), multiplicative factor or diversity parameter in the asset selection stage $\gamma$, and $-1\%\leq T\leq -0.5\%$. s-RBFN with radial basis functions.}
    \label{fig:large_figure1mmin}
\end{figure}

\section{Sensitivity Analysis of Diversity Parameters across different Regimes, Asset Classes and Performance Metrics}
\label{appendixD}

\subsection{Performance Analysis of the s-RBFN Model Across Different Regimes and Asset Classes as a Function of the Diversity Parameter $\varepsilon$ and the Number of Portfolio Constituents $n$}
\label{Sensisection}

\begin{table}[H]
\centering
\caption{Performance metrics across varying values of $\varepsilon$ and number of assets $n$ during the U.S. equities market rally spanning 2009 to 2016.}
\label{tab:compact_metric_summary}
\captionsetup{labelfont={color=blue},textfont={color=blue}}

\renewcommand{\arraystretch}{1.1}

\begin{tabular}{l|c|cccccc}
\toprule
\textbf{Metric} & \textbf{$n$} & \textbf{0} & \textbf{0.1} & \textbf{0.35} & \textbf{0.5} & \textbf{0.75} & \textbf{0.95} \\
\midrule
\multirow{5}{*}{Sharpe} 
 & 5   & 0.645 & 0.629 & 0.399 & \textbf{1.035} & 0.685 & 0.691 \\
 & 10  & 1.093 & \textbf{1.208} & 0.960 & 0.687 & 1.023 & 1.220 \\
 & 20  & \textbf{1.369} & 1.135 & 1.135 & 1.452 & 1.135 & 0.852 \\
 & 100 & 1.020 & 1.182 & 1.061 & 1.231 & \textbf{1.229} & 1.186 \\
 & \textbf{Avg.} & 1.032 & 1.039 & 0.889 & 1.101 & 1.018 & 0.987 \\
\midrule
\multirow{5}{*}{Max Drawdown} 
 & 5   & -0.307 & \textbf{-0.286} & -0.219 & -0.328 & -0.336 & -0.285 \\
 & 10  & -0.352 & -0.252 & -0.258 & -0.348 & -0.322 & \textbf{-0.215} \\
 & 20  & -0.261 & -0.235 & \textbf{-0.220} & -0.259 & -0.248 & -0.289 \\
 & 100 & -0.260 & -0.259 & -0.261 & \textbf{-0.231} & -0.258 & -0.252 \\
 & \textbf{Avg.} & -0.295 & -0.258 & -0.240 & -0.292 & -0.291 & -0.260 \\
\midrule
\multirow{5}{*}{Sortino} 
 & 5   & 0.923 & 0.871 & 0.536 & \textbf{1.417} & 0.893 & 0.983 \\
 & 10  & 1.403 & \textbf{1.626} & 1.336 & 0.934 & 1.397 & 1.763 \\
 & 20  & 1.897 & 1.532 & 1.541 & \textbf{1.900} & 1.453 & 1.125 \\
 & 100 & 1.328 & 1.541 & 1.385 & \textbf{1.520} & 1.611 & 1.557 \\
 & \textbf{Avg.} & 1.388 & 1.393 & 1.200 & 1.443 & 1.339 & 1.357 \\
\midrule
\multirow{5}{*}{Omega} 
 & 5   & 1.123 & 1.118 & 1.073 & \textbf{1.205} & 1.130 & 1.132 \\
 & 10  & 1.221 & \textbf{1.242} & 1.185 & 1.129 & 1.198 & 1.237 \\
 & 20  & 1.270 & 1.224 & 1.224 & \textbf{1.296} & 1.211 & 1.165 \\
 & 100 & 1.201 & 1.238 & 1.211 & 1.231 & \textbf{1.246} & 1.237 \\
 & \textbf{Avg.} & 1.204 & 1.206 & 1.173 & 1.215 & 1.196 & 1.193 \\
\bottomrule
\end{tabular}
\end{table}

\begin{table}[H]
\centering
\caption{Performance metrics across varying values of the diversity parameter $\varepsilon$ and number of assets $n$, averaged over $M$, during the U.S. Credit Crunch Crisis (2007--2009).}

\label{tab:compact_metric_summary_withM}
\renewcommand{\arraystretch}{1.1}

\begin{tabular}{l|c|cccccc}
\toprule
\textbf{Metric} & \textbf{$n$} & \textbf{0} & \textbf{0.1} & \textbf{0.35} & \textbf{0.5} & \textbf{0.75} & \textbf{0.95} \\
\midrule
\multirow{5}{*}{Sharpe} 
 & 5   & -0.454 & 0.007 & \textbf{0.380} & -0.027 & -0.744 & 0.585 \\
 & 10  & -0.457 & -0.027 & -0.225 & -0.203 & -0.484 & \textbf{-0.112} \\
 & 20  & \textbf{0.129} & -0.041 & -0.034 & 0.215 & -0.284 & -0.226 \\
 & 100 & -0.186 & -0.170 & -0.209 & \textbf{-0.124} & -0.264 & -0.298 \\
 & \textbf{Avg.} & -0.242 & -0.058 & -0.022 & -0.035 & -0.429 & \textbf{-0.013} \\
\midrule
\multirow{5}{*}{Max Drawdown} 
 & 5   & -0.646 & \textbf{-0.690} & -0.605 & -0.645 & -0.790 & -0.488 \\
 & 10  & -0.713 & -0.564 & -0.604 & -0.607 & -0.640 & \textbf{-0.534} \\
 & 20  & \textbf{-0.498} & -0.570 & -0.551 & -0.382 & -0.578 & -0.575 \\
 & 100 & -0.532 & -0.540 & -0.577 & -0.545 & -0.581 & \textbf{-0.606} \\
 & \textbf{Avg.} & -0.597 & -0.591 & -0.584 & \textbf{-0.545} & -0.647 & -0.551 \\
\midrule
\multirow{5}{*}{Sortino} 
 & 5   & -0.623 & \textbf{0.008} & 0.542 & -0.036 & -1.077 & 0.846 \\
 & 10  & -0.625 & -0.038 & -0.319 & -0.270 & -0.749 & \textbf{-0.167} \\
 & 20  & \textbf{0.191} & -0.053 & -0.047 & 0.309 & -0.401 & -0.313 \\
 & 100 & -0.267 & -0.238 & -0.291 & \textbf{-0.173} & -0.372 & -0.412 \\
 & \textbf{Avg.} & -0.331 & -0.080 & -0.029 & -0.043 & -0.900 & \textbf{-0.011} \\
\midrule
\multirow{5}{*}{Omega} 
 & 5   & 0.921 & 1.001 & 1.073 & 0.995 & 0.878 & \textbf{1.111} \\
 & 10  & 0.912 & \textbf{0.995} & 0.960 & 0.964 & 0.916 & 0.981 \\
 & 20  & 1.024 & 0.992 & 0.994 & \textbf{1.039} & 0.949 & 0.960 \\
 & 100 & 0.967 & 0.970 & 0.963 & \textbf{0.978} & 0.954 & 0.947 \\
 & \textbf{Avg.} & 0.956 & 0.990 & 0.998 & \textbf{0.994} & 0.924 & \textbf{1.000} \\
\bottomrule
\end{tabular}
\end{table}

\begin{figure}[H]
    \centering

    \begin{subfigure}[t]{0.49\textwidth}
        \includegraphics[width=\linewidth]{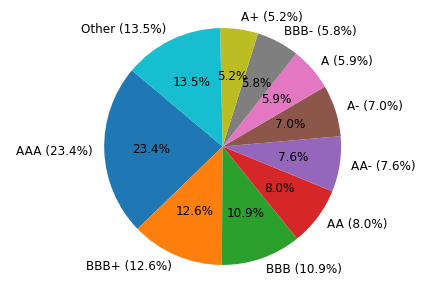}
        \caption{Rating}
    \end{subfigure}
    \hfill
    \begin{subfigure}[t]{0.49\textwidth}
        \includegraphics[width=\linewidth]{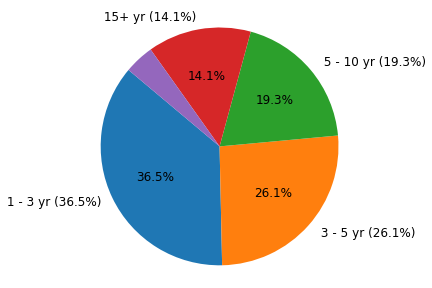}
        \caption{Tenor (curve)}
    \end{subfigure}
    \hfill
    \begin{subfigure}[t]{0.49\textwidth}
        \includegraphics[width=\linewidth]{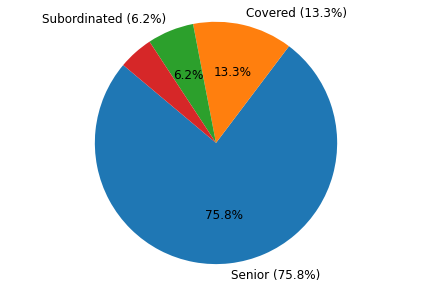}
        \caption{Seniority}
    \end{subfigure}

    \vspace{3mm}

    \begin{subfigure}[t]{0.4\textwidth}
        \includegraphics[width=\linewidth]{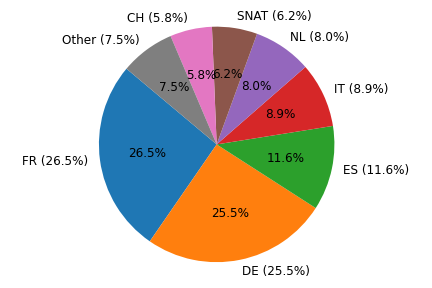}
        \caption{Country (ISO)}
    \end{subfigure}
    \hfill
    \begin{subfigure}[t]{0.59\textwidth}
        \includegraphics[width=\linewidth]{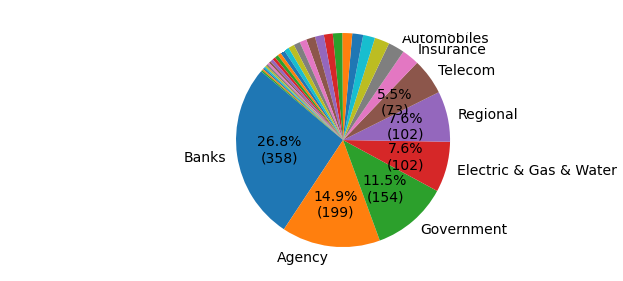}
        \caption{Sector}
    \end{subfigure}

\caption{Distribution of the 1,336 bonds in the data set across key categories: rating, curve (tenor), seniority, country (ISO), and sector.}

    \label{fig:bond_pie_subplots}
\end{figure}

\begin{table}[H]
\centering
\caption{Performance metrics across varying values of $\varepsilon$ and number of assets $n$ on a fixed income data set covering the global bond market (public, private, institutional, corporate, municipal sectors) from 2014 to 2018.}
\label{tab:compact_metric_summary_fixed_income}
\renewcommand{\arraystretch}{1.1}
\begin{tabular}{l|c|cccccc}
\toprule
\textbf{Metric} & \textbf{$n$} & \textbf{0} & \textbf{0.1} & \textbf{0.35} & \textbf{0.5} & \textbf{0.75} & \textbf{0.95} \\
\midrule
\multirow{5}{*}{Sharpe} 
 & 5   & -0.25 & -0.28 &  \textbf{0.01} & -0.68 & -0.31 &  0.03 \\
 & 10  &  0.48 & -0.57 & -0.20 &  \textbf{0.51} &  0.15 & -0.44 \\
 & 20  & -0.31 &  \textbf{0.51} & -0.54 & -0.55 & -0.33 & -0.20 \\
 & 100 &  0.41 &  0.31 &  \textbf{0.52} & -0.48 &  0.39 &  0.35 \\
 & \textbf{Avg.} & \textbf{0.08} & -0.01 & -0.05
 & -0.30 & -0.02 & -0.06 \\
\midrule
\multirow{5}{*}{Max Drawdown} 
 & 5   & -0.10 & -0.21 & \textbf{-0.07} & -0.19 & -0.17 & -0.09 \\
 & 10  & -0.14 & -0.10 & \textbf{-0.06} & -0.15 & -0.07 & -0.10 \\
 & 20  & -0.06 & -0.12 & -0.10 & -0.08 & -0.09 & \textbf{-0.05} \\
 & 100 & -0.08 & -0.06 & -0.06 & -0.07 & \textbf{-0.05} & -0.06 \\
 & \textbf{Avg.} & -0.10 & -0.12 & \textbf{-0.07} & -0.12 & -0.10 & -0.08 \\
\midrule
\multirow{5}{*}{Sortino} 
 & 5   & -0.29 & -0.29 &  \textbf{0.01} & -0.75 & -0.32 &  0.04 \\
 & 10  &  4.52 & -0.67 & -0.24 &  \textbf{9.34} &  0.18 & -0.49 \\
 & 20  & -0.36 & \textbf{12.24} & -0.61 & -0.64 & -0.43 & -0.24 \\
 & 100 &  2.97 &  1.36 & \textbf{16.16} & -0.62 &  2.17 &  1.17 \\
 & \textbf{Avg.} & 1.21 & 3.16 & \textbf{3.33} & 1.83 & 0.40 & 0.12 \\
\midrule
\multirow{5}{*}{Omega} 
 & 5   & 0.93 & 0.91 & \textbf{1.00} & 0.84 & 0.91 & 1.01 \\
 & 10  & 2.84 & 0.88 & 0.96 & \textbf{5.01} & 1.03 & 0.90 \\
 & 20  & 0.94 & \textbf{4.76} & 0.89 & 0.90 & 0.91 & 0.96 \\
 & 100 & 1.44 & 1.20 & \textbf{3.79} & 0.91 & 1.31 & 1.18 \\
 & \textbf{Avg.} & 1.54 & \textbf{1.94} & 1.66 & 1.91 & 1.29 & 1.26 \\
\bottomrule
\end{tabular}
\end{table}

\subsection{Comparative Analysis of Different Portfolio Optimization Methods}

\label{compdifreg}

\begin{table}[H]
\centering
\caption{Performance metrics by strategy and number of assets during the U.S. Equity Rally regime.}
\label{tab:performance_metrics_rally}
\resizebox{\textwidth}{!}{%
\begin{tabular}{|c|c|c|c|c|c|c|c|}
\hline
\textbf{n} & \textbf{s-RBFN} & \textbf{1\_N} & \textbf{CVaR RP} & \textbf{HERC} & \textbf{HRP} & \textbf{IV} & \textbf{MD} \\
\hline
\multicolumn{8}{|c|}{\textbf{Sharpe Ratio}} \\
\hline
5   & 1.11 & 0.51 & 0.58 & 0.08 & 0.48 & 0.58 & 0.49 \\
10  & 1.03 & 1.11 & 1.10 & 1.20 & 1.17 & 1.10 & 1.16 \\
20  & 1.16 & 1.12 & 1.20 & 1.00 & 1.18 & 1.20 & 1.11 \\
100 & 1.15 & 1.13 & 1.19 & 0.97 & 1.23 & 1.19 & 0.83 \\
\hline
\textbf{Avg} & \textbf{1.11} & 0.97 & 1.02 & 0.81 & 1.02 & 1.02 & 0.90 \\
\hline
\multicolumn{8}{|c|}{\textbf{Maximum Drawdown}} \\
\hline
5   & -0.30 & -0.31 & -0.28 & -0.41 & -0.29 & -0.28 & -0.26 \\
10  & -0.31 & -0.34 & -0.27 & -0.27 & -0.25 & -0.27 & -0.27 \\
20  & -0.26 & -0.22 & -0.20 & -0.23 & -0.20 & -0.20 & -0.19 \\
100 & -0.25 & -0.26 & -0.21 & -0.23 & -0.21 & -0.21 & -0.32 \\
\hline
\textbf{Avg} & -0.28 & -0.28 & \textbf{-0.24} & -0.29 & \textbf{-0.24} & \textbf{-0.24} & -0.26 \\
\hline
\multicolumn{8}{|c|}{\textbf{Sortino Ratio}} \\
\hline
5   & 1.25 & 0.71 & 0.80 & 0.11 & 0.65 & 0.80 & 0.68 \\
10  & 1.46 & 1.39 & 1.41 & 1.67 & 1.52 & 1.41 & 1.51 \\
20  & 1.64 & 1.50 & 1.61 & 1.26 & 1.57 & 1.61 & 1.52 \\
100 & 1.55 & 1.47 & 1.53 & 1.26 & 1.58 & 1.53 & 1.07 \\
\hline
\textbf{Avg} & \textbf{1.48} & 1.27 & 1.34 & 1.08 & 1.33 & 1.34 & 1.20 \\
\hline
\multicolumn{8}{|c|}{\textbf{Omega Ratio}} \\
\hline
5   & 1.00 & 0.94 & 0.94 & 0.98 & 0.94 & 0.94 & 0.94 \\
10  & 1.17 & 0.91 & 0.90 & 0.92 & 0.89 & 0.90 & 0.87 \\
20  & 1.02 & 0.96 & 0.95 & 0.93 & 0.94 & 0.95 & 0.93 \\
100 & 0.99 & 0.97 & 0.95 & 0.95 & 0.95 & 0.95 & 0.95 \\
\hline
\textbf{Avg} & \textbf{1.04} & 0.95 & 0.93 & 0.95 & 0.93 & 0.93 & 0.92 \\
\hline
\end{tabular}
}
\end{table}

\begin{table}[H]
\centering
\caption{Performance metrics by strategy and number of assets during the U.S. Credit Crunch (2007--2009).}
\label{tab:metrics_credit_crunch}
\resizebox{\textwidth}{!}{%
\begin{tabular}{|c|c|c|c|c|c|c|c|}
\hline
\textbf{Assets} & \textbf{s-RBFN} & \textbf{1/N} & \textbf{CVaR RP} & \textbf{HERC} & \textbf{HRP} & \textbf{IV} & \textbf{MD} \\
\hline
\multicolumn{8}{|c|}{\textbf{Sharpe Ratio}} \\
\hline
5   & -0.64 & -0.33 & -0.37 & -0.11 & -0.34 & -0.37 & -0.35 \\
10  &  0.82 & -0.53 & -0.59 & -0.42 & -0.65 & -0.59 & -0.72 \\
20  & -0.15 & -0.22 & -0.28 & -0.38 & -0.34 & -0.28 & -0.38 \\
100 & -0.07 & -0.18 & -0.28 & -0.23 & -0.31 & -0.28 & -0.28 \\
\hline
\textbf{Avg.} & \textbf{-0.01} & -0.32 & -0.38 & -0.29 & -0.41 & -0.38 & -0.43 \\
\hline
\multicolumn{8}{|c|}{\textbf{Maximum Drawdown}} \\
\hline
5   & -0.80 & -0.60 & -0.57 & -0.54 & -0.58 & -0.57 & -0.56 \\
10  & -0.60 & -0.68 & -0.61 & -0.69 & -0.62 & -0.61 & -0.62 \\
20  & -0.64 & -0.55 & -0.52 & -0.59 & -0.54 & -0.52 & -0.53 \\
100 & -0.56 & -0.53 & -0.51 & -0.50 & -0.51 & -0.51 & -0.48 \\
\hline
\textbf{Avg.} & -0.65 & -0.59 & \textbf{-0.55} & -0.58 & -0.56 & -0.55 & -0.55 \\
\hline
\multicolumn{8}{|c|}{\textbf{Sortino Ratio}} \\
\hline
5   & -0.12 & -0.49 & -0.55 & -0.17 & -0.50 & -0.55 & -0.52 \\
10  &  0.07 & -0.72 & -0.80 & -0.52 & -0.86 & -0.80 & -0.91 \\
20  & -0.21 & -0.31 & -0.39 & -0.50 & -0.48 & -0.39 & -0.50 \\
100 & -0.11 & -0.25 & -0.40 & -0.31 & -0.44 & -0.40 & -0.37 \\
\hline
\textbf{Avg.} & \textbf{-0.09} & -0.44 & -0.54 & -0.38 & -0.57 & -0.54 & -0.58 \\
\hline
\multicolumn{8}{|c|}{\textbf{Omega Ratio}} \\
\hline
5   & 0.89 & 0.94 & 0.94 & 0.98 & 0.94 & 0.94 & 0.94 \\
10  & 1.01 & 0.91 & 0.90 & 0.92 & 0.89 & 0.90 & 0.87 \\
20  & 0.97 & 0.96 & 0.95 & 0.93 & 0.94 & 0.95 & 0.93 \\
100 & 0.99 & 0.97 & 0.95 & 0.95 & 0.95 & 0.95 & 0.95 \\
\hline
\textbf{Avg.} & \textbf{0.97} & 0.95 & 0.94 & 0.95 & 0.93 & 0.94 & 0.92 \\
\hline
\end{tabular}
}
\end{table}

\begin{table}[H]
\centering
\caption{Performance metrics by strategy and number of assets on a data set of 1,336 global bonds, including public, corporate, institutional, municipal, and other sectors (2014--2018).}
\label{tab:all_metrics_stackedbonds}
\renewcommand{\arraystretch}{1.15}
\resizebox{\textwidth}{!}{%
\begin{tabular}{|c|c|c|c|c|c|c|}
\hline
\textbf{Assets} & \textbf{s-RBFN} & \textbf{1/N} & \textbf{IV} & \textbf{MD} & \textbf{HRP} & \textbf{HERC} \\
\hline
\multicolumn{7}{|c|}{\textbf{Sharpe Ratio}} \\
\hline
5   & -0.25 & -0.48 & -0.80 & -0.65 & -0.73 & -0.78 \\
10  &  0.48 &  0.47 &  0.29 & -0.14 &  0.32 &  0.19 \\
20  & -0.31 & -0.37 & -0.42 &  0.08 & -0.16 & -0.37 \\
100 &  0.41 &  0.39 &  0.32 &  0.06 &  0.36 & -0.02 \\
\hline
\textbf{Avg.} & \textbf{0.08} & 0.00 & -0.15 & -0.16 & -0.05 & -0.25 \\
\hline
\multicolumn{7}{|c|}{\textbf{Maximum Drawdown}} \\
\hline
5   & -0.10 & -0.11 & -0.12 & -0.10 & -0.11 & -0.12 \\
10  & -0.14 & -0.12 & -0.12 & -0.13 & -0.13 & -0.14 \\
20  & -0.06 & -0.06 & -0.09 & -0.17 & -0.11 & -0.09 \\
100 & -0.08 & -0.08 & -0.15 & -0.07 & -0.12 & -0.18 \\
\hline
\textbf{Avg.} & \textbf{-0.10} & \textbf{-0.09} & -0.12 & -0.12 & -0.12 & -0.13 \\
\hline
\multicolumn{7}{|c|}{\textbf{Sortino Ratio}} \\
\hline
5   & -0.29 & -0.54 & -0.73 & -0.59 & -0.67 & -0.73 \\
10  &  4.52 &  4.42 &  1.91 & -0.25 &  2.19 &  0.80 \\
20  & -0.36 & -0.48 & -0.58 &  0.11 & -0.23 & -0.50 \\
100 &  2.97 &  2.85 &  4.09 &  0.09 &  3.87 & -0.13 \\
\hline
\textbf{Avg.} & \textbf{1.21} & 1.06 & 1.17 & -0.16 & 1.29 & 0.11 \\
\hline
\multicolumn{7}{|c|}{\textbf{Omega Ratio}} \\
\hline
5   & 0.93 & 0.90 & 0.80 & 0.82 & 0.81 & 0.81 \\
10  & 2.84 & 2.62 & 1.53 & 0.91 & 1.69 & 1.20 \\
20  & 0.94 & 0.91 & 0.79 & 1.11 & 0.88 & 0.84 \\
100 & 1.44 & 1.40 & 1.65 & 1.07 & 1.79 & 0.98 \\
\hline
\textbf{Avg.} & \textbf{1.53} & 1.46 & 1.19 & 0.98 & 1.29 & 0.96 \\
\hline
\end{tabular}
}
\end{table}

\end{document}